\documentclass{jfm}

\usepackage{graphicx}
\usepackage{newtxtext}
\usepackage{newtxmath}
\usepackage{natbib}
\usepackage{hyperref}
\hypersetup{
    colorlinks = true,
    urlcolor   = blue,
    citecolor  = black,
}

\newcommand{\RomanNumeralCaps}[1]

\usepackage[caption=false]{subfig}
\usepackage{multirow}
\usepackage{cases}
\usepackage{booktabs}
\usepackage{multicol}
\usepackage{booktabs}
\usepackage{tabularx}
\usepackage{anyfontsize}

\usepackage{algorithm2e,epsfig,natbib,listings,color,cleveref}
\usepackage{subfig,float}
\usepackage{adjustbox}
\usepackage{xcolor}

\usepackage{color}
\definecolor{orange}{rgb}{1,0.5,0}
\definecolor{amethyst}{rgb}{0.6,0.4,0.8}
\definecolor{aureolin}{rgb}{0.99,0.93,0.0}
\definecolor{awesome}{rgb}{1.0,0.13,0.32}
\definecolor{ao-green}{rgb}{0.0, 0.5, 0.0}

\newcommand{\review}[1]{\textcolor{black}{#1}}

\title{Data-Augmented Resolvent Analysis of Wall-Bounded High-Pressure Transcritical Flow}

\author{Marc Bernades\aff{1,2}\corresp{\email{marc.bernades@upc.edu}}, Jonathan M.~O. Massey\aff{2}\corresp{\email{masseyj@stanford.edu}}, Beverley J. McKeon\aff{2}\corresp{\email{bjmckeon@stanford.edu}} \and Llu\'is Jofre\aff{1}\corresp{\email{lluis.jofre@upc.edu}}}

\affiliation{\aff{1}Department of Fluid Mechanics, Universitat Polit\`ecnica de Catalunya $\cdot$ BarcelonaTech (UPC), Barcelona 08019, Spain,
\aff{2}Department of Mechanical Engineering, Stanford University, Stanford 94305, CA, USA}

\begin{document}
\maketitle

\begin{abstract}
High-pressure transcritical fluid flows are central to modern energy and propulsion systems. A key challenge lies in confined configurations, where optimizing performance requires a deep understanding of the coupled hydrodynamic and thermodynamic nonlinearities that govern such flows. In this regard, low-order decomposition techniques, particularly resolvent analysis, \review{offer an interpretable linear input–output framework to identify and quantify dominant amplification mechanisms of coherent flow structures.}
This work, thus, pursues two main objectives: (i) establish a resolvent-based framework tailored to high-pressure transcritical fluid flows, and (ii) characterize the spatio-temporal sensitivity of the resolvent operator using data-driven turbulent base flows. These analyses identify the flow responses and forcings that optimally enhance mixing and heat transfer, along with their characteristic scales.
The results of the resolvent framework reveals that amplification is dominated by streamwise-elongated structures with spanwise periodicity, associated with peak singular values near (normalized) spanwise wavenumbers of order unity. Unlike ideal-gas or incompressible/isothermal flows, the dominant forcings arise from thermodynamic fluctuations in the pseudo-boiling region. Moreover, when the system \review{is linearized about the turbulent mean flow}, this results in intensified responses manifesting as coherent counter-rotating vortex pairs.
The energetic scale motions are constrained by the nature of the low-Reynolds-number regime \review{and non-isothermal flow conditions} considered, with a single dominant spectral mode reaching streamwise lengths comparable to those in the instantaneous fields. Data-driven analyses further reveal coherent structures propagating at phase speeds absent from \review{classical} incompressible wall-bounded turbulence. These structures are intensified near the pseudo-boiling region and constrained toward the hot wall, in \review{phenomenological} agreement with resolvent-mode predictions of near-wall scale motions.
\end{abstract}



\section{Introduction}  \label{sec:introduction}	

On the one hand, high-pressure transcritical fluids operate within thermodynamic regimes where supercritical gas-like and liquid-like states can be distinguished across the pseudo-boiling line~\citep{Simeoni2010-A,Banuti2015-A,Raju2017-A,Jofre2020-A,Jofre2021-A,Li2024-A}. This transition region is characterized by a pronounced nonlinear phase change, particularly within the pressure range $1 < P/P_c < 3$~\citep{Banuti2015-A}, where $P_c$ corresponds to critical pressure.
Unlike subcritical two-phase flows, single-component high-pressure transcritical fluids undergo a continuous (second-order) phase transition between gas-like and liquid-like states. This behavior arises due to the small Knudsen numbers (i.e., $K\!n \ll 1$) typical of supercritical fluid conditions, which result from significantly reduced mean free paths~\citep{Dahms2013-A}.
Supercritical fluids have long been utilized in thermo-fluid engineering applications, including liquid rocket engines, gas turbines, and supercritical water-cooled reactors~\citep{Yoo2013-A}. More recently, direct numerical (DNS) and large-eddy simulation (LES) studies have been employed to explore the potential of high-pressure transcritical fluids to generate microconfined turbulence~\citep{Bernades2022-A,Bernades2023-A,Bernades2024-A,Bernades2025-A}, which can significantly enhance mixing and heat transfer in microfluidic systems.
The flow physics in these regimes differs markedly from \review{classical} incompressible turbulent wall-bounded flows due to the large variations of thermophysical properties and the emergence of localized baroclinic torques. These torques lead to enhanced vorticity generation, destabilizing the flow, and transferring rotational energy across a broad range of scales via vortex stretching.
However, the underlying mechanisms governing these phenomena are still not fully comprehended neither systematically characterized.

On the other hand, a detailed understanding of the driving mechanisms in turbulent flows, particularly regarding the statistical scaling and structural coherence of their fluctuations and base flow, remains incomplete. Moving beyond the \review{classical} near-wall scaling hypothesis, recent studies have proposed that the nature of scale motions depends significantly on outer-scale variables~\citep{Mckeon2010-A}.
In this regard, instead of relying on complex models that attempt to predict the large degrees of freedom inherent in fluid dynamics, modal analysis techniques offer an efficient alternative by decomposing the flow field into a linear combination of modes. This approach effectively captures the dominant dynamical structures of the system.
Modal analysis methods can be broadly categorized into data-driven and operator-based techniques. Data-driven methods include low-order approaches such as proper orthogonal decomposition (POD), spectral POD (SPOD), and dynamic mode decomposition (DMD). In contrast, operator-based techniques derive from linearizations of the equations of fluid motion, including linear stability analysis, which relies on eigenvalue decompositions, and resolvent analysis accounting for the effect of nonlinear ``forcing'' terms.
With growing interest in using resolvent analysis to elucidate input-output relationships across various flow configurations, it has been widely applied to numerous internal and external flow problems. Examples include channel flows~\citep{Jovanovic2005-A,Moarref2013-A,Symon2021-A}, pipe flows~\citep{Mckeon2010-A}, boundary layers~\citep{Nogueira2021-A,Gomez2022-A}, airfoil wakes~\citep{Thomareis2018-A}, turbulent jets~\citep{Schmidt2018-A,Pickering2021-A}, and combustion applications~\citep{Skene2019-A}.
Fundamentally, resolvent analysis corresponds to the state-space system representation widely used in dynamical systems and control theory. It naturally provides a framework for designing active and passive flow control strategies, leveraging spatially or temporally periodic actuation to optimize performance metrics such as drag reduction, lift enhancement, mixing, or noise suppression. Moreover, resolvent analysis serves as a basis for estimating flow dynamics and identifying dominant coherent structures.
Thus, this methodology represents a valuable complement to both computational simulations and experimental investigations by offering profound insights into the fundamental physics and modeling of fluid flows. Furthermore, understanding the processes that sustain and amplify distinct turbulent length scales can improve comprehension of the laminar-to-turbulent transition, inform the selection of relevant length and time scales for flow control, and enhance the development of reduced-order models.

In general terms, the resolvent operator is defined as a linear transfer function that relates control inputs (namely, the nonlinear terms of the linearized equations, referred to as forcing) to their associated output responses through uniform Fourier transforms in both space and time. Specifically, the linear operator captures energy amplification, while the nonlinear forcing represents the mechanisms responsible for energy redistribution.
The singular value decomposition (SVD) of this operator reveals modes corresponding to both forcing and response, ranked by the magnitude of their singular values, which represent the operator’s linear gain. Consequently, the response can be reconstructed if the forcing terms are known. Even relying solely on the mean flow, several important features can be extracted; most notably, the gain of the leading mode, often described as low-rank behavior~\citep{Pickering2021-A}. This property facilitates a low-dimensional representation of turbulent flows~\citep{Moarref2013-A, Sharma2013-A, Mckeon2017-A}. For example, \citet{Bennedine2016-A} predicted the streamwise velocity spectra by utilizing the largest response mode amplitude, thereby reducing the discrepancy between measured and predicted profiles; however, caution is warranted when truncating energy contributions from the remaining modes~\citep{Morra2021-A}.
Moreover, \citet{Moarref2013-A} surveyed coherent turbulent motions evident in the energy spectra, such as near-wall streaks (NWS), large-scale motions (LSMs), and very-large-scale motions (VLSMs). In incompressible wall-bounded turbulence, NWSs typically consist of quasi-streamwise structures and counter-rotating vortices located approximately $15$ inner units above the wall, responsible for the highest turbulent kinetic energy (TKE) production. LSMs are characterized by hairpin vortices~\citep{Jimenez1998-A} and, particularly at low Reynolds numbers, extend from the wall to the boundary layer edge. \review{VLSMs are inertial layer structures with characteristic lengths of roughly $10-30$ outer units with pronounced velocity-fluctuation signatures in the logarithmic region \citep{Kim1999-A}}. Additionally, \citet{Saxton-Fox2017-A} developed a model consistent with the streamwise energy spectrum and amplification gains from the linearized equations of fluid motion, identifying instantaneous turbulent structures associated with different scale motions.

Coherent structures play a fundamental role in the dynamics of turbulent flows, influencing momentum transfer, energy cascades, and scalar transport across a wide range of scales~\citep{Robinson1991-A,Adrian2007-A}. In wall-bounded turbulence, these structures often organize into wave-like packets that propagate with characteristic phase speeds, driving key phenomena such as bursting, energy amplification, and large-scale modulation~\citep{Alamo2006-A,Smits2011-A}. Accurately capturing and analyzing these structures is especially critical in high-pressure transcritical fluid regimes, where turbulence is further complicated by strong nonlinear interactions arising from steep thermodynamic gradients, pseudo-boiling transitions, and non-ideal fluid behavior due to tightly coupled pressure, temperature, and density fields~\citep{Banuti2015-A,Oschwald2006-A}. These factors introduce new challenges to \review{classical} turbulence modeling and modal decomposition, motivating the development of more specialized analytical frameworks.
One effective strategy for extracting coherent structures involves applying Fourier transforms in both space and time, often combined with phase speed filtering to isolate traveling wave packets at specific convective velocities~\citep{Alamo2006-A}. This approach facilitates the study of wave-like motions and their amplification mechanisms. In particular, spectral filtering at fixed phase speeds enables the decomposition of the flow into dynamically relevant components without relying on strict statistical convergence.
Alternatively, data-driven methods such as DMD and SPOD have been widely employed to approximate linear operator behavior and extract dynamically significant modes from simulations or experiments~\citep{Schmid2010-A,Towne2018-A}. While DMD is often used as a surrogate for computing resolvent modes~\citep{Hermann2021-A}, its sensitivity to broadband content and lack of an explicit forcing; response framework can limit accuracy, particularly in compressible flows~\citep{Schmid2010-A,Symon2019-A} and high-pressure transcritical fluid regimes. In contrast, SPOD provides frequency-resolved orthogonal modes consistent with the resolvent operator under assumptions of stochastic forcing~\citep{Towne2018-A}, and has been successfully applied in both incompressible and compressible flows~\citep{Schmidt2018-A,Pickering2021-A}.
In this context, POD remains a widely used tool for identifying the most energetic modes in turbulence~\citep{Lumley1967-A}, particularly in complex multiphysics flows. For example, POD has been applied to study coherent structures in jets~\citep{Arndt1997-A}, channel flows~\citep{Sirovich1987-A}, and high-speed flows with significant thermodynamic coupling~\citep{Grinstein2007-B}. Although POD does not explicitly resolve frequency content, its strength in capturing spatial energy distributions makes it a valuable complement to spectral filtering techniques.

Although resolvent analysis has long been an attractive tool for flow modeling and control, research efforts have predominantly focused on incompressible flows. Recently, \citet{Rolandi2024-A} provided a comprehensive summary of a resolvent analysis framework tailored for compressible flows, detailing both physics-based and data-driven approaches to constructing and analyzing the resolvent operator.
The intrinsic limitations of \review{linear transient growth analysis} have motivated extensions of resolvent methods to more accurately capture the influence of nonlinear terms on flow responses. Specifically, transient growth approaches rely solely on identifying the initial disturbance that leads to maximum amplification~\citep{Bernades2025b-A,Bernades2024c-A}, effectively using these forcing quantities to predict the optimal input in resolvent analysis that maximizes the output response. 
\review{However, the resolvent framework typically assumes that the nonlinear forcing is white in time for each fixed set of spatial wavenumbers. While early formulations often adopt spatially uncorrelated forcing for simplicity, spatial correlations can in principle be incorporated through an appropriate forcing covariance. The assumption of temporally white forcing, together with limited knowledge of the forcing statistics, motivates complementary techniques based on flow statistics for improved accuracy~\citep{Zare2017-A}.}
Accordingly, the primary objective of this work is to develop a resolvent analysis framework for wall-bounded, high-pressure transcritical fluid flows driven by turbulent base states. The study focuses on a non-isothermal Poiseuille flow defined using a similar setup as a DNS of a high-pressure transcritical turbulent channel flow~\citep{Bernades2023-A}. Within this setup, the resolvent operator is characterized by analyzing gains, forcings, and response sensitivities relative to the ideal-gas case, with particular emphasis on the sensitivity of singular values in wavenumber space for a fixed frequency.
Furthermore, a detailed coherent structure analysis of the high-pressure transcritical turbulent channel flow is conducted using a hybrid approach. This approach combines Fourier decomposition in space and time with phase speed filtering to isolate coherent structures traveling at specific convective velocities~\citep{Alamo2006-A,Towne2018-A}, followed by variable-specific POD mode extraction. This filtering isolates wave-like structures that are otherwise obscured in the full flow field, enabling identification and characterization of traveling structures as well as assessment of their spatial organization and energy content. Subsequently, POD is applied to different thermodynamic and kinematic variables to localize the energetic content within these filtered modes. This systematic decomposition enables physically interpretable isolation of traveling wave components.
\review{It should be noted that the present study focuses on assessing the spatial organization of resolvent response modes relative to DNS-resolved structures, rather than providing a full quantitative validation against SPOD modes. While SPOD provides a natural framework for comparison (particularly given its connection to the white-in-time forcing assumption inherent in resolvent analysis) such a quantitative projection remains beyond the scope of this work. Instead, the investigation aims to establish that the resolvent framework, when applied to turbulent mean profiles, successfully captures the dominant wall-normal and streamwise structural features of high-pressure transcritical flows.}
Thus, the proposed methodology offers a physically grounded framework for mode extraction in high-pressure transcritical flows, where conventional methods often face challenges. The resulting insights advance understanding of coherent dynamics influenced by real-fluid effects and provide valuable guidance for reduced-order modeling in complex thermodynamic regimes.
\review{Finally, in the present work, the term ``data-augmented'' refers to the utilization of DNS data to validate and interpret structures predicted by the analytical resolvent, rather than to modify the operator itself. The resolvent framework identifies low-phase-speed structures residing between the hot wall and the pseudo-boiling region; a domain where coherence is typically absent in classical turbulent wall-bounded flows. Fourier-based spectral filtering and POD analysis are subsequently employed to verify the existence of these structures and to characterize their spatial correlation with the associated thermodynamic fields.}

To this end, the structure of this work is outlined as follows.
First, Section~\ref{sec:flow_modeling} introduces the flow physics modeling, covering the equations of fluid motion, real-fluid thermodynamics, and high-pressure transport coefficients.
Next, Section~\ref{sec:resolvent_framework} presents the resolvent analysis framework, detailing the linearized equations and normalization, the discretization method, the resolvent operator, and the SVD approach used to compute gains, responses, and forcing quantities.
In addition, data-driven methods (including spatio-temporal Fourier transforms and POD) are theoretically introduced in Section~\ref{sec:data_driven}.
Following, the setup and parameters of the computational experiments performed (non-isothermal Poiseuille flow and wall-bounded high-pressure transcritical turbulent flow) are described in Section~\ref{sec:computational_experiments}.
The results of the resolvent operator sensitivity and data-driven analyses are shown and discussed in Section~\ref{sec:results}. This section defines the resolvent operator for high-pressure transcritical fluid flows, followed by a sensitivity analysis and coherent structure examination. These results are compared and validated using spatio-temporal Fourier transform techniques to identify near-wall scale motions. Additionally, energetic modes obtained through POD are evaluated.
Finally, Section~\ref{sec:conclusions} offers concluding remarks and outlines future research directions arising from this work.

\section{Flow Physics Modeling}  \label{sec:flow_modeling}	

The framework utilized in terms of (i) equations of fluid motion, (ii) real-fluid thermodynamics, and (iii) high-pressure transport coefficients is detailed below.

\subsection{Equations of fluid motion}  \label{sec:compressible_eq}	

The flow motion of supercritical fluids is described by the following set of dimensionless equations of mass, momentum, and total energy.
\begin{align}
 \frac{\partial \rho^\star}{\partial t^\star} + \nabla^\star \cdot  \left( \rho^\star \mathbf{u}^\star \right) & = 0,    \label{eq:mass} \\
 \frac{\partial \left( \rho^\star \mathbf{u}^\star \right)}{\partial t^\star} + \nabla^\star \cdot  \left( \rho^\star \mathbf{u}^\star  \mathbf{u}^\star \right) & = -\nabla^\star P^\star + \frac{1}{{Re}} \nabla^\star \cdot \boldsymbol{\tau^\star} + \mathbf{F}^\star,  \label{eq:momentum} \\
 \frac{\partial \left( \rho^\star E^\star \right) }{\partial t^\star} + \nabla^\star \cdot  \left( \rho^\star \mathbf{u}^\star E^\star  \right) & = - \frac{1}{{Re} {Br}} \nabla^\star \cdot \mathbf{q}^\star -\nabla^\star \cdot  (P^\star \mathbf{u}^\star)\label{eq:energytransport} \\
 & \quad + \frac{1}{{Re}}\nabla^\star \cdot (\boldsymbol{\tau^\star} \cdot \mathbf{u}^\star) + \mathbf{F}^\star \cdot \mathbf{u}^\star, \nonumber 
\end{align}
where superscript $(\cdot)^\star$ denotes dimensionless quantities, and the corresponding dimensional variables are the density $\rho$, the velocity vector $\mathbf{u}$, the pressure $P$, the viscous stress tensor $\boldsymbol{\tau} = \mu ( \nabla \mathbf{u} + \nabla \mathbf{u}^{T} ) + \lambda(\nabla \cdot \mathbf{u})\mathbf{I}$ with $\mu$ the dynamic viscosity and $\lambda = - ( 2\mu/3 )$ the bulk viscosity, the total energy $E = e + |\mathbf{u}|^{2}/2$ with $e$ the internal energy, the Fourier heat flux $\mathbf{q} = - {\kappa} \nabla T$ with $\kappa$ the thermal conductivity and $T$ the temperature, and \review{$\mathbf{F}$} a body force to drive the flow in the streamwise direction.

The resulting set of scaled equations includes two dimensionless numbers: (i) the Reynolds number $Re = \rho_{b}U_{r}\delta/\mu_{b}$ characterizing the ratio between inertial and viscous forces, where subscript $(\cdot)_b$ refers to bulk quantities, $\delta$ is the channel half-height, and $U_r$ is a reference velocity, defined as the centerline streamwise velocity for the isothermal laminar base flow, or as the bulk velocity for the turbulent flow obtained from DNS results; and (ii) the Brinkman number $Br = \mu_b U_{r}^{2} / (\kappa_b T_b) = Pr Ec$ quantifying the ratio of viscous heat generation to external heating through the walls.
The Brinkman number can also be expressed as the combination of Prandtl number $Pr =\mu_{b} {c_{P}}_{b} / \kappa_{b}$ expressing the ratio between momentum and thermal diffusivity, where $c_{P}$ is the isobaric specific heat capacity, and Eckert number $Ec = U_{r}^{2}/({c_{P}}_{b} T_{b})$, which accounts for the ratio between advective mass transfer and heat dissipation potential.
The set of dimensionless equations is based on the following set of inertial scalings~\citep{Jofre2020b-A,Jofre2023-A}
\begin{align}
 \textbf{x}^{\star} & = \frac{\textbf{x}}{\delta}, \quad \textbf{u}^{\star} = \frac{\textbf{u}}{U_{r}}, \quad \rho^{\star} = \frac{\rho}{\rho_{b}}, \quad T^{\star} = \frac{T}{T_{b}}, \quad P^{\star} = \frac{P}{\rho_{b} U_{r}^{2}}, \nonumber \\ 
 E^{\star} & = \frac{E}{U_{r}^{2}}, \quad \mu^{\star} = \frac{\mu}{\mu_{b}}, \quad \kappa^{\star} = \frac{\kappa}{\kappa_{b}}, \quad \boldsymbol{F}^{\star} = \frac{U_{r} {\rho_b}^2 \boldsymbol{F}}{\delta}, \quad t^{\star} = \frac{\delta}{U_r}, \label{eq:scalings}
\end{align}
where $\textbf{x}$ represents the position vector.

\subsection{Real-fluid thermodynamics}  \label{sec:Peng_Robinson}	

The thermodynamic space of solutions for the state variables pressure, temperature, and density is described by an equation of state.
At supercritical pressures, a rapid transition from supercritical liquid-like to gas-like behavior occurs at the pseudo-boiling temperature $T_{pb}$, which corresponds to the point where a second-order (continuous) phase transition occurs characterized by a peak of $c_P$.
Therefore, a popular choice for systems at high pressures is the \cite{Peng1976-A} equation of state.
In general form, the relation can be expressed in terms of the compressibility factor $Z = P/(\rho R^{\prime} T)$, where $R^{\prime}$ is the specific gas constant, which in dimensionless form reads
\begin{equation}
  P^{\star} = \frac{Z \rho^{\star} T^{\star}}{\hat{\gamma}_b M\!a^{2}},    \label{ch8:eq:real-gas_state} 
\end{equation}
where $\hat{\gamma} \approx Z (c_{P} / c_{V}) [(Z + T (\partial Z / \partial T)_\rho)/(Z+T(\partial Z / \partial T)_P)]$ is an approximated real-fluid heat capacity ratio~\citep{Firoozabadi2016-B} with $c_{V}$ the isochoric specific heat capacity.
\review{The formulation introduces the dimensionless Mach number $M\!a = U_r / c_b$}, where $c_b$ is the bulk speed of sound, \review{representing the ratio between the characteristic flow velocity and the acoustic propagation speed}.
In addition, real-fluid thermodynamics need to be supplemented with the corresponding high-pressure thermodynamic variables (e.g., internal energy, heat capacities) based on departure functions~\citep{Reynolds2019-B} calculated as a difference between two states.
These functions operate by transforming the thermodynamic variables from ideal-gas conditions (low pressure - only temperature dependent) to supercritical conditions (high pressure).
The ideal-gas components are calculated by means of the NASA 7-coefficient polynomial~\citep{Burcat2005-TR}, while the analytical departure expressions to high pressures are derived from the Peng-Robinson equation of state as detailed, for example, in~\citet{Jofre2021-A}.

However, within the resolvent analysis framework, empirically based models are preferred due to the sensitivity of modal analysis to nonlinear thermodynamic effects~\citep{Bernades2025b-A}. Notable examples include the NIST~\citep{NIST-M} and the open-source CoolProp~\citep{Bell2014-A} libraries, which offer accurate representations of the non-ideal thermodynamic behavior characteristic of high-pressure transcritical fluids.
Thermodynamic properties in these libraries are derived from the Helmholtz free energy equation of state. The accuracy of CoolProp has been benchmarked against NIST reference data~\citep{NIST-M}, and its influence on linear stability results has been evaluated by~\citet{Bernades2025b-A}.
Accordingly, this study employs the CoolProp library for the resolvent analysis, while the Peng–Robinson equation of state is used for the data-driven analysis to maintain consistency with the model adopted in the DNS computations.

\subsection{High-pressure transport coefficients}  \label{sec:transport_coefficients}

The high pressures considered in this work preclude the use of simplified relations for calculating dynamic viscosity and thermal conductivity. Standard methods for Newtonian fluids at high pressures typically rely on empirical correlations proposed by~\citet{Chung1984-A,Chung1988-A}, which depend on properties such as the critical temperature $T_c$, critical density $\rho_c$, molecular weight $W$, acentric factor $\omega$, association factor $\kappa_a$, and dipole moment $\mathcal{M}$, along with thermophysical data from the NASA 7-coefficient polynomial~\citep{Burcat2005-TR}. Additional details can be found in works such as \citet{Jofre2021-A} and \citet{Poling2001-B}.
However, consistent with the treatment of thermodynamic quantities, transport properties in the resolvent analysis framework are modeled using the open-source CoolProp library~\citep{Bell2014-A}, while the empirical correlations by Chung \textit{et al.} are adopted for the data-driven analyses. Comparative assessments of transport models (specifically those from Chung \textit{et al.}, NIST, and CoolProp) have been thoroughly discussed by \citet{Bernades2025b-A}.

\section{Resolvent analysis framework}  \label{sec:resolvent_framework}	

The objective of this section is to outline the resolvent operator. First, the linearized equations and linear theory operators are presented based on previous modal stability studies~\citep{Bernades2025b-A,Bernades2024c-A} in conjunction with the discretized Chebyshev method. Consequently, the resolvent operator and its weighted normalization are elaborated upon. Lastly, the SVD approach is introduced to define the gains, forcings (inputs) and responses (outputs) of the transfer function.

\subsection{Linearized equations}

The flow field can be decomposed into the base state and the fluctuation [hereafter denoted with subscript $(\cdot)_0$ and superscript $(\cdot)^\prime$, respectively] as
\begin{equation}{}
  \mathbf{q} = \mathbf{q}_0 + \mathbf{q}^\prime, \label{eq:Decomposition}
\end{equation}
where vector $\mathbf{q}$ is composed of the velocity components ($u$, $v$ and $w$) and the independent thermodynamic variables ($\rho$ and $T$).
This decomposition yields a perturbation vector $\mathbf{q}^\prime = (\rho^\prime, u^\prime, v^\prime, w^\prime, T^\prime)^T$ whose corresponding steady-state flow vector is assumed to be parallel to the walls, and consequently only function of the wall-normal direction $y$ as $\mathbf{q}_0 = [\rho_0(y), u_0(y), 0, 0, T_0(y)]^T$. From this point forward, superscript $(\cdot)^\star$ is omitted and assumed that all equations and variables are in dimensionless form for the brevity of exposition.

The selection of the two main thermodynamic variables $\rho$ and $T$ enforces the perturbations of the remaining thermophysical variables ($E, P, \mu, \kappa)$ to be expressed as a function of this pair of quantities by means of a Taylor expansion with respect to the base flow.
For example, by neglecting higher-order terms, the pressure perturbation can be approximated by expanding the base flow pressure as
\begin{equation}
P^\prime = \left.\frac{\partial P_0}{\partial \rho_0}\right\vert_{T_0} \rho^\prime + \left.\frac{\partial P_0}{\partial T_0}\right\vert_{\rho_0} T^\prime + \mathcal{O}(\epsilon), 
\end{equation}
with the resulting error of the approximation to be of the order of the amplitude ($\epsilon$) of the infinitesimal perturbations~\citep{Alves2016-A}.
Thus, by substituting Eq.~\ref{eq:Decomposition} into the dimensionless equations of fluid motion (Eqs.~\ref{eq:mass}-\ref{eq:energytransport}), linear stability equations are derived as a function of the perturbation vector $\mathbf{q}^\prime$; extensive derivation of the equations is covered in~\cite{Bernades2025b-A}.
They can be cast in compact form as
\begin{align}
     \mathbf{L_t} \frac{\partial\mathbf{q}^\prime}{{\partial t}} & + \mathbf{L_x} \frac{\partial\mathbf{q}^\prime}{{\partial x}} + \mathbf{L_y} \frac{\partial\mathbf{q}^\prime}{{\partial y}} + \mathbf{L_z} \frac{\partial\mathbf{q}^\prime}{{\partial z}} +  \mathbf{L_q} \mathbf{q}^\prime \nonumber \\
     & + \mathbf{V_{xx}} \frac{\partial^2 \mathbf{q}^\prime}{{\partial x^2}} +  \mathbf{V_{yy}} \frac{\partial^2 \mathbf{q}^\prime}{{\partial y^2}} +  \mathbf{V_{zz}} \frac{\partial^2 \mathbf{q}^\prime}{{\partial z^2}} + 
      \mathbf{V_{xy}} \frac{\partial^2 \mathbf{q}^\prime}{{\partial x \partial y}} + \mathbf{V_{xz}} \frac{\partial^2 \mathbf{q}^\prime}{{\partial x \partial z}} + \mathbf{V_{yz}} \frac{\partial^2 \mathbf{q}^\prime}{{\partial y \partial z}} = \mathbf{f}, \label{eq:LST}
\end{align}
where the nonlinear terms not presented in Eq.~\ref{eq:LST} (i.e., partial derivatives of multiple perturbation variables) have been grouped into $\mathbf{f} = (f_\rho, f_u, f_v, f_w, f_T)^T$, and $\mathbf{L_t}$, $\mathbf{L_x}$, $\mathbf{L_y}$, $\mathbf{L_z}$, $\mathbf{L_q}$, $\mathbf{V_{xx}}$, $\mathbf{V_{yy}}$, $\mathbf{V_{zz}}$, $\mathbf{V_{xy}}$, $\mathbf{V_{xz}}$, and $\mathbf{V_{yz}}$ correspond to the Jacobian matrices of the base flow and thermophysical properties.
In detail, the Jacobian matrices are of size $5\times5$, corresponding to each field of the perturbation vector $\mathbf{q}^\prime$. 
The particular form of these matrices are detailed in related works~\citep{Bernades2025b-A,Bernades2024-B}, which are obtained by inspection of the linearized equations derived.

\subsection{Discretization method} \label{sec:LST_discretization}

The discretization of the linearized equations is based on Chebyshev collocation~\citep{Trefethen2000-B} with a domain spanning the interval $0 \leq y/\delta \leq 2$ and discretized as
\begin{equation}
    y_j = \delta \left( 1 - \textrm{cos} \frac{\pi j}{N} \right), \quad j = 0, \dots , N,
\end{equation}
where $N$ corresponds to the total number of collocation points.
In this regard, Chebyshev differentiation matrices are utilized to obtain the discretized equations and define the operators of the linear stability eigenvalue problem.
Particularly, the mesh size selected for this work is $N = 200$, which provides grid-independent results based on the convergence of the S-shaped Mack branches (viz. eigenspectrum). The error of the results scales with $\mathcal{O}(h^2)$ where $h = H / N$ and $H = 2 \delta$ is the channel height. In particular, further increasing the grid size by $50\%$ improves the accuracy by $\mathcal{O}(10^{-4})$; for brevity of exposition, the corresponding grid-convergence results are not shown in this paper and are covered in~\citet{Bernades2025b-A}.
Moreover, the system of equations is subjected to $u^\prime = v^\prime  = w^\prime  = T^\prime = 0$ boundary conditions for both walls.

\subsection{Resolvent operator}

By assuming a homogeneous mean flow in streamwise and spanwise directions and time, Fourier decomposition is applied to the perturbation vector
\begin{equation}
    \textcolor{red}{\mathbf{q}(x,y,z,t) = \int_{-\infty}^{\infty} \int_{-\infty}^{\infty} \int_{-\infty}^{\infty} \hat{\mathbf{q}}(y; k_x, k_z, \omega) e^{i(\kappa_x x + \kappa_z z - \omega t)} dk_x dk_z d\omega}, \label{eq:Fourier_decomp}
\end{equation}
\review{where $\hat{\mathbf{q}}(y; k_x, k_z, \omega)$ denotes the Fourier coefficient. This representation corresponds to the superposition of all spectral components in the flow}. \review{For the analysis of an individual Fourier mode at a specific wavenumber–frequency combination, the perturbation reduces to the normal-mode form as}
\begin{equation}
    \textcolor{red}{\mathbf{q}(x,y,z,t) = \hat{\mathbf{q}}(y) e^{(i \kappa_x x + i \kappa_z z - i \omega t)}}, \label{eq:LST_normalmode}
\end{equation}
where $\kappa_x$ and $\kappa_z$ are the prescribed streamwise and spanwise wavenumbers, respectively, $\omega$ is the normalized temporal frequency [which can be written in terms of normalized phase speed $c$ as $\omega = c \cdot \kappa_x$~\citep{Moarref2013-A,Mckeon2019-A}] whose real and imaginary parts correspond, respectively, to the wall-normal angular frequency and its local growth rate; the superindex $\hat{(\cdot)}$ denotes Fourier transform.
Hence, by substituting Eq.~\ref{eq:LST_normalmode} into Eq.~\ref{eq:LST}, the following response equation is obtained
\begin{align}
    (- i \omega \mathbf{L_t} & + i \kappa_x \mathbf{L_x} + \mathbf{L_y} D + i \kappa_z \mathbf{L_z} + \mathbf{L_q} \nonumber \\
    & - \kappa_x^2 \mathbf{V_{xx}} + i \kappa_x  \mathbf{V_{xy}} D - \kappa_x\kappa_z \mathbf{V_{xz}} +  \mathbf{V_{yy}} D^2 + i \kappa_z  \mathbf{V_{yz}} D - \kappa_z^2  \mathbf{V_{zz}})\hat{\mathbf{q}} =  \hat{f}, \label{eq:discrete_linear_equation}
\end{align}
where $D = d/dy$ is the derivative operator of the Chebyshev discretization.
Typically, in linear stability theory (LST), Eq.~\ref{eq:discrete_linear_equation} can be solved either in the temporal \review{(prescribed $\omega$)} or spatial \review{(prescribed $\kappa_x$ and $\kappa_z$)} domain, where $\hat{f} = 0$.
However, in resolvent analysis, there is no distinction, and the operator is built in the same manner for both cases.
In particular, the transport equations presented above have been rearranged \review{so that the nonlinear terms, which are neglected in the linearized system, appear as} a forcing $\mathbf{f}$, allowing for the derivation of the transfer function resolvent operator ($\mathbf{H}$) for the flow in input-output format.
This forcing term represents the nonlinear terms of the equations.
Additionally, $\mathbf{q}^\prime$ is considered homogeneous in the streamwise and spanwise directions, and in time for the Fourier transforms, leading to the general expression form~\citep{Madhusudanan2022-A}
\begin{equation}
 \mathbf{\hat{q}} = \underbrace{[ -i \omega \mathbf{B}(k_x, k_z,\omega) +  \mathbf{A}(k_x, k_z,\omega) ]^{-1}}_{\mathbf{H}(k_x, k_z,\omega)} \mathbf{\hat{f}}, \label{eq:dynamic_system_pof}
\end{equation}
where $\mathbf{A}(k_x, k_z,\omega)$ $\in \mathbb{C}^{5Nx5N}$ and $\mathbf{B}(k_x, k_z,\omega)$ $\in \mathbb{C}^{5Nx5N}$ are the linear operators, and the transfer kernel $\mathbf{H}$($k_x, k_z, \omega$) $\in \mathbb{C}^{5Nx5N}$ is the resolvent operator of the flow that maps the nonlinear terms $\mathbf{\hat{f}}$ with the state variables $\mathbf{\hat{q}}$; viz. it is the spatio-temporal frequency response of the system.
Matrix $\mathbf{A}$ contains the finite-dimensional approximations of the linearized equations, which can be obtained by inspection of Eq.~\ref{eq:discrete_linear_equation} to map the temporal partial derivatives of Eq.~\ref{eq:dynamic_system_pof} and $\mathbf{B}$ in the form
\begin{numcases}{}
 {\mathbf{A}} = \thinspace i \kappa_x \mathbf{L_x} + \mathbf{L_y} D + i \kappa_z \mathbf{L_z} + \mathbf{L_q} \nonumber \\
 \quad \quad - \kappa_x^2 \mathbf{V_{xx}} + i \kappa_x  \mathbf{V_{xy}} D - \kappa_x\kappa_z \mathbf{V_{xz}} +  \mathbf{V_{yy}} D^2 + i \kappa_z  \mathbf{V_{yz}} D - \kappa_z^2  \mathbf{V_{zz}}, \\ \label{eq:linear_operator}
 \mathbf{B} = \mathbf{L_t}, \nonumber
\end{numcases}
where the linear operator $\mathbf{L_t}$ defining $\mathbf{B}$ reads 
\begin{equation}
\mathbf{L_t} = 
\begin{pmatrix}
1   & 0      & \cdots & \cdots & 0 \\
u_0 & \rho_0 & \ddots &  & \vdots \\
0   & 0      & \rho_0 & \ddots & \vdots  \\
0   & \vdots & \ddots    & \rho_0 & 0  \\
e_0 + \rho_0 \frac{\partial e_0}{\partial T_0} & 0 & \cdots & 0 & \rho_0 \frac{\partial e_0}{\partial T_0}
\end{pmatrix}.
\end{equation} \label{eq:B_L_t}

Before performing the resolvent analysis, a norm needs to be selected. It is known that for compressible flows the Chu norm~\citep{Chu1965-A,Hanifi1996-A} is a well-suited candidate~\citep{Ren2019b-A} defined as
\begin{equation}
    \mathcal{E}(q) = \int \left[  ({u^\prime}^\dagger {u^\prime} + {v^\prime}^\dagger {v^\prime} + {w^\prime}^\dagger {w^\prime}) + m_\rho {\rho^\prime}^\dagger {\rho^\prime} + m_T {T^\prime}^\dagger {T^\prime} \right]  dV, \label{eq:E_norm}
\end{equation}
where $(\cdot)^\dagger$ denotes the complex conjugate. For ideal-gas thermodynamics, selecting the Mack's energy norm with $m_\rho = T_0 / (\rho_0^2 \gamma {M\!a}^2)$ and $m_T = 1/[\gamma(\gamma - 1 )T_0 {M\!a}^2]$ is commonly employed.
However, for supercritical fluid flows the results are robust when $m_\rho = m_T = 1$~\citep{Ren2019b-A}.
Hence, the Chu norm is incorporated with a weight matrix $\mathbf{W}$ giving the discrete inner product \review{$\langle\mathbf{\hat{q}_1},\mathbf{\hat{q}_2}\rangle = \mathbf{\hat{q}_1}^\dagger \mathbf{W}  \mathbf{\hat{q}_2}^\dagger$}, assuming that the response and forcing weight matrices are equivalent~\citep{Gomez2022-A}; i.e., \review{$\langle\mathbf{\hat{f}_1},\mathbf{\hat{f}_2}\rangle = \mathbf{\hat{f}_1}^\dagger \mathbf{W}  \mathbf{\hat{f}_2}^\dagger$}. Thus, Eq.~\ref{eq:dynamic_system_pof} can be rewritten as
\begin{equation}
 \mathbf{\hat{q}} = \underbrace{\left[ \mathbf{W}^{1/2} \thinspace \mathbf{H}(k_x, k_z,\omega) \thinspace \mathbf{W}^{-1/2} \right]}_{\mathbf{H_w}(k_x, k_z,\omega)} \mathbf{\hat{f}}, \label{eq:ARA_input_output}
\end{equation}
where ${\mathbf{H_w}(k_x, k_z,\omega)}$ is the normalized resolvent operator.
Typically, the numerical integration weights are chosen to be equal, and based on the state variable vector $\mathbf{\hat{q}}$ and the selected norm, the matrix $\mathbf{W}$ can be expressed as
\begin{equation}
    \mathbf{W} = \textrm{diag}\left(m_\rho, 1, 1, 1, m_T \right). \label{eq:weight_matrix}
\end{equation}

\subsection{Singular value decomposition} \label{sec:SVD}

A SVD is utilized to analyze the resolvent operator $\mathbf{H_w}(k_x, k_z,\omega)$. This technique identifies the inputs and their corresponding outputs. It also offers a low-rank approximation of the input-output dynamics of the entire system; high-gain inputs are useful for flow control applications as they indicate where the actuation produces the largest impact on the flow~\citep{Taira2017-A,Pickering2021-A}.
In detail, the SVD of the resolvent operator can be expressed as
\begin{equation}
    \mathbf{H_w}(k_x, k_z,\omega) = \sum_{i=1}^{5N} \Psi_i(y) \sigma_i \phi_i(y), \label{eq:operator_SVD}
\end{equation}
where $\sigma_i$ are the singular values arranged in descending order, $\psi_i$ are the singular vectors of the resolvent response modes, and $\phi_i$ are the singular vectors of the forcing modes.
Consequently, a forcing to the resolvent operator along $\phi_i$ gives a response along $\Psi_i$ amplified by a factor $\sigma_i$. Based on the ordering criteria, the most sensitive forcing direction corresponds to $i = 1$.

Therefore, based on this operator, for a particular frequency $\omega$, the maximum amplification $\mathcal{G}$ of any forcing is determined as~\citep{Mckeon2017-A}
\begin{equation}
    \mathcal{G} = \textrm{max} \thinspace \frac{\mathbf{\hat{q}}^\dagger \mathbf{W}  \mathbf{\hat{q}}^\dagger}{\mathbf{\hat{f}}^\dagger \mathbf{W}  \mathbf{\hat{f}}^\dagger}.
\end{equation}
\review{For a harmonic forcing of the form $\mathbf{\hat{f}}(y,t) = \mathbf{\hat{f}}(y) e^{-i \omega t}$, and under the assumption of linear stability, the flow response asymptotically becomes harmonic, which can be expressed as $\mathbf{\hat{q}}(y,t) = \mathbf{\hat{q}}(y) e^{-i \omega t}$.}

\section{Data-driven methods}  \label{sec:data_driven}	

This section presents the mathematical framework underlying the data-driven methods employed in this work. A total of $2000$ flow field snapshots were collected at equidistant time intervals; representing a trade-off between computational cost and statistical convergence, consistent with prior studies~\citep{Towne2018-A,Berkooz1993-A}. This sampling captures approximately $7.5$ flow-through time units, defined as $t_c = L_x / u_b$.
\review{The time interval between DNS snapshots is $\Delta t^+ = \Delta t \thinspace u_\tau / \nu = 0.375$, which corresponds to a fraction of the eddy turnover time $\Delta t \thinspace u_\tau / \delta = 0.00375$. Therefore, the total recorded flow time span ensures adequate temporal resolution and statistical sampling of the dominant coherent structures.}
First, the spatio-temporal fast Fourier transform approach is described. Due to memory constraints, it is applied to two-dimensional ($x$–$y$) slices of the domain. Subsequently, the POD framework is introduced, in which the energetic modes are extracted in three dimensions.

\subsection{Spatio-temporal Fourier transform}

To isolate coherent structures convecting at a specific phase speed, a spatio-temporal \review{Fourier transform} of the velocity field is employed. This approach is motivated by the challenging nuances encountered when trying to reconstruct a data-driven resolvent operator using the method proposed by~\citet{Hermann2021-A}; developing a framework suitable for such nonlinear flows is left for future work.
Accordingly, the conventional \review{Fourier transform} is briefly described. Consider a scalar component \( u(x, y, z, t) \) defined over a time interval \( [0, T] \) and a spatial domain \( \Omega = [0, L_x] \times [0, L_z] \), corresponding to the homogeneous directions \( x \) and \( z \). The \review{standard} Fourier transform is then defined as
\begin{equation}
\hat{u}(\kappa_x, y, \kappa_z, \omega) = 
\frac{1}{T L_x L_z}
\int_0^T \int_0^{L_x} \int_0^{L_z}
u(x, y, z, t) \,
e^{-i (\kappa_x x + \kappa_z z - \omega t)} 
\, d{z} d{x} d{t},
\label{eq:fft}
\end{equation}
where \( \kappa_x \) and \( \kappa_z \) are the streamwise and spanwise wavenumbers, and \( \omega \) is the angular frequency.
Therefore, in \review{Fourier} space, the phase speed can be masked by computing the local phase speed for each spectral component, given by the following expression
\begin{equation}
c(\kappa_x, \kappa_z, \omega) = 
\frac{\omega}{\sqrt{\kappa_x^2 + \kappa_z^2}}.
\end{equation}
To isolate components convecting within a target phase speed band \( c \in [c_{\min}, c_{\max}] \), a spectral mask is applied in the form
\begin{equation}
\mathcal{M}(\kappa_x, \kappa_z, \omega) =
\begin{cases}
1, & \text{if } c_{\min} \leq \dfrac{\omega}{\kappa} \leq c_{\max}, \quad \kappa \neq 0, \\
0, & \text{otherwise},
\end{cases}
\end{equation}
with \( \kappa = \sqrt{\kappa_x^2 + \kappa_z^2} \). The filtered field in spectral space then becomes
\begin{equation}
\hat{u}_{\text{filtered}}(\kappa_x, y, \kappa_z, \omega) = 
\mathcal{M}(\kappa_x, \kappa_z, \omega) \cdot \hat{u}(\kappa_x, y, \kappa_z, \omega).
\end{equation}
Finally, the inverse Fourier transform reconstructs the filtered signal in physical space as
\begin{equation}
u_{\text{filtered}}(x, y, z, t) = 
\textcolor{red}{\frac{1}{(2 \pi)^3} \int_{0}^{\frac{2 \pi} {L_x}} \int_{0}^{\frac{2 \pi} {L_z}} \int_{0}^{\frac{2 \pi} {T}}} 
\hat{u}_{\text{filtered}}(\kappa_x, y, \kappa_z, \omega)
\, e^{i (\kappa_x x + \kappa_z z - \omega t)} \textcolor{red}{d{\kappa_x} d{\kappa_z} d{\omega}}.
\end{equation}
This operation extracts spatio-temporal structures associated with the prescribed phase speed range, allowing further analysis such as modal decomposition or conditional averaging.
In practice, the spatial Fourier transforms were performed along the homogeneous directions with wavenumbers defined from the discrete Fourier frequencies. The temporal transform was applied to the complete sequence of snapshots over the interval $[0,T]$, without additional windowing (i.e., employing a rectangular window). 
This procedure extracts spatio-temporal structures associated with the prescribed phase-speed range, allowing further analysis such as modal decomposition or conditional averaging.

\subsection{Proper orthogonal decomposition}

POD has become a fundamental tool for extracting coherent structures in complex fluid flows~\citep{Lumley1967-A,Berkooz1993-A}. The method relies on matrix diagonalization to decompose the flow field into a linear combination of time-dependent coefficients and space-dependent eigenvectors.
In this work, a POD-based analysis is applied to high-pressure transcritical wall-bounded flow with the aim of identifying dominant energetic regions within the domain and investigating their interaction with the pseudo-boiling region. Specifically, the flow field is projected using the energy normalization matrix defined in Eq.~\ref{eq:E_norm}, which incorporates energy-weighted information across all relevant physical variables.
POD can be performed using either eigenvalue decomposition or SVD. The former is more general and numerically robust, making it the preferred method when the full snapshot matrix is available. In contrast, SVD is typically employed in the ``method of snapshots'', which is more computationally efficient when the number of snapshots is significantly smaller than the size of the state vector.

To extract energetically significant modes, an inner product that reflects the physical energy of the system is defined for two states \( \mathbf{q}_i \) and \( \mathbf{q}_j \). In detail, the energy-weighted inner product reads
\begin{equation}
\label{eq:energy_inner_product}
\langle \mathbf{q}_i, \mathbf{q}_j \rangle_\mathcal{E}
= \int_\Omega \mathbf{q}_i^\top(\mathbf{x}) \, \mathbf{W}(\mathbf{x}) \, \mathbf{q}_j(\mathbf{x}) \, d{\mathbf{x}},
\end{equation}
where \( \mathbf{W}(\mathbf{x}) \in \mathbb{R}^{5 \times 5} \) is a positive symmetric weighting matrix that reflects local kinetic and thermal energy, whose definition is presented in Eq.~\ref{eq:weight_matrix} with $( m_{\rho}, m_u, m_v, m_w, m_T )$ non-dimensional unitary scaling parameters resulting from the energy norm defined in Eq.~\ref{eq:E_norm}.

The definition of the snapshot matrix is constructed for the number of snapshots $M$ as
\begin{equation}
\label{eq:snapshot_matrix}
\mathbf{Q} =
\begin{bmatrix}
\mathbf{q}^1 & \mathbf{q}^2 & \cdots & \mathbf{q}^M
\end{bmatrix}
\in \mathbb{R}^{N \times M},
\end{equation}
where each column \( \mathbf{q}^k \in \mathbb{R}^N \) is the discretized state vector at time \( t_k \), and \( N = 5 \times N_x \times N_y \times N_z \) with \( N \) the number of spatial grid points in each spatial direction.
Therefore, the energy-weighted covariance matrix is then expressed as
\begin{equation}
\label{eq:covariance_matrix}
\mathbf{C} = \mathbf{Q}^\top \mathbf{W} \mathbf{Q} \in \mathbb{R}^{M \times M}.
\end{equation}
Based on these matrices, the eigenvalue problem is solved to obtain the POD modes defined as
\begin{equation}
\label{eq:eigenproblem}
\mathbf{C} \, \mathbf{a}_k = \lambda_k \, \mathbf{a}_k,
\end{equation}
where \( \lambda_k \) is the energy associated with the \( k \)-th mode, and \( \mathbf{a}_k \in \mathbb{R}^M \) is the corresponding time coefficient vector.
The spatial POD modes \( \boldsymbol{\phi}_k \in \mathbb{R}^N \) are then recovered in the form
\begin{equation}
\label{eq:mode_reconstruction}
\boldsymbol{\phi}_k = \frac{1}{\sqrt{\lambda_k}} \, \mathbf{Q} \, \mathbf{a}_k.
\end{equation}
These modes are orthonormal with respect to the energy inner product defined in Eq.~\ref{eq:energy_inner_product}, and they capture the most energetic structures of the flow field.
For completeness, the low-rank approximation of the POD operator is evaluated in Appendix~\ref{sec:Appendix_C}. Based on the datasets analyzed, it is found that $k = 379$ modes are required to capture $95\%$ of the total energy. This result is further validated through a flow field reconstruction using $450$ modes.
All POD modes presented in this work follow this decomposition described and were computed under the same flow conditions (DNS dataset described in Section~\ref{sec:computational_experiments}).

\section{Computational experiments} \label{sec:computational_experiments}

The resolvent operator is first examined using a non-isothermal Poiseuille flow under similar working conditions as the high-pressure transcritical turbulent channel flow investigated by~\citet{Bernades2023-A}. Subsequently, the corresponding DNS data are used for the data-driven analyses.
Accordingly, this section describes both: (i) the non-isothermal Poiseuille flow configuration, and (ii) the DNS setup, from which the turbulent mean (base) flow is extracted.

\subsection{Non-isothermal Poiseuille flow}

\begin{figure*}
	\centering
    \subfloat[\vspace{-8mm}]{\includegraphics[width=0.325\linewidth]{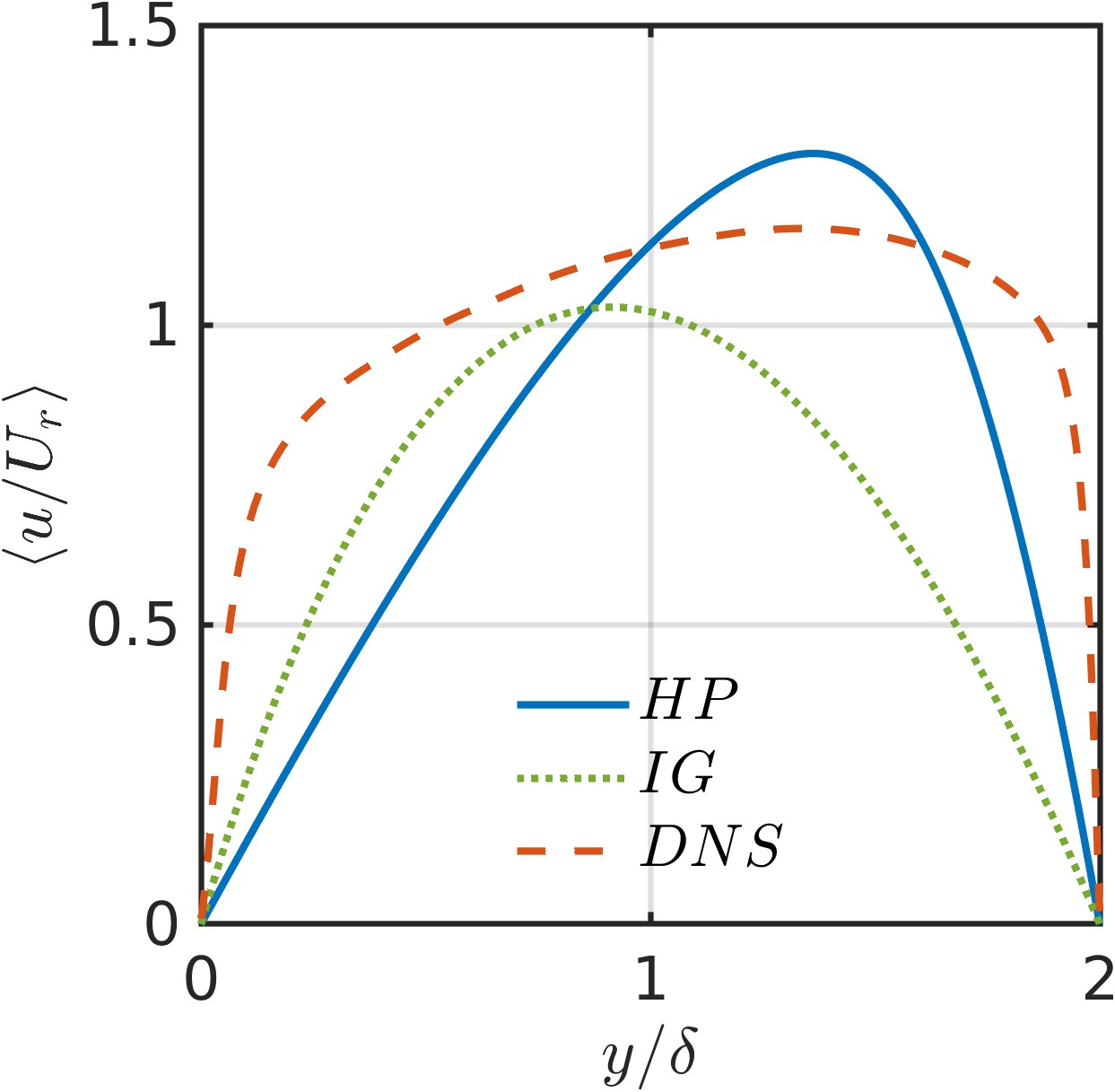}} \hfill
    \subfloat[\vspace{-8mm}]{\includegraphics[width=0.332\linewidth]{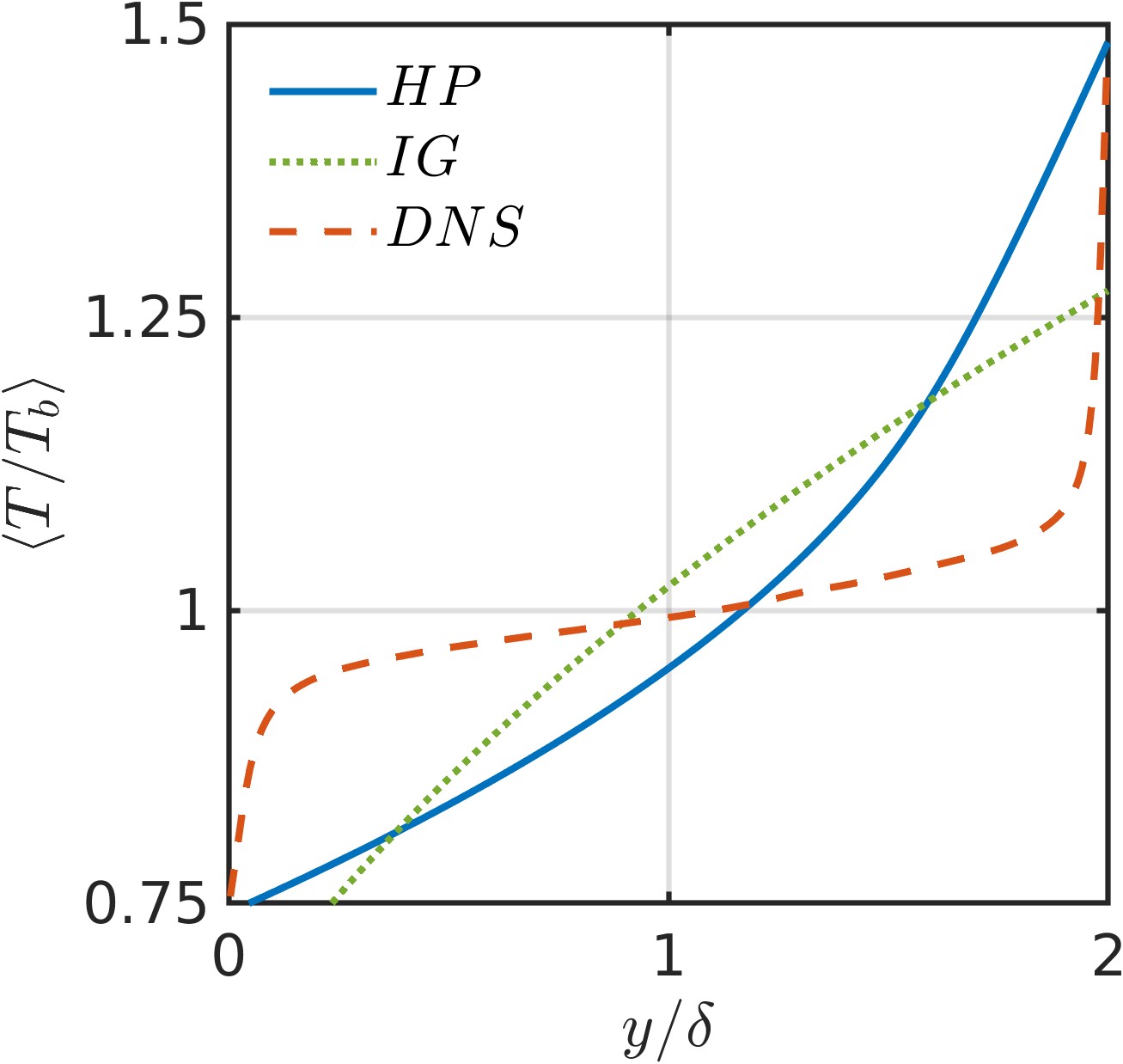}} \hfill
    \subfloat[\vspace{-8mm}]{\includegraphics[width=0.322\linewidth]{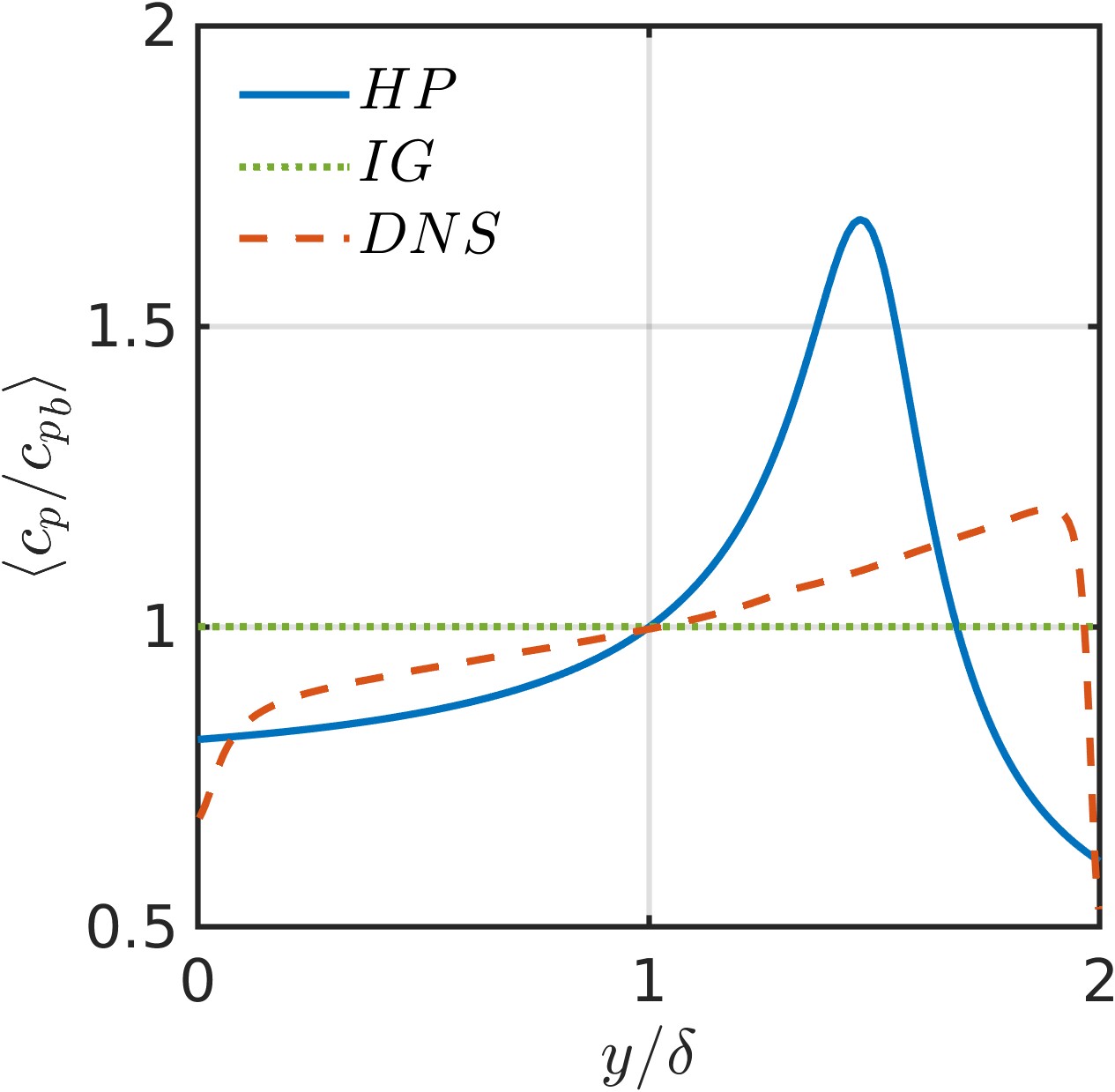}} \\ \vspace{5mm}
	\caption{Comparison of base flows at high-pressure (HP) and ideal-gas (IG) conditions, including laminar Poiseuille solutions and ensemble-averaged turbulent mean flow from DNS, for (a) streamwise velocity, (b) temperature, and (c) isobaric specific heat capacity, all normalized using bulk scaling parameters.} 
 \label{fig:Baseflow}
\end{figure*}

The plane Poiseuille flow is employed to isolate oblique three-dimensional effects from the analysis. A temperature difference is imposed between the walls to ensure the fluid operates along transcritical thermodynamic trajectories, analogous to those in the DNS case.
Figure~\ref{fig:Baseflow} presents the converged, analytically derived laminar base flows for both high-pressure (HP) and ideal-gas (IG) conditions, along with the ensemble-averaged DNS results, referred to as the turbulent mean flow (i.e., the base flow used for the analytical resolvent analysis in this work). 
Notable differences are observed across the cases. The Poiseuille flows are based on laminar profiles, whereas the DNS provides a turbulent mean profile. As it can be observed, the ideal-gas case does not present important thermophysical variations. Additionally for the high pressure cases, the pseudo-boiling region, identified by a peak in isobaric specific heat capacity, appears at $y/\delta \approx 1.5$ for the Poiseuille flow and at $y/\delta \approx 1.9$ for the DNS case. Further details regarding the derivation of these base flows and the corresponding validation can be found in~\citet{Bernades2025b-A}.
Finally, to match the flow conditions of the DNS, the Brinkman number was adjusted to $Br \approx 5.6 \cdot 10^{-6}$, resulting in a reference velocity of $U_r \approx 1\thinspace\textrm{m/s}$.

\subsection{DNS of high-pressure transcritical channel flow} \label{sec:DNS}

\begin{figure}
	\centering    {\includegraphics[width=0.9\linewidth]{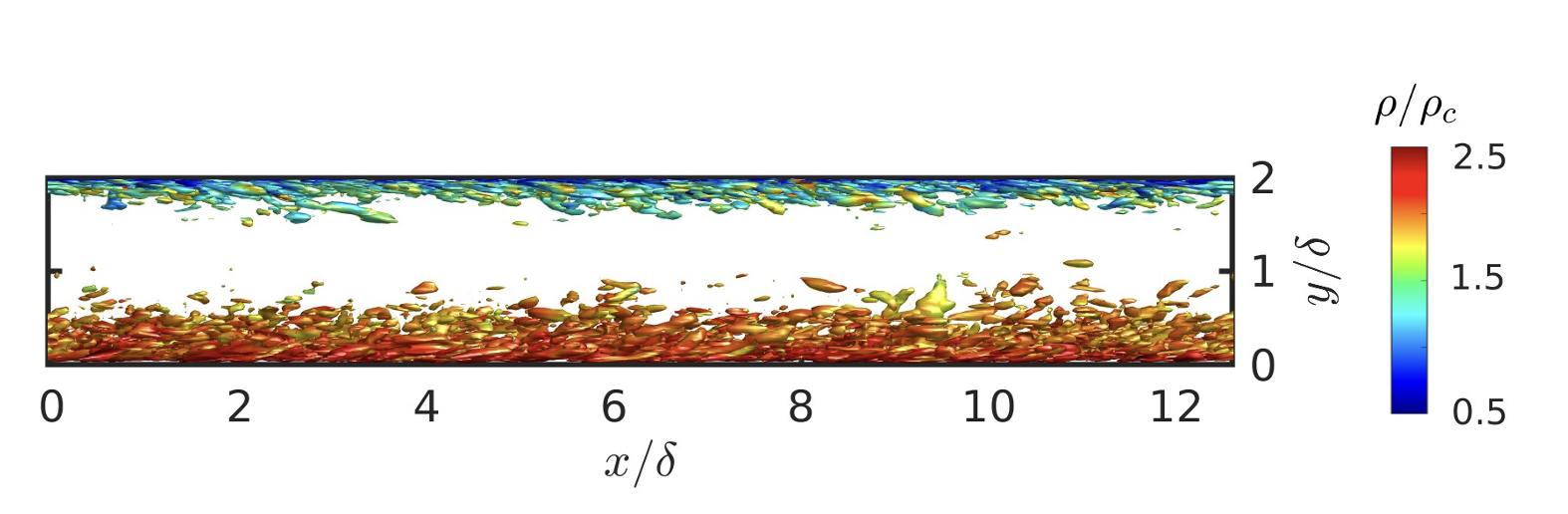}} \\ \vspace{-3mm}
 \vspace{1mm}
	\caption{Schematic of the DNS channel flow showing isosurfaces of normalized Q-criterion at $Q = 2 \cdot 10^9$, colored by reduced density, on an $x$–$y$ plane view.} \label{fig:DNS_setup}
\end{figure}

The DNS of high-pressure transcritical channel flow, conducted following the methodology outlined by~\citet{Bernades2023-A} using the RHEA flow solver~\citep{RHEA2023-A,Abdellatif2023-A}, is briefly introduced. The system operates with N$_2$ at a supercritical bulk pressure of $P_b/P_c = 2$, confined between isothermal cold/bottom ($cw$) and hot/top ($hw$) walls; the system is periodic in the streamwise and spanwise directions. These walls are separated by a distance $H = 2\delta$, with $\delta = 100\thinspace{\mu\textrm{m}}$ representing the channel half-height. The wall temperatures are set to $T_{cw}/T_c = 0.75$ and $T_{hw}/T_c = 1.5$, where subscript $(\cdot)_c$ denotes critical conditions.
Time-averaged DNS results are used to extract the turbulent mean flow fields, which serve as the base state for the subsequent \review{resolvent analysis}. A schematic of the setup is shown in Figure~\ref{fig:DNS_setup}.

The non-isothermal configuration considered forces the fluid to follow a transcritical thermodynamic trajectory across the pseudo-boiling region, as shown in Figure~\ref{fig:Baseflow}(c).
The friction Reynolds number at the cold wall ($\text{Re}_{\tau,cw}$) is set to $100$ to ensure fully developed turbulent conditions. The corresponding dimensional parameters and computational domain represent a DNS with spatial resolutions of $\Delta x^{+} = 9.8$ and $\Delta z^{+} = 3.3$ (normalized by the cold-wall friction velocity), and non-uniform stretching in the wall-normal direction. The vertical grid spacing ranges between $0.2 \le \Delta y^{+} \le 2.1$ at the cold wall and $0.5 \le \Delta y^{+} \le 5.6$ at the hot wall.
The flow is initialized following the procedure described by~\citet{Bernades2023-A}, and statistical data is collected after reaching statistically steady conditions (defined by convergence in the first- and second-order flow statistics and the skin-friction coefficient) over approximately 50 flow through time units, defined as $t_{b} = L_x/u_{b} \sim \delta/u_{\tau,bw}$.
The computational solution of the equations of fluid motion follows the numerical scheme developed by~\citet{Bernades2023b-A}. For time integration, a CFL number of approximately 0.9 is selected to ensure stability of the solution and proper extraction of data snapshots.

\begin{figure*}
	\centering
	\subfloat[\vspace{-8mm}]{\includegraphics[width=0.49\linewidth]{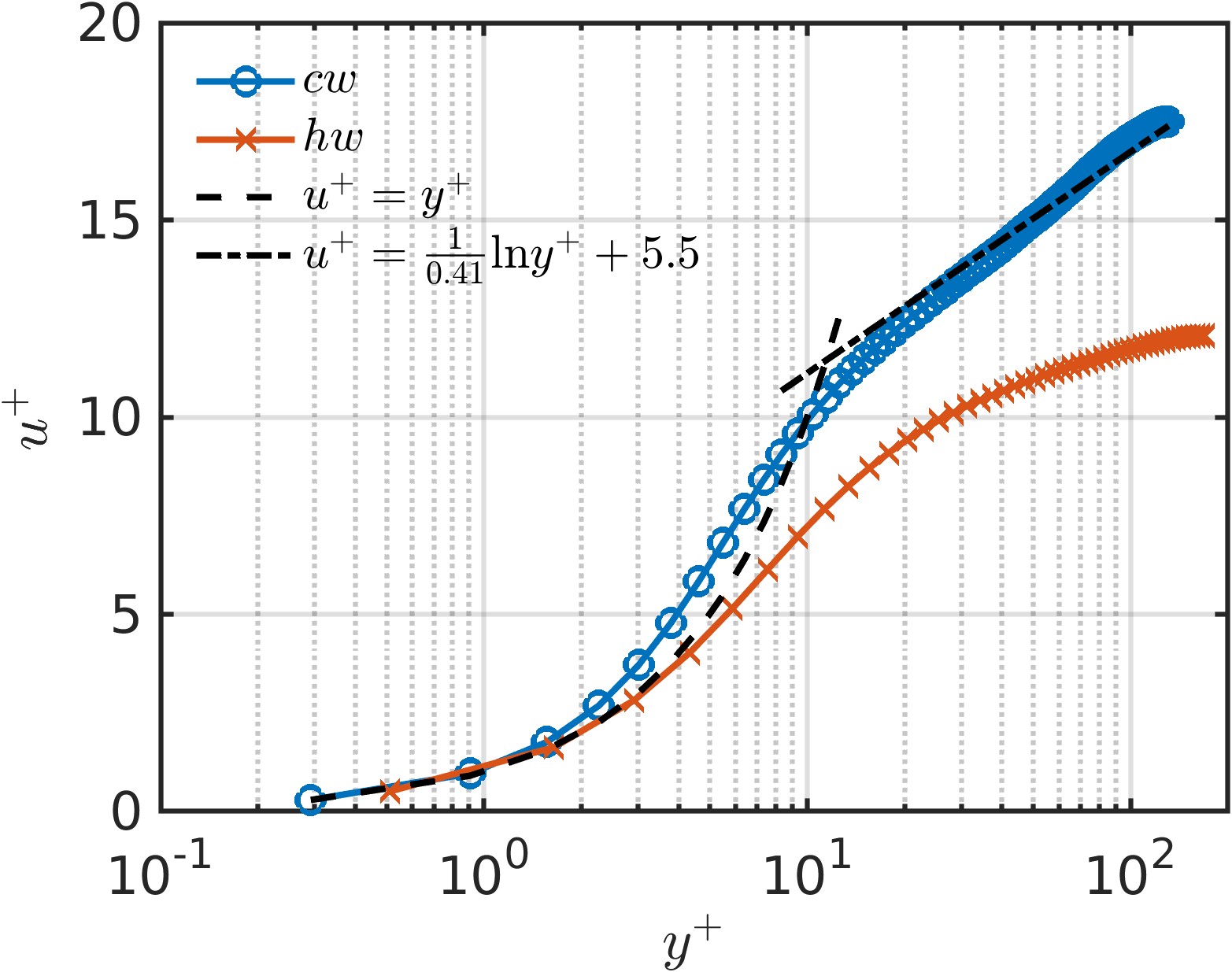}} \hfill
    \subfloat[\vspace{-8mm}]{\includegraphics[width=0.495\linewidth]{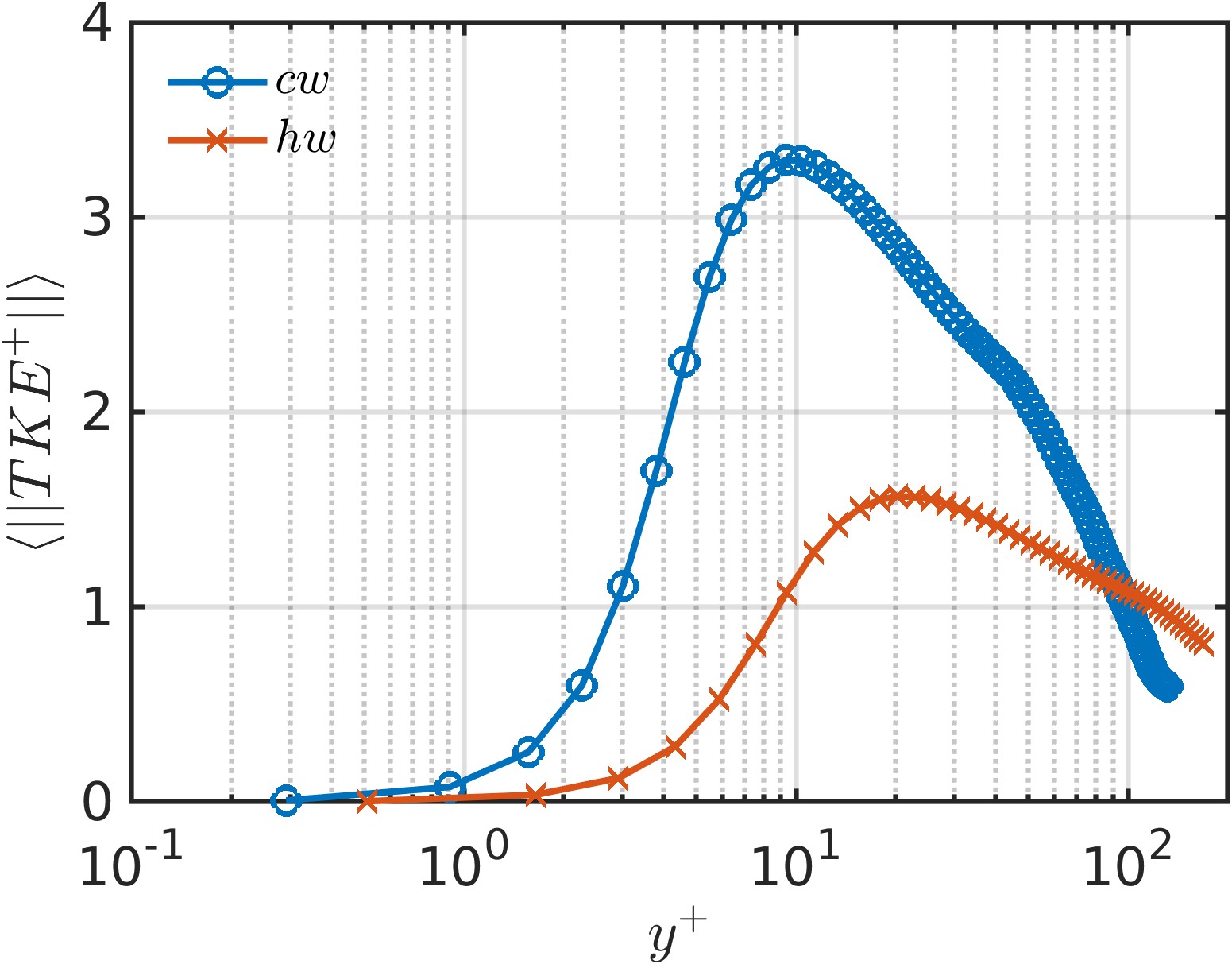}} \\ \vspace{5mm}
	\caption{Normalized wall-normal profiles of (a) mean streamwise velocity and (b) TKE for the cold and hot walls.} 
 \label{fig:DNS_statistics}
\end{figure*}

Figure~\ref{fig:DNS_statistics} depicts the normalized mean streamwise velocity and TKE, calculated from Favre-averaged fluctuations, along the wall-normal direction for the cold and hot walls. As it can be observed, the velocity profile at the hot (top) wall deviates from canonical turbulent boundary layer behavior, with suppressed velocities relative to the cold (bottom) wall (distance from the wall expressed in viscous units). In this regard,~\citet{Ma2018-A} noted limited success near the hot (top) wall with violent density fluctuations.
Moreover, the locations of maximum TKE for the cold and hot walls correspond to $y_{cw}^+ \approx 10$ and $y_{hw}^+ \approx 20$, respectively.
Finally, although all mean-base flow quantities are time-averaged and denoted by $\overline{(\cdot)}$, the overbar notation is omitted in the following sections for clarity in the presentation of the resolvent analysis results.

\section{Results \& Discussion}  \label{sec:results}


This section presents and discusses the results.
It begins with an introduction to the near-wall scale motions identified through the analytical resolvent operator.
Subsequently, turbulent structures at various phase speed levels (revealed by the resolvent framework) are examined using Fourier transform-based data-driven techniques, alongside the most energetic modes obtained from POD.

\subsection{Analytical resolvent analysis} \label{sec:analytical_resolvent}

This subsection presents the results of the analytical resolvent analysis, with a focus on: (i) the dominant resolvent modes associated with the highest amplification, and (ii) the corresponding coherent structures arising at specific wavelengths and phase speeds.
\review{The present analysis builds on the foundation established in~\citet{Bernades2024-B}, where the resolvent operator’s rank, sensitivity, and suboptimal modes were systematically characterized for similar flow conditions. Particularly, in this work, the focus is placed on the leading response modes to interpret their structural correspondence with DNS-based scale motions.}

\subsubsection{Operator-driven resolvent modes} \label{sec:resolvent_modes}

\begin{figure*}
	\centering
	{\includegraphics[width=0.52\linewidth]{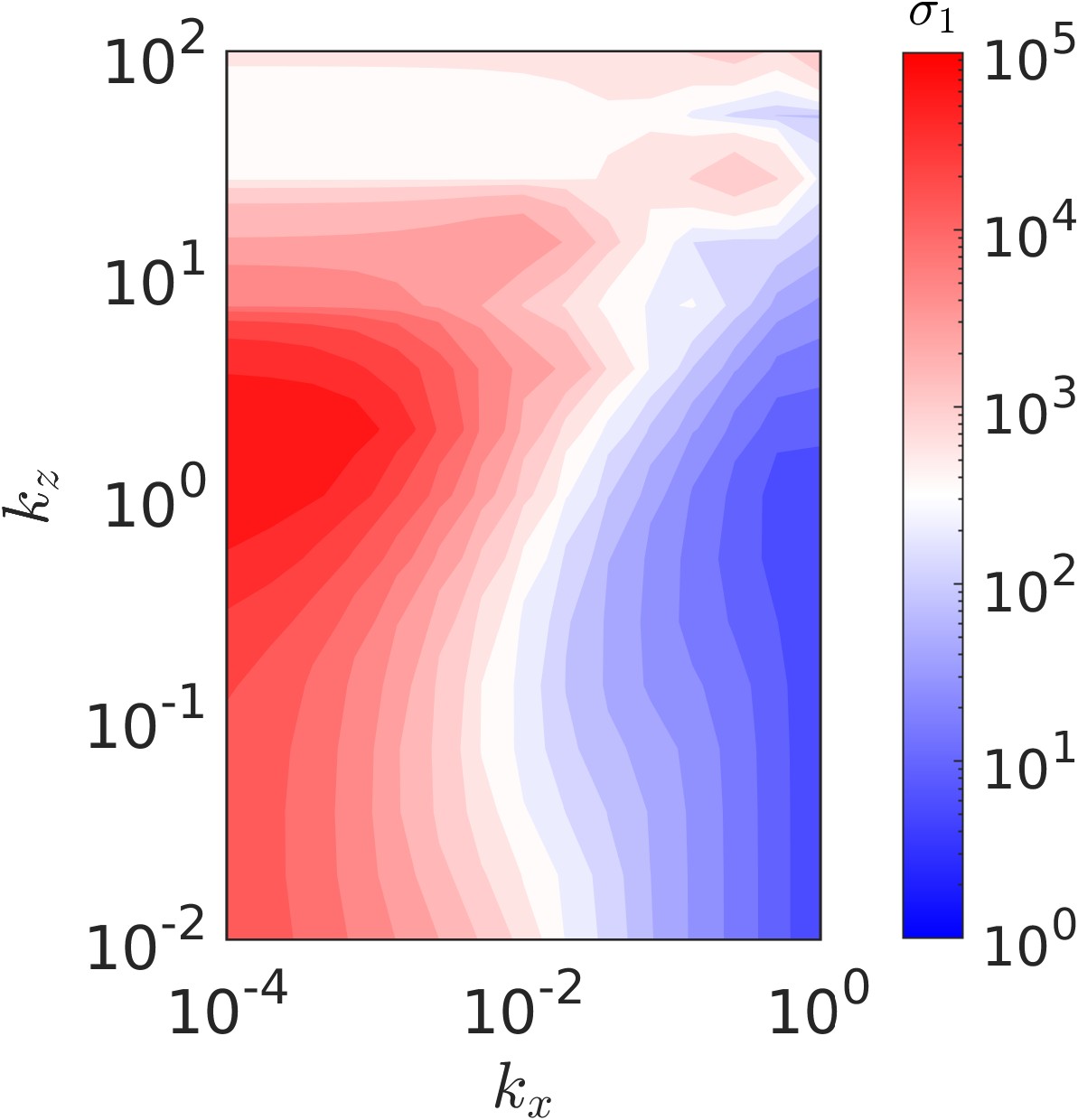}} 
	\caption{Map of the maximum singular values ($\sigma_1$) at phase speed $c = 1$ (normalized by bulk velocity) for the turbulent mean flow.} 
 \label{fig:sigma_maps_turbulent}
\end{figure*}

\begin{figure*}
	\centering
	{\includegraphics[width=0.488\linewidth]{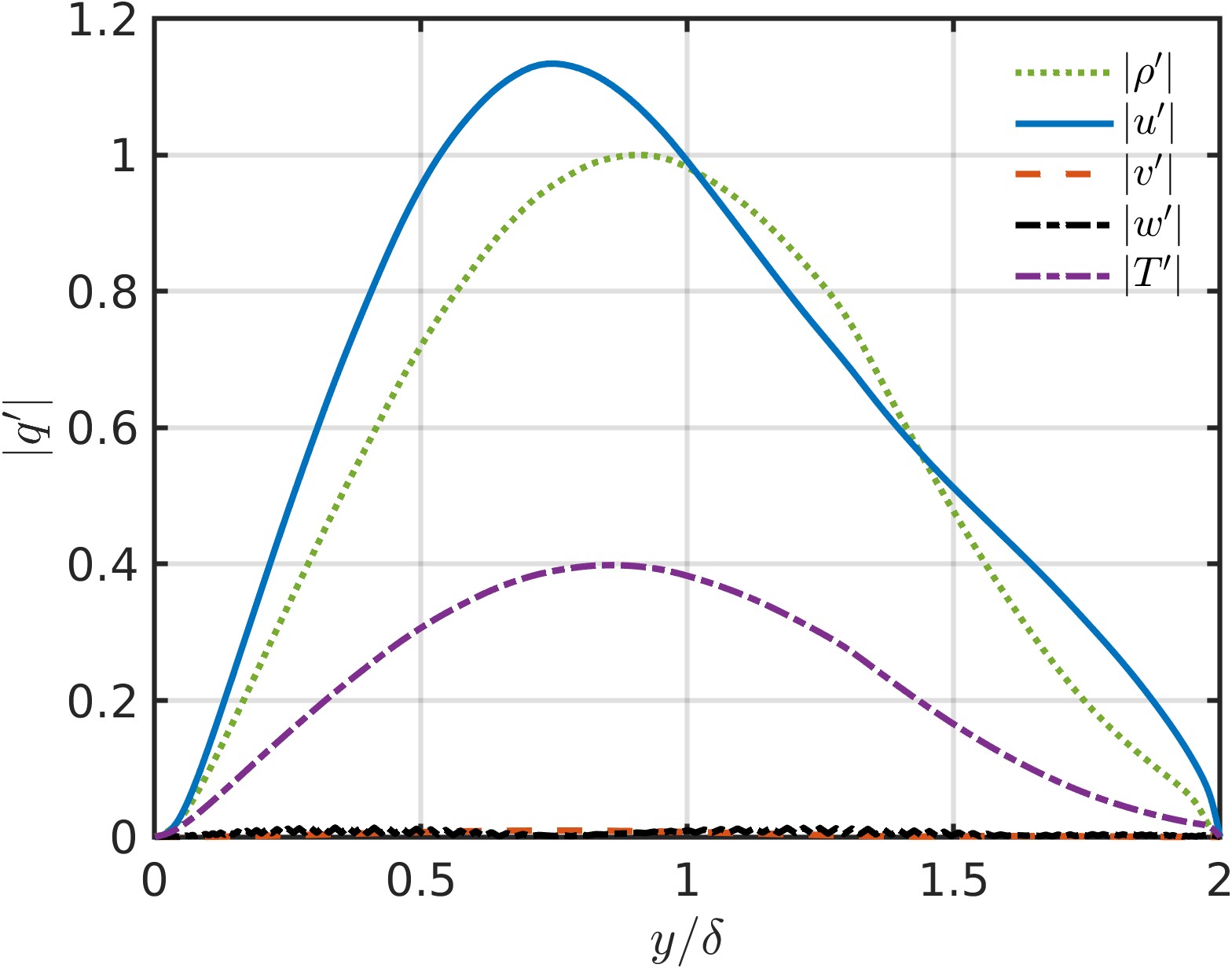}} \hfill
    {\includegraphics[width=0.502\linewidth]{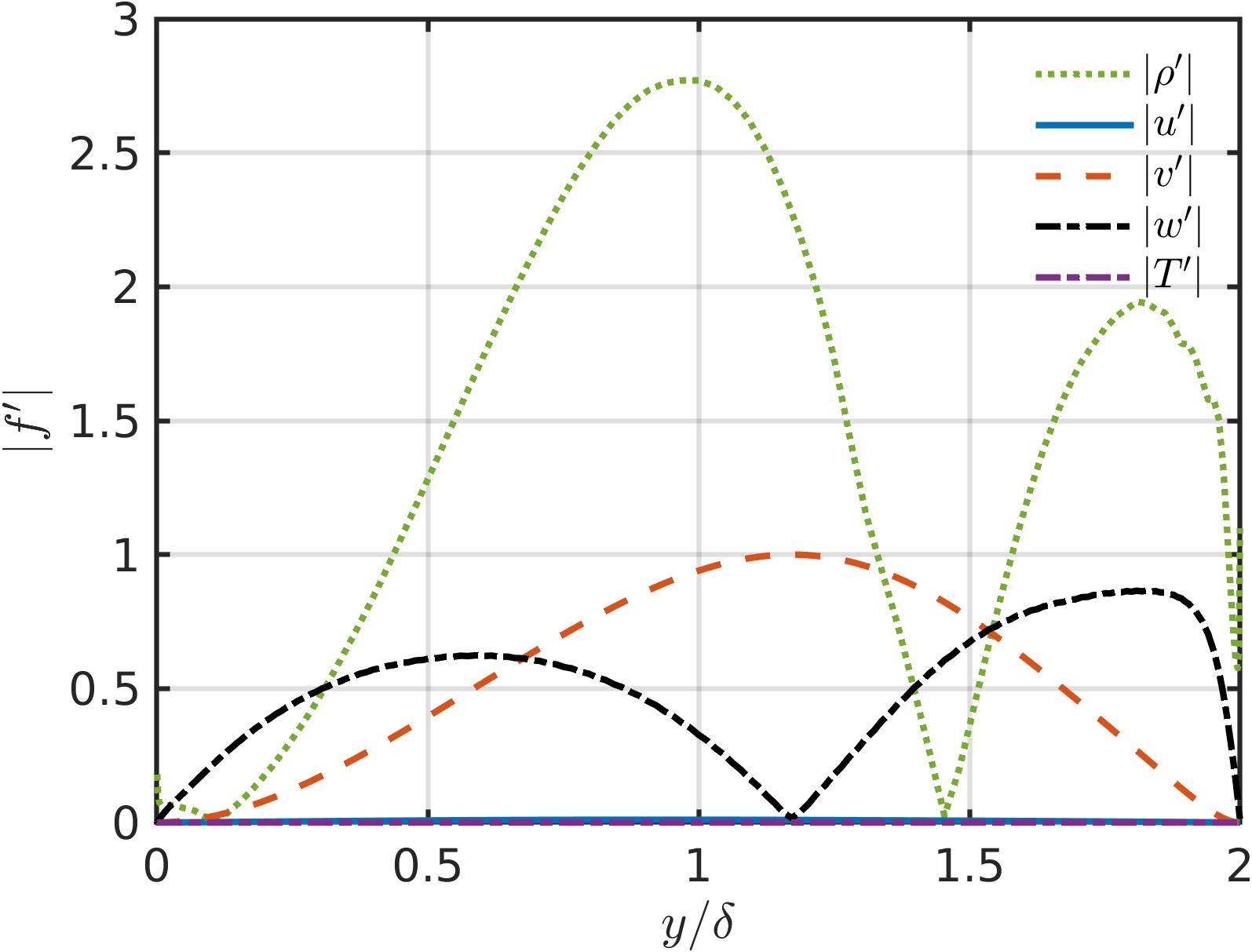}} \\ \vspace{-2mm}
	\caption{Responses (a) and forcing (b) along wall-normal direction at the wavenumbers parameter-space of maximum amplification for data-driven turbulent mean flow.} 
 \label{fig:response_forcing_sigma_1_HP_turbulent}
\end{figure*}

The resolvent operator is constructed by linearizing the equations of fluid motion around the turbulent mean flow, obtained from ensemble-averaged DNS data.
In this case, the Reynolds number used for the operator is the bulk value extracted from the DNS data, ${Re}_b = 3625$, with the corresponding mean flow yielding a Brinkman number of ${Br}_b \approx 5 \cdot 10^{-5}$.
It is important to note that the body force driving the flow is set to $F = 0$, as it is intrinsically handled by the feedback loop controller within the simulation.
Figure~\ref{fig:sigma_maps_turbulent} presents the amplification associated with the leading singular value. While the amplification patterns exhibit slight variations compared to the laminar base flow case, the optimal wavenumber locations remain largely consistent.
\review{At $c^+=1$, the normalized phase speed corresponds to the average maximum value near the pseudo-boiling region. This illustrates resolvent amplification at a physically relevant phase speed rather than at the wall-normal coordinate of maximum kinetic energy.}
However, both the optimal forcings and their corresponding responses differ markedly from those observed in the high-pressure laminar-based resolvent analysis, as shown in Figure~\ref{fig:response_forcing_sigma_1_HP_turbulent}. Near the hot/top wall, the forcing is influenced by the pseudo-boiling region, where strong density and spanwise velocity fluctuations give rise to a pair of counter-rotating vortices elongated in the wall-normal direction.
\review{Specifically, the topology of the structures is biased toward the cold wall, with the response being dominated by a single, streamwise-aligned mode. This structure conforms to the wall-normal mean profile and is advected downstream.}

\subsubsection{Near-wall scale motions} \label{sec:scale_motions}

Coherent structures provide a foundational framework for analyzing turbulent flows, particularly through SVD applied to \review{ ensembles of spatial flow realizations, often after spectral filtering,} to isolate motions at distinct scales~\citep{Moarref2013-A,Saxton-Fox2017-A,Morra2021-A}. In this study, the relevant flow scales are first identified via resolvent analysis, using representative wavenumber triplets corresponding to characteristic motions [namely: large-scale motions (LSMs), very-large-scale motions (VLSMs), and near-wall small-scale motions (NWSMs)].
The spectral signatures of these structures are evident in the premultiplied energy spectra presented in Appendix~\ref{sec:Appendix_A}, supported by one-dimensional streamwise and spanwise spectra, \review{which explicitly labels each wavenumber–frequency combination with its associated scale, enabling readers to directly link the modes discussed in the text to their corresponding "physical" structures.}
These spectra highlight distinct energy peaks: small-scale features dominate the near-wall region, while larger-scale structures emerge in the outer layer. However, in the low-Reynolds-number transcritical regime studied here, partial overlap between these bimodal features is \review{observed; triplet definitions $(\lambda_x,\lambda_z,c)$ in outer and viscous units are detailed in Appendix~\ref{sec:Appendix_A} based on Table~\ref{tab:energetic_wavenumbers}.}

\begin{figure*}
	\centering
    {\includegraphics[width=0.78\linewidth]{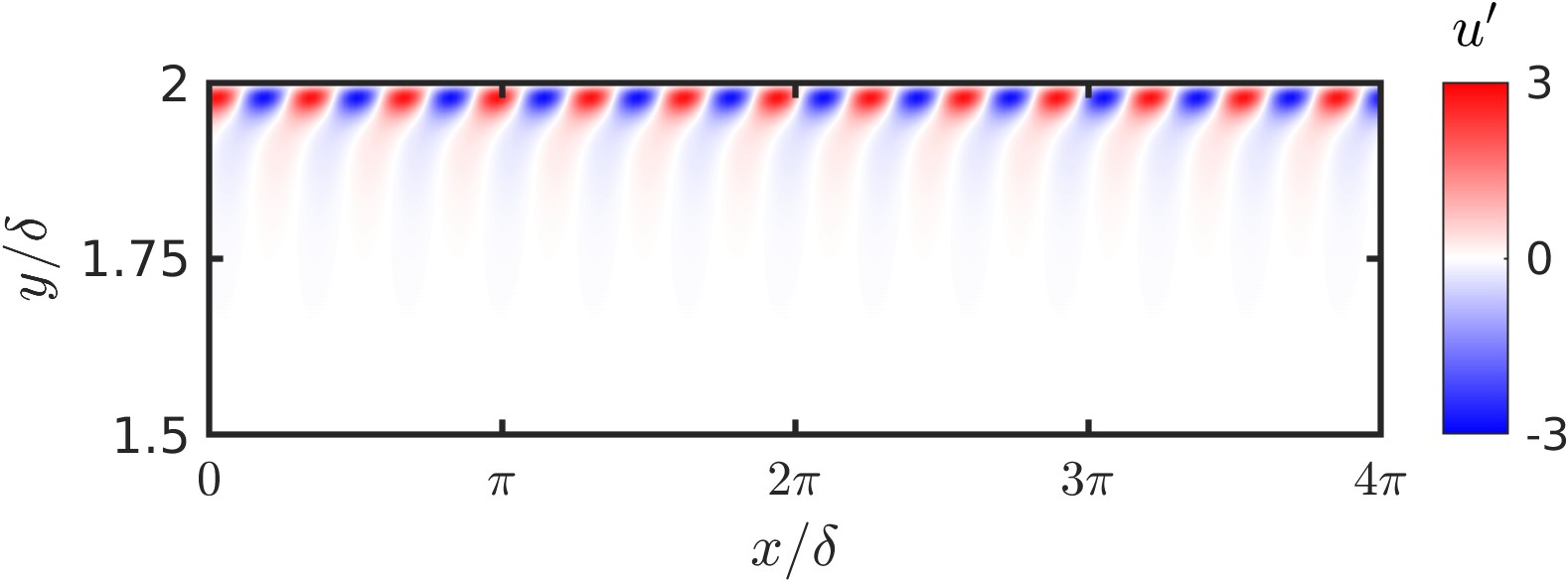}} \\ \vspace{-1mm}
    {\includegraphics[width=0.78\linewidth]{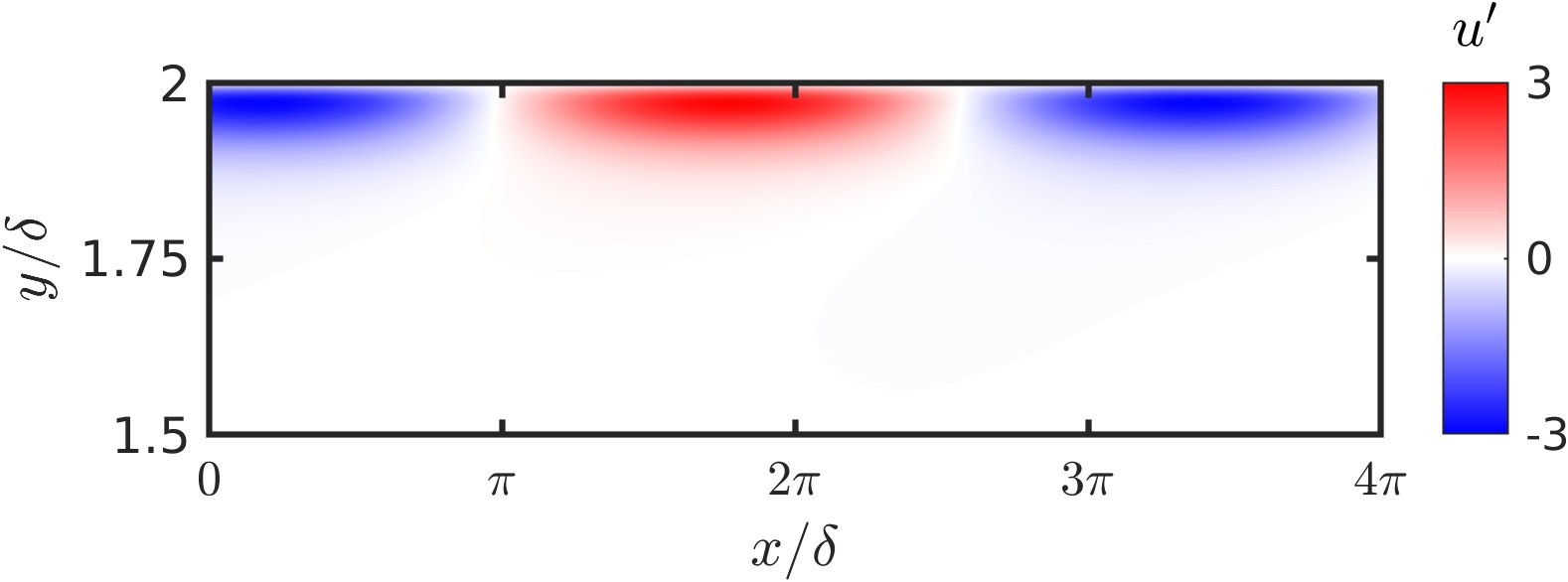}} \\ \vspace{-2mm}
	\caption{Normalized velocity fluctuation patterns corresponding to LSM (top), \review{whose triplet is $(\lambda_x,\lambda_z,c) = (1,0.5,0.95)$,} and VLSM (bottom) \review{with the corresponding triplet $(\lambda_x,\lambda_z,c) = (10,1,1.09)$} shown within the upper quarter of the channel height.} 
 \label{fig:scale_motions_LSM_VLSM}
\end{figure*}

\review{Figure~\ref{fig:scale_motions_LSM_VLSM} displays representative LSM and VLSM structures in the streamwise-wall-normal (x-y) plane, emphasizing the streamwise organization of these coherent motions.}

Notably, both LSMs and VLSMs align with the same physical wavelength, indicating a shared optimal wavenumber associated with peak energy content in the turbulent spectra. It is important to note that LSMs and VLSMs are classically distinguished by their streamwise wavelengths, with VLSMs extending $6$–$10$ $\times$ longer than the $\mathcal{O}(\delta)$ scales associated with LSMs~\citep{Jimenez1998-A,Kim1999-A}. In the present study, however, the resolvent analysis at low-Reynolds numbers yields only a single energetic wavelength, consistent with the peak of the turbulent spectra. This is interpreted as a Reynolds-number limitation, reflecting insufficient spectral separation to clearly distinguish between LSM and VLSM amplification scales, rather than evidence that both structures collapse to the same physical wavelength. \review{Nevertheless, the resolvent operator is interrogated at two distinct streamwise wavelengths within these overlapping ranges to examine whether it highlights different coherent structures.}
\review{The identified scale-specific motions are concentrated near the hot/top wall and are confined by the pseudo-boiling region, highlighting a physically significant mechanism localized around $y/\delta \approx 1.9$. Given the low Reynolds number of this study, the turbulent spectra exhibit a single energetic wavelength; consequently, the streamwise and spanwise wavenumbers are selected based on this spectral peak and on representative wall-normal locations where energetic motions typically emerge.
The corresponding phase speeds are chosen at physically significant locations, such as the channel centerline and the pseudo-boiling height. Notably, the wall-normal and spanwise structures of the resolvent response and forcing modes are determined entirely by the linearized operator and the turbulent mean profile, while the streamwise dependence is prescribed by the selected Fourier mode. Unlike symmetric isothermal flows, the mean shear in this configuration does not support large-scale motions at the channel centerline, reflecting the profound influence of the fluid’s thermophysical properties and the specific dynamics of the high-pressure transcritical regime.
While the resolvent modes share the same streamwise scale by construction, their wall-normal and spanwise organization varies depending on the targeted wall-normal location. Modes associated with the pseudo-boiling region exhibit pronounced confinement and wall-normal displacement, whereas those closer to the hot/top wall are more spatially compact and display distinct variations in vortical inclination. These features indicate a redistribution of energy and momentum consistent with the strongly modified mean shear and thermophysical gradients. Thus, to avoid confusion with canonical LSMs or VLSMs, typically observed at higher Reynolds numbers, these structures are referred to as scale-specific motions.}

\begin{figure*}
	\centering
    {\includegraphics[width=0.80\linewidth]{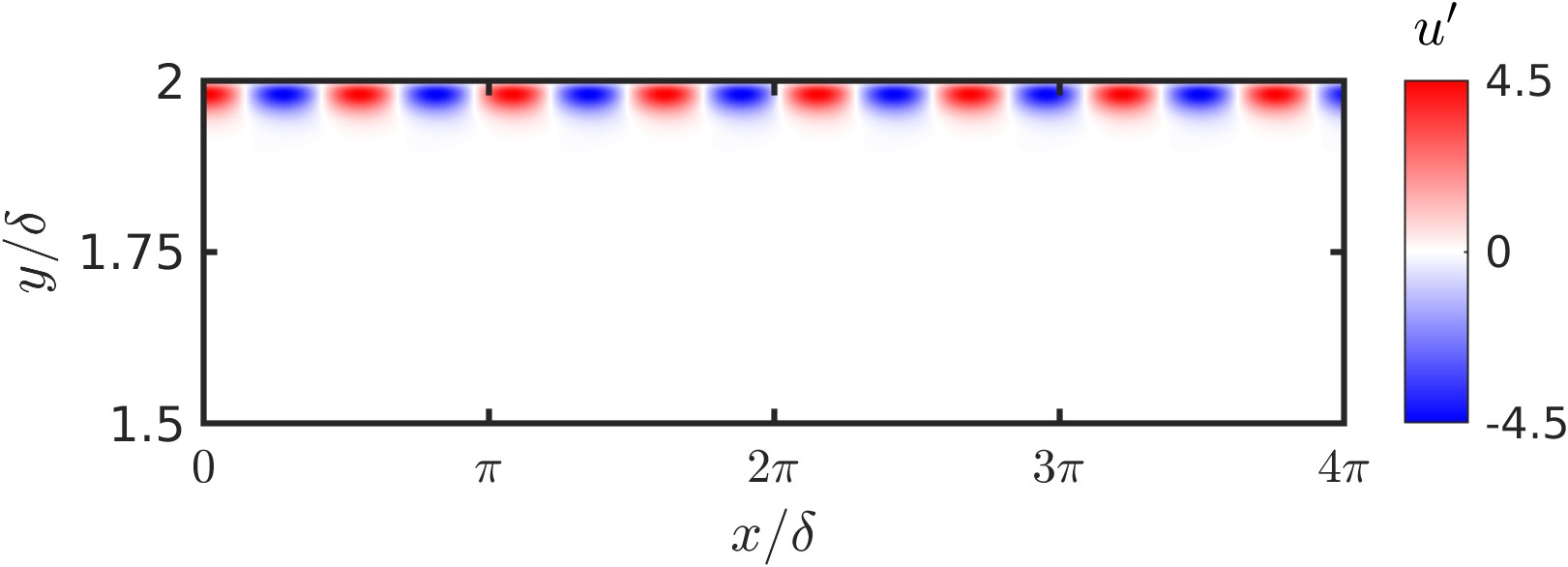}} \\ \vspace{-1mm}
    {\includegraphics[width=0.80\linewidth]{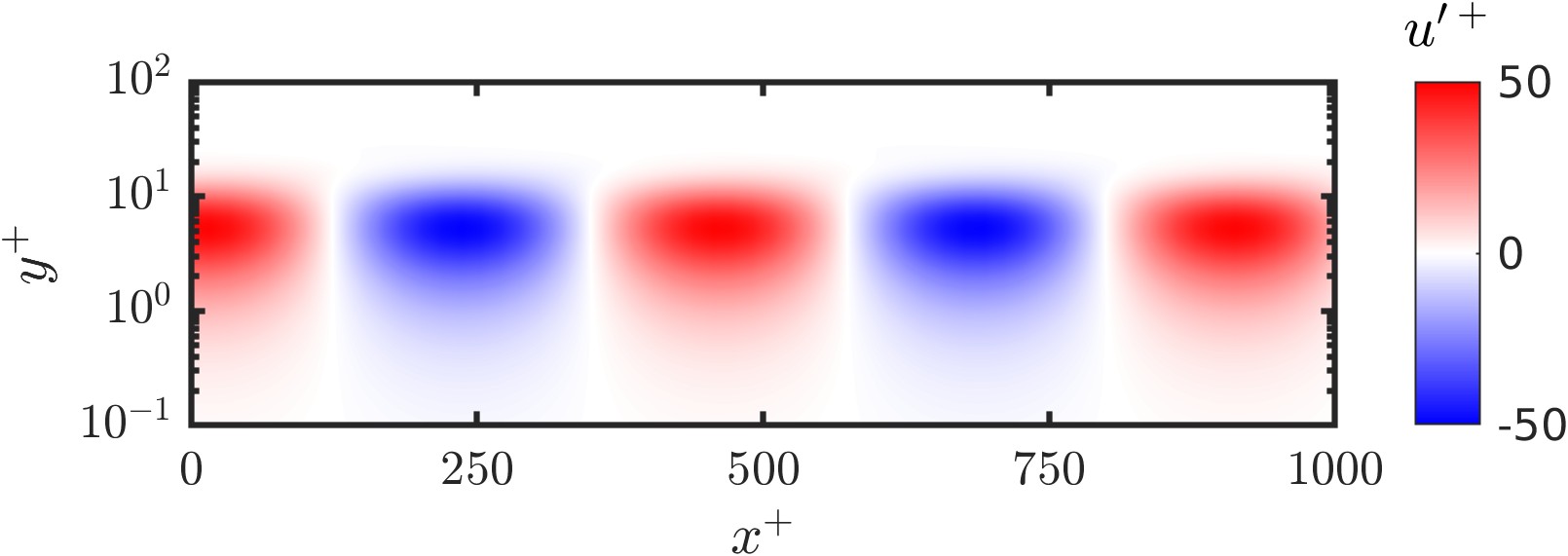}} \\ \vspace{-2mm}
	\caption{Normalized near-wall scale motions at the hot/top wall: (top) view in the $x$–$y$ plane, magnified to the upper quarter of the channel height; (bottom) view in the $x^{+}$–$y^{+}$ plane with a semi-logarithmic $y$-axis whose triplet is \review{($\lambda_x^+,\lambda_z^+,c^+) = (450,100,10)$}.} 
 \label{fig:scale_motions_NWSM}
\end{figure*}

Figure~\ref{fig:scale_motions_NWSM} presents NWSMs structures at the hot/top wall, shown in both inner and outer scaling. At the cold/bottom wall, NWSMs are only weakly captured, consistent with the energy spectra, and linked to wavenumbers prone to numerical instability. Nevertheless, the second resolvent mode reveals structures at scales comparable to those near the hot/top wall, suggesting a degree of symmetry in the resolved dynamics despite the asymmetric energy distribution.
Notably, the NWSM structures exhibit qualitative similarities to the VLSMs, particularly in their pattern of alternating high- and low-speed streaks centered around $y^+ \approx 5$ at the hot/top wall. This indicates that the pseudo-boiling region significantly influences the dynamics of coherent structures, especially in low-viscosity zones where such motions are more readily sustained.
To further explore these features, a data-driven analysis is proposed below to characterize the modal properties and energy content of the scale-specific motions. This approach aims to verify the presence of coherent structures propagating at phase speeds atypical of \review{classical} isothermal wall-bounded turbulence~\citep{Schmidt2018-A,Towne2018-A,Moarref2013-A}, thereby offering new insights into the unique dynamics of high-pressure transcritical flows.

\subsection{Spatio-temporal characterization of turbulent structures} \label{sec:data_driven_results}

This subsection presents the analysis of turbulent structures using data-driven methods, focusing on: (i) coherent structures extracted via \review{Fourier transform}-based filtering techniques, and (ii) energized flow regions identified through POD to highlight dominant energetic modes.
Notably, amplified responses occur at a relatively slow wavespeed, $c^+=1$. Although such low-speed responses are not typically associated with energetic motions in canonical wall turbulence, the resolvent analysis reveals that they are amplified in the transcritical case. The corresponding singular values ($\sigma_1 \approx 2.9$) \review{quantifies the input–output gain of the leading resolvent mode, indicating that the selected forcing produces a locally significant flow response and confirming the enhanced amplification observed in the spatial structure of this mode}, emphasizing the altered scaling and energetic pathways characteristic of this regime.

\subsubsection{\review{Fourier transform}-based filtered coherent structures}

\begin{figure*}
	\centering
	{\includegraphics[width=0.9\linewidth] {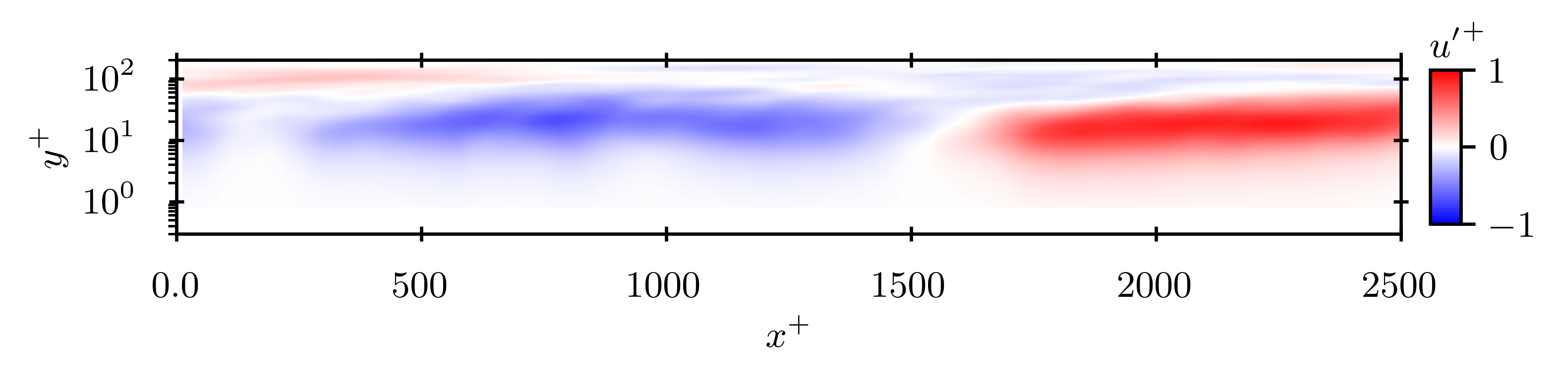}} \\ \vspace{-6mm}
    {\includegraphics[width=0.9\linewidth]{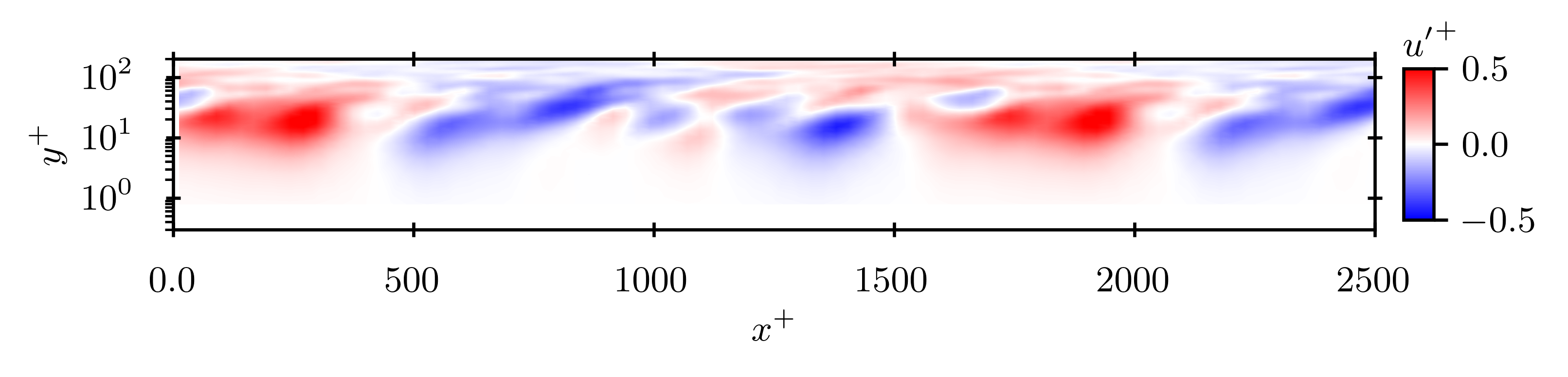}} \\ \vspace{-6mm}
    {\includegraphics[width=0.9\linewidth]{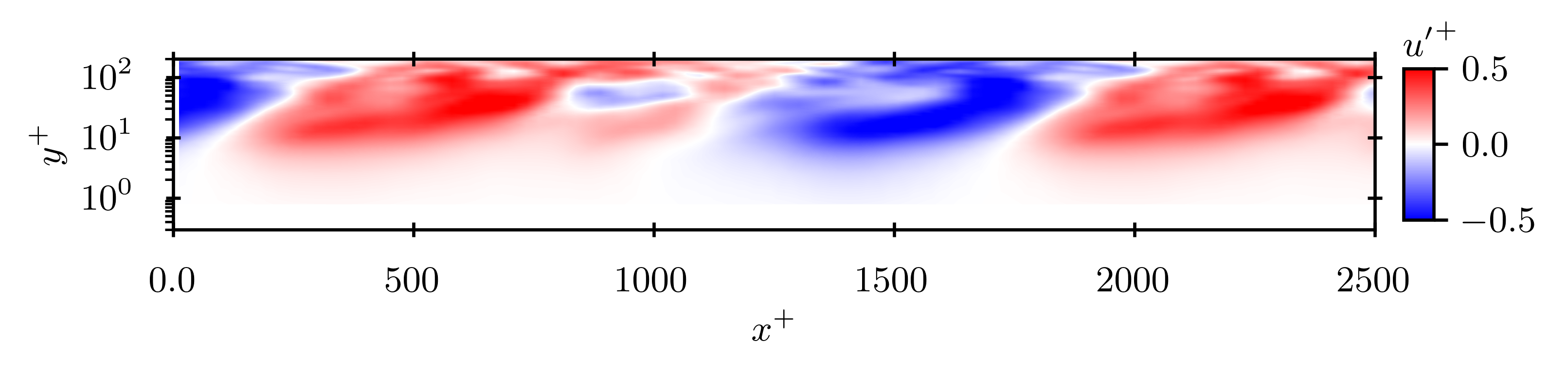}} \\ \vspace{-6mm}
    {\includegraphics[width=0.9\linewidth]{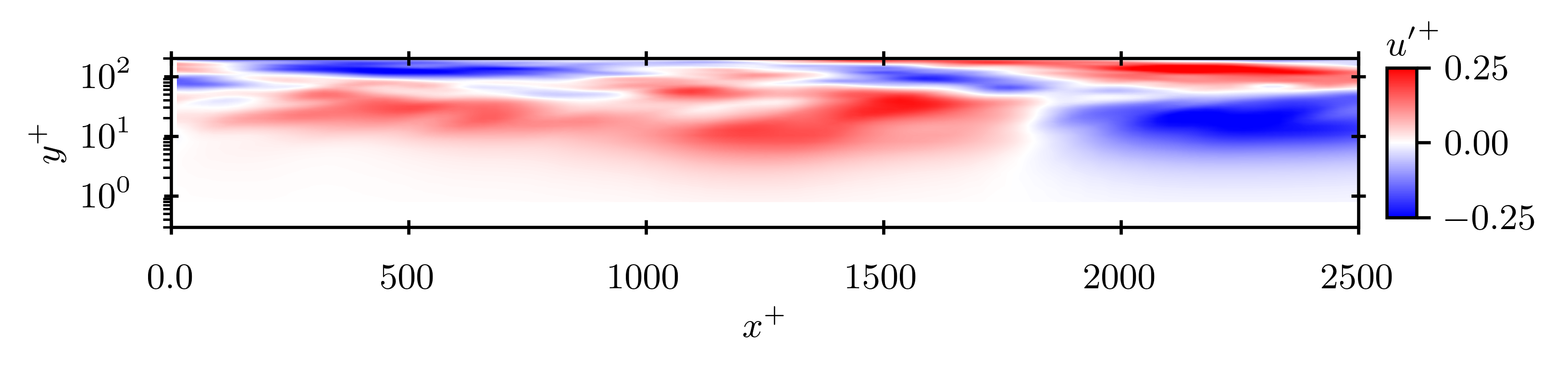}} \\ \vspace{-6mm}
    {\includegraphics[width=0.9\linewidth]{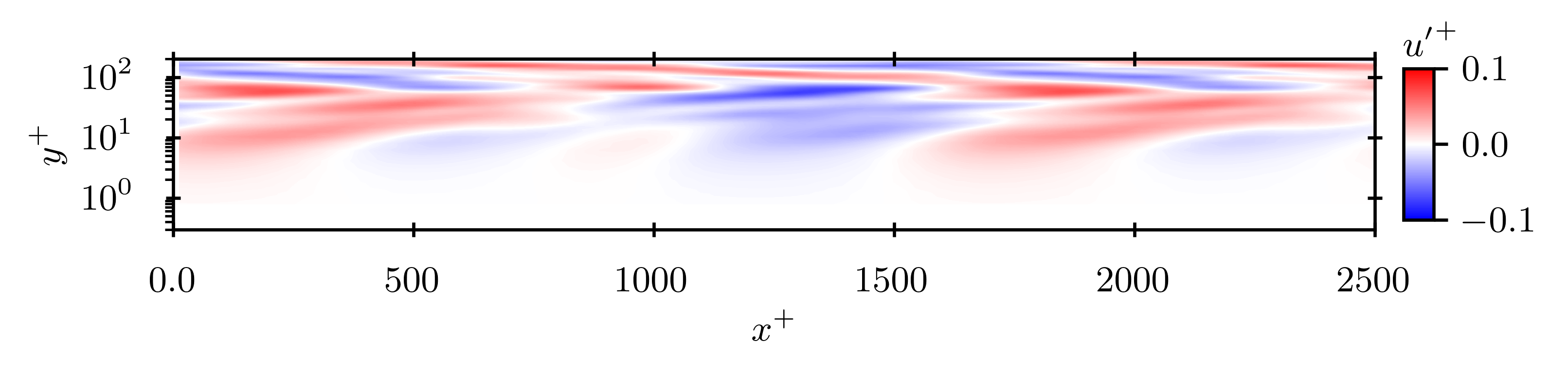}} \\ \vspace{0mm}
	\caption{Filtered coherent structures on the $x$-$y$ plane at $c^+ = 1$, $5$, $10$, $15$, and $20$, colored by streamwise velocity fluctuations normalized by hot/top wall units (from top to bottom). The vertical axis is referenced from the hot/top wall, with increasing $y^+$ denoting positions toward the channel center. Filtered coherent structures  shown below follow the same convention. Corresponding movies illustrating the time evolution of these structures are provided in the supplementary material.} 
 \label{fig:FFT_u}
\end{figure*}

The filtered fields presented in Section~\ref{sec:scale_motions} are obtained via a data-driven decomposition of the full space–time turbulent dataset using targeted phase-speed windows. 
\review{The width of the phase-speed window is implicitly defined through a physically motivated \textit{argmin} criterion, ensuring that the resulting structures correspond to the energetically dominant motions at each location. By avoiding the imposition of an arbitrary bandwidth, this approach allows the intrinsic scales of the flow to dictate the window's extent.}
By applying spatio-temporal Fourier transforms, the data are selectively reconstructed in physical space to isolate coherent structures propagating at prescribed convective velocities ($c^+ = 1,5,10,15,20$), expressed in viscous units relative to the hot/top wall. This methodology is intended to \review{isolate flow features in prescribed wavenumber–frequency bands consistent with resolvent-based predictions, which are subsequently assessed for their energetic and dynamical relevance using modal decompositions}~\citep{Sharma2013-A,Saxton-Fox2017-A}, which are often obscured in conventional, unfiltered turbulence fields.
Figure~\ref{fig:FFT_u} presents the streamwise velocity fluctuations reconstructed at each phase speed. The wall-normal axis is scaled in hot/top wall viscous units, referencing the thermodynamically active side of the channel; viz. nearest to the pseudo-boiling region.
At the lowest phase speed, $c^+ = 1$, the filtered fields reveal near-wall, streak-like structures confined below the pseudo-boiling layer. These features emphasize the dominance of hot/top-wall thermodynamic effects in this regime. Structurally, they resemble the self-sustaining motions characteristic of the viscous sublayer and buffer region. However, here they are identified on the basis of propagation speed rather than spatial wavelength, enabling the detection of slow-moving modes that may arise from strong local thermodynamic forcing. Their tight confinement to the hot/top wall vicinity supports the presence of non-canonical, low-speed motions that are not typically observed in \review{classical} incompressible wall-bounded turbulence~\citep{Alamo2006-A,Hutchins2007b-A}.

As the phase speed increases, the filtered structures become increasingly elongated in the streamwise direction and shift upward in the wall-normal coordinate, while remaining confined below the pseudo-boiling region. The coherence and spatial organization of these motions highlight the role of the pseudo-boiling layer as a dynamic modulator, which delineates the vertical extent of energetic structures and potentially acts as a phase-selective barrier or waveguide.
For intermediate phase speeds ($c^+=10–15$), the structures resemble large-scale wavepackets extending into the \review{inertial layer}, yet they remain bounded above by the pseudo-boiling zone. At the highest phase speed ($c^+=20$), the reconstructed motions occupy a broad wall-normal range, consistent with outer layer resolvent modes. Nonetheless, a pronounced hot/top wall asymmetry persists, shaped by the geometry and location of the pseudo-boiling region.

\begin{figure*}
	\centering
	{\includegraphics[width=0.89\linewidth] {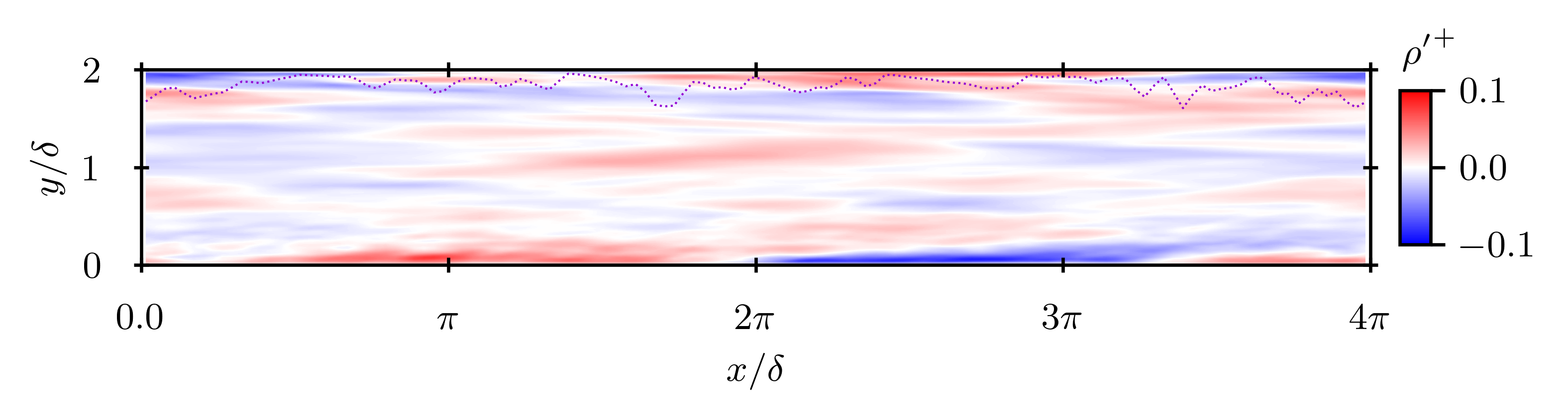}} \\ \vspace{-6mm}
    {\includegraphics[width=0.89\linewidth]{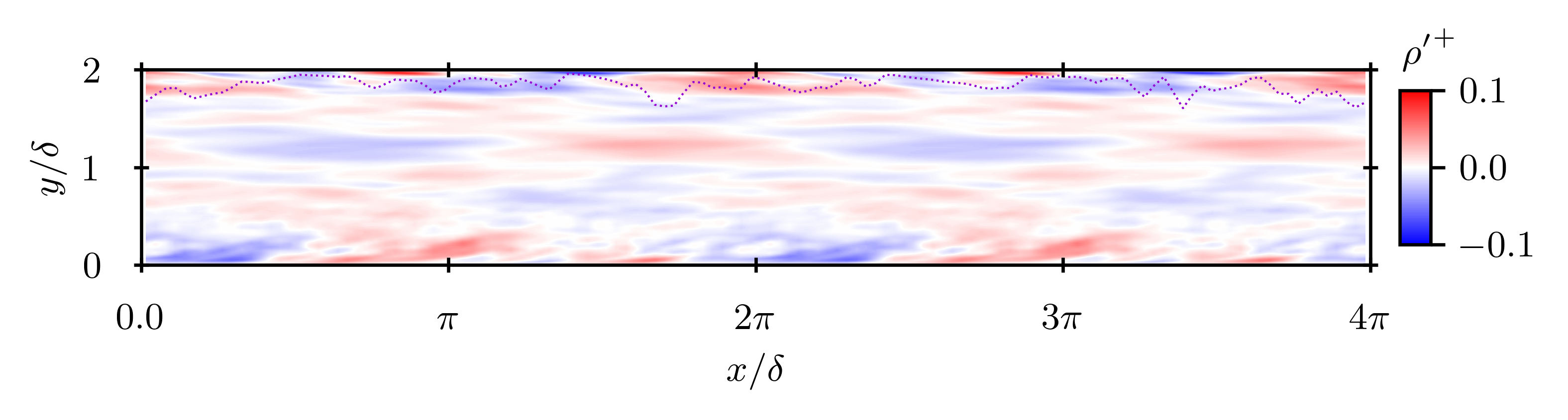}} \\ \vspace{-6mm}
    {\includegraphics[width=0.89\linewidth]{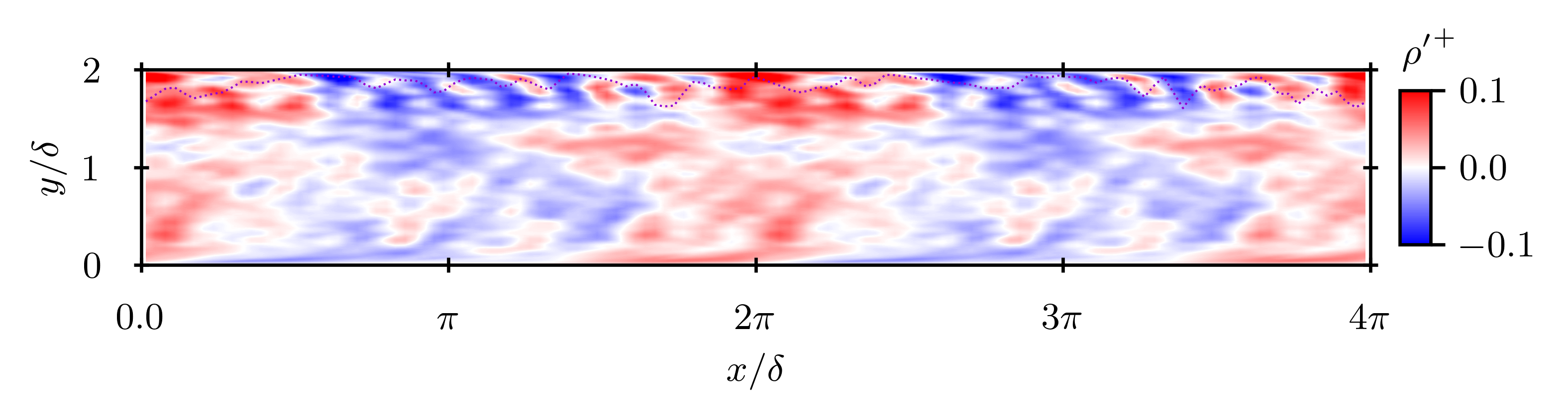}} \\ \vspace{-6mm}
    {\includegraphics[width=0.89\linewidth]{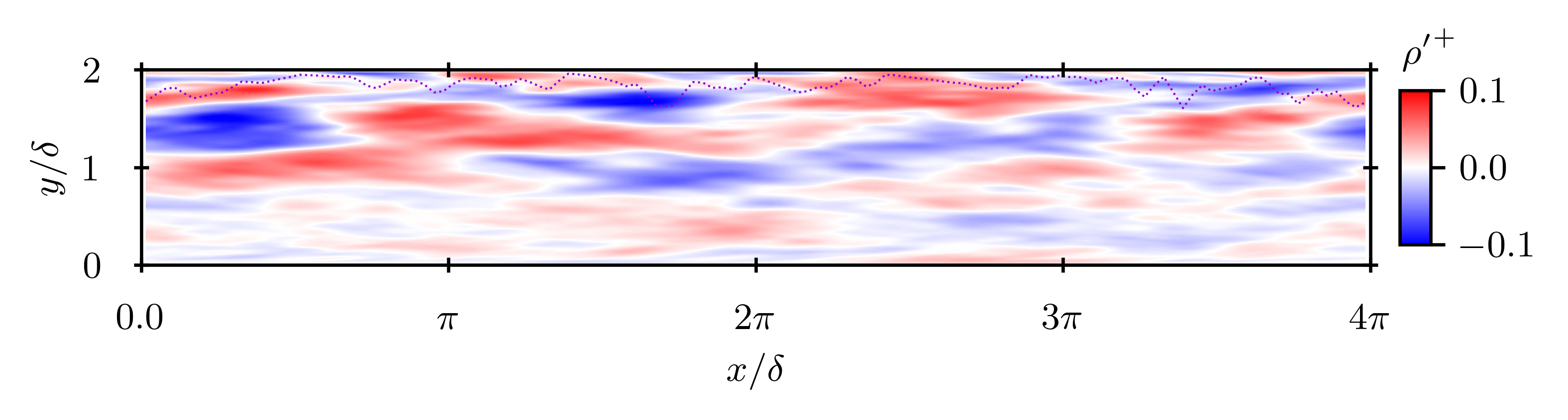}} \\ \vspace{-6mm}
    {\includegraphics[width=0.89\linewidth]{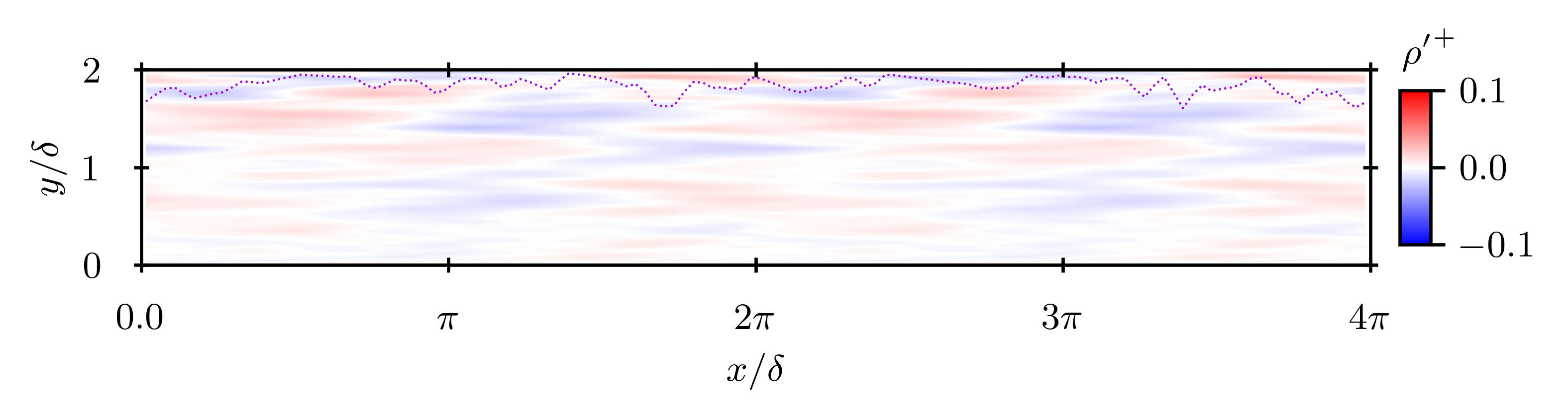}} \\ \vspace{0mm}
	\caption{Filtered coherent structures on the $x$-$y$ plane at $c^+ = 1$, $5$, $10$, $15$, and $20$, colored by normalized density fluctuations (from top to bottom). Corresponding movies showing the time evolution of these structures are available in the supplementary material.} 
 \label{fig:FFT_rho}
\end{figure*}

In contrast to the velocity field, Figure~\ref{fig:FFT_rho} shows the filtered density fluctuations expressed in outer units, revealing the full channel height and highlighting the flow’s asymmetric configuration. The pseudo-boiling line, marked by a dotted contour in each subplot, is situated closer to the hot/top wall. At the lowest phase speed ($c^+=1$), density fluctuations are highly localized near the hot/top wall and display small-scale, thermally driven features, likely associated with baroclinic modes that are highly sensitive to steep thermodynamic gradients near the pseudo-boiling line.
For intermediate phase speeds ($c^+=5-10$), the structures become more elongated and coherent in the streamwise direction, aligning with the pseudo-boiling line geometry and suggesting coupling between density fluctuations and the local thermodynamic state. At $c^+=15$, wave-like density perturbations emerge that closely follow the pseudo-boiling contour, indicating potential baroclinic interaction with the stratified layer. At the highest phase speed ($c^+=20$), the fluctuations appear broader and more dispersed across the channel height, consistent with fast-propagating outer-layer modes. Despite their increased spatial extent, these fluctuations remain biased toward the hot/top wall, underscoring the persistent influence of the pseudo-boiling layer on the organization and vertical distribution of turbulent energy.

\begin{figure*}
	\centering
	{\includegraphics[width=0.87\linewidth] {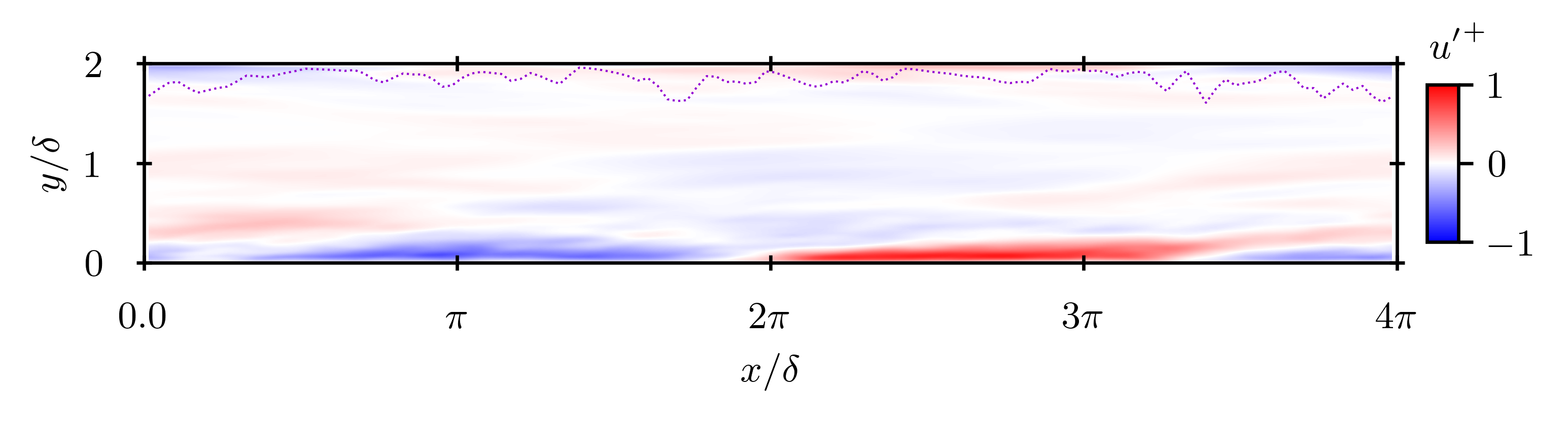}} \\ \vspace{-6mm}
    {\includegraphics[width=0.89\linewidth]{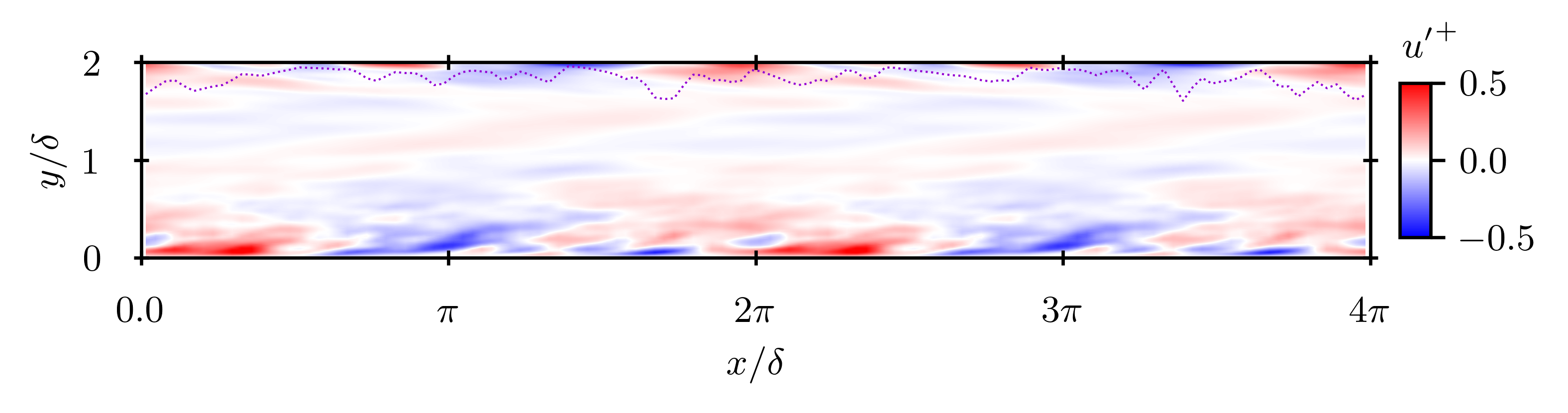}} \\ \vspace{-6mm}
    {\includegraphics[width=0.89\linewidth]{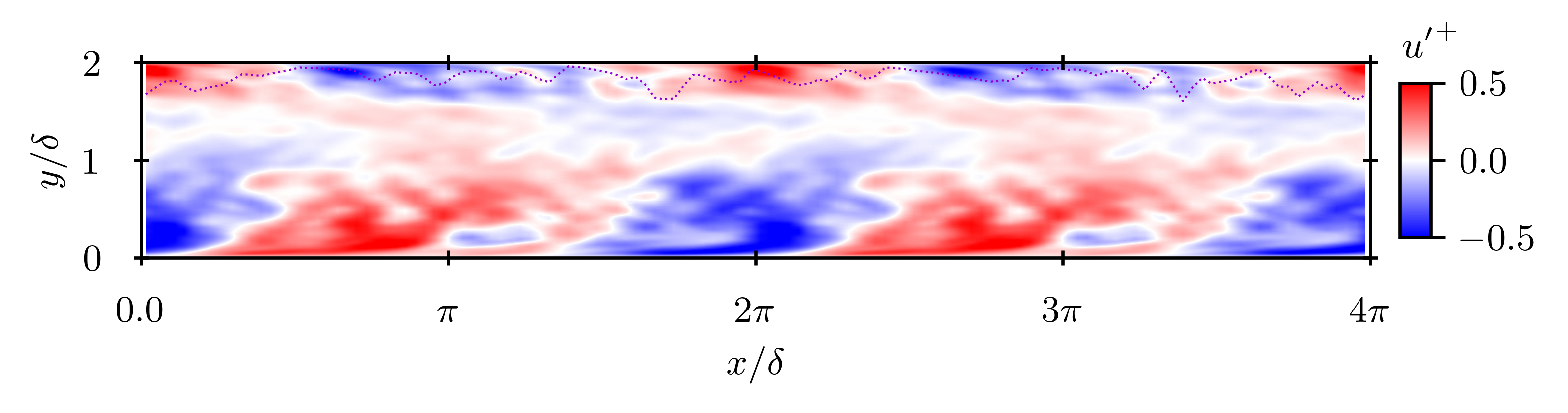}} \\ \vspace{-6mm}
    {\includegraphics[width=0.91\linewidth]{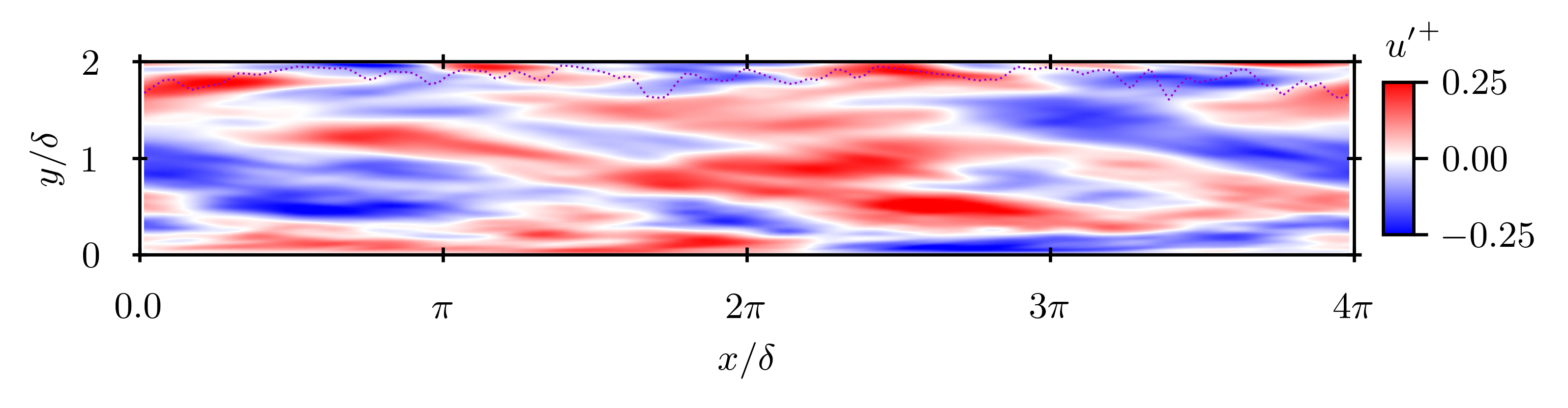}} \\ \vspace{-6mm}
    {\includegraphics[width=0.91\linewidth]{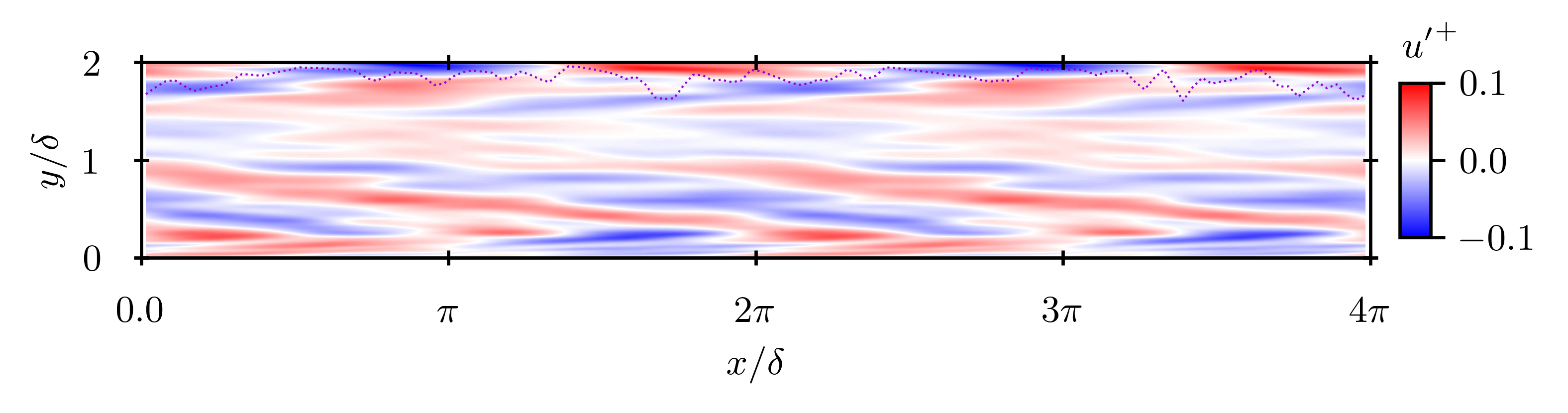}} \\ \vspace{0mm}
	\caption{\review{Filtered coherent structures on the $x$–$y$ plane at $c^+ = 1$, 5, 10, 15, and 20, colored by streamwise velocity fluctuations in outer scaling (from top to bottom). Corresponding movies showing the temporal evolution of these structures are provided in the supplementary material.}} 
 \label{fig:FFT_u_outer}
\end{figure*}

Overall, the velocity and density reconstructions across phase speeds reveal a continuous spectrum of convecting coherent motions, spanning from slow, near-wall structures to fast, outer-layer modes, all asymmetrically shaped by the pseudo-boiling region. This region acts as a phase-selective boundary, modulating or confining structures according to their convective velocity and thermodynamic sensitivity. Such behavior reflects the distinctive physics of high-pressure transcritical turbulence, where \review{classical} wall-bounded flow mechanisms are intricately coupled with strong real-fluid effects. For completeness, analogous reconstructions of streamwise velocity in outer scalings are provided in \review{Figure~\ref{fig:FFT_u_outer}}, offering a global perspective on the spatial extent and asymmetry of the filtered motions.

\subsubsection{POD analysis of energetic locations}

\begin{figure*}
	\centering
	{\includegraphics[width=0.9\linewidth] {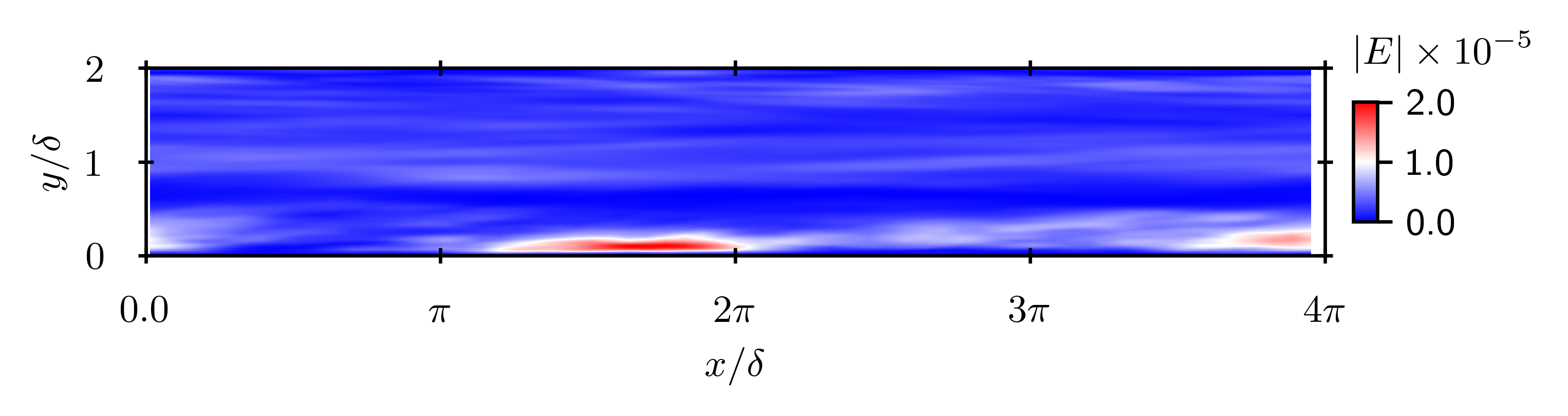}} \\ \vspace{-6mm}
    {\includegraphics[width=0.9\linewidth]{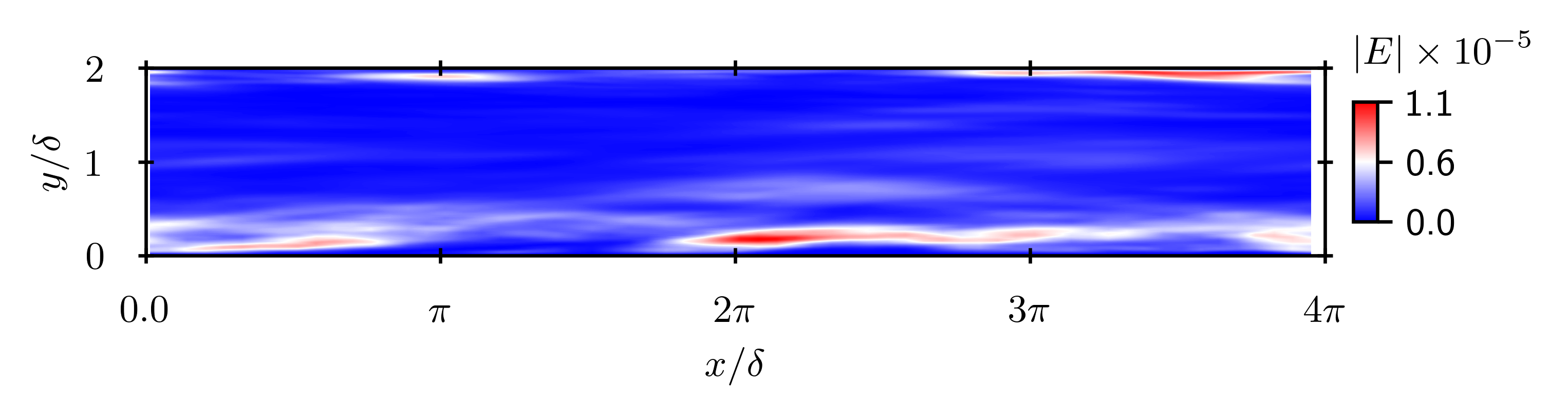}} \\ \vspace{-6mm}
    {\includegraphics[width=0.9\linewidth] {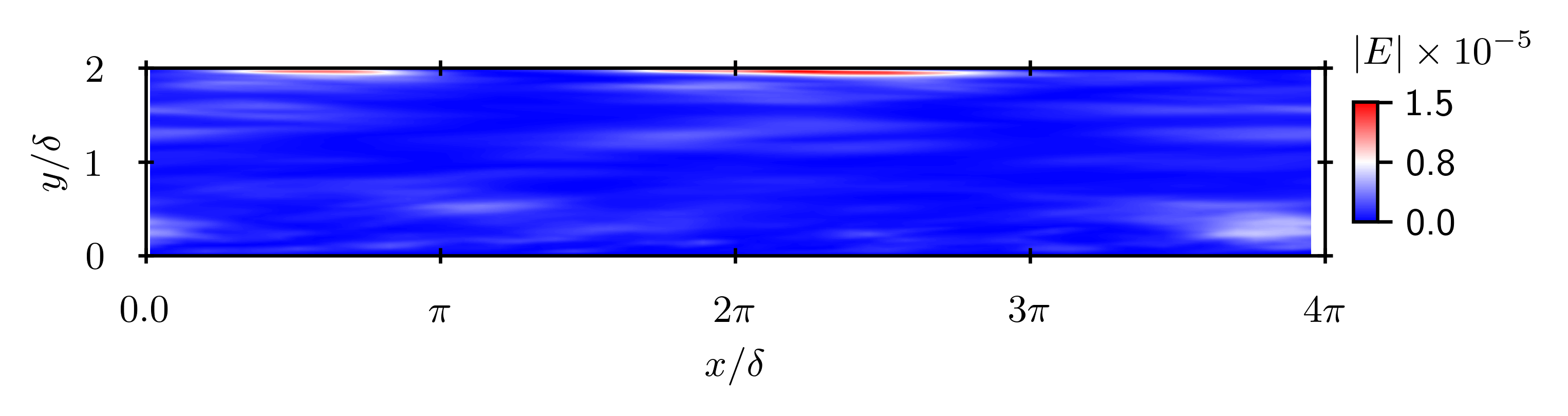}} \\ \vspace{-6mm}
    {\includegraphics[width=0.9\linewidth]{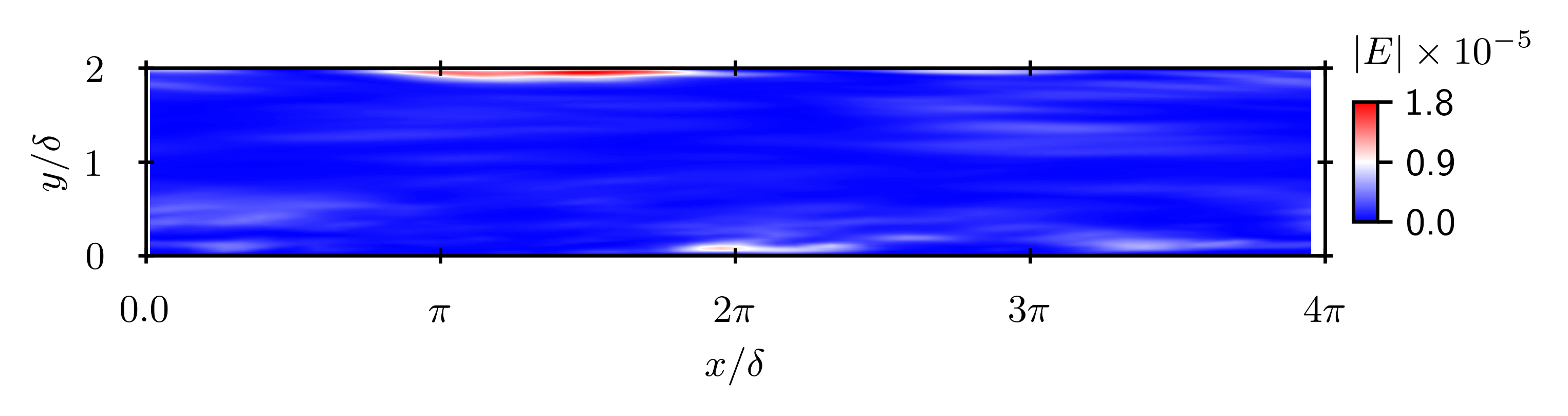}} \\ \vspace{-6mm}
    {\includegraphics[width=0.9\linewidth] {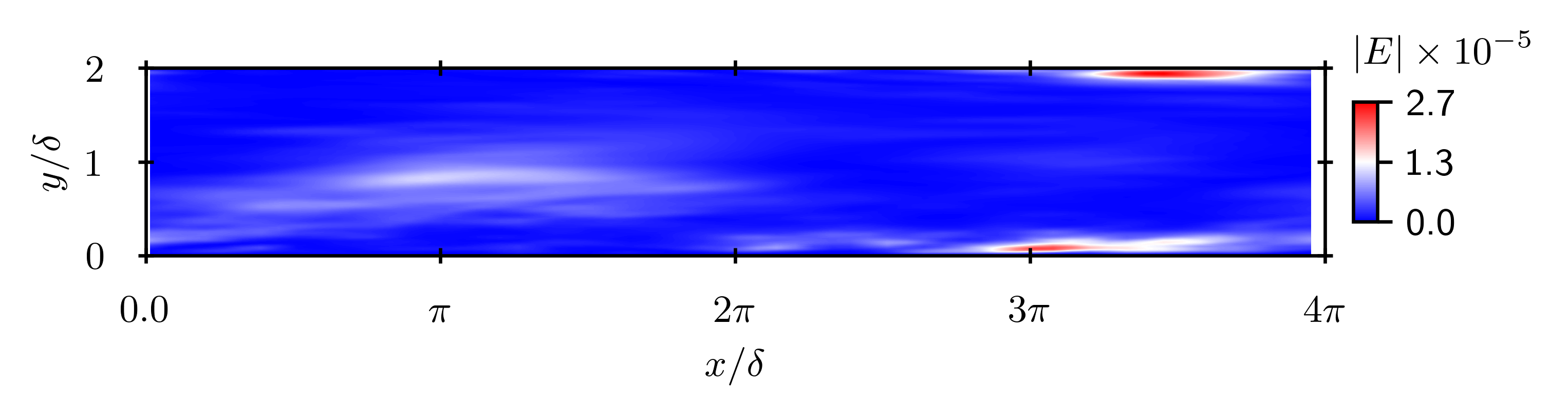}}  \vspace{-5mm}
	\caption{Normalized POD energy modes on the $x$-$y$ plane, shown from the first through fifth mode.} 
 \label{fig:POD_energy_mode_1_5}
\end{figure*}

\begin{figure*}
	\centering
    {\includegraphics[width=0.9\linewidth]{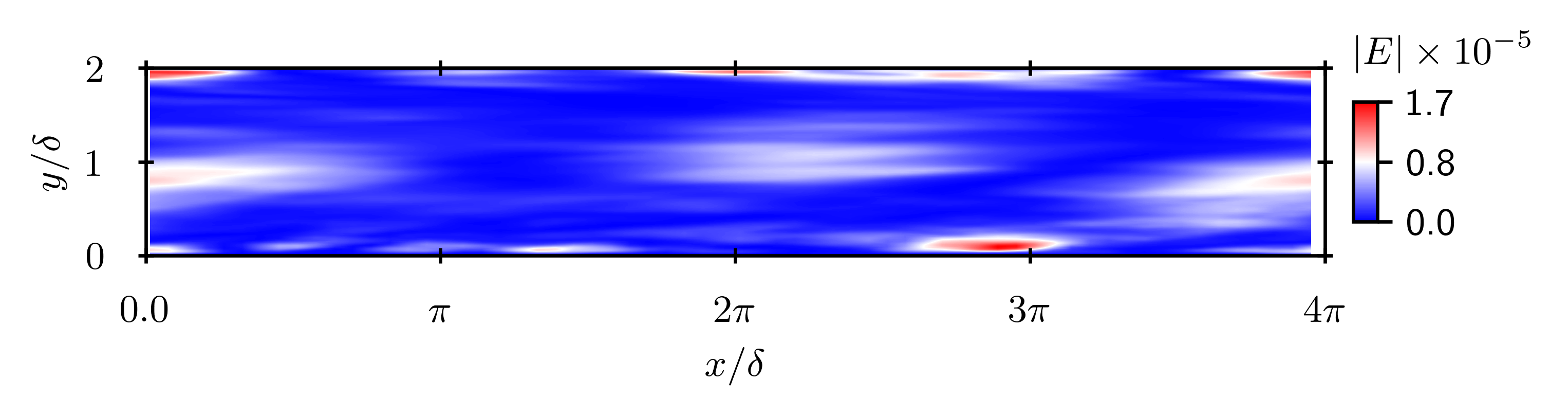}} \\ \vspace{-6mm}
    {\includegraphics[width=0.9\linewidth] {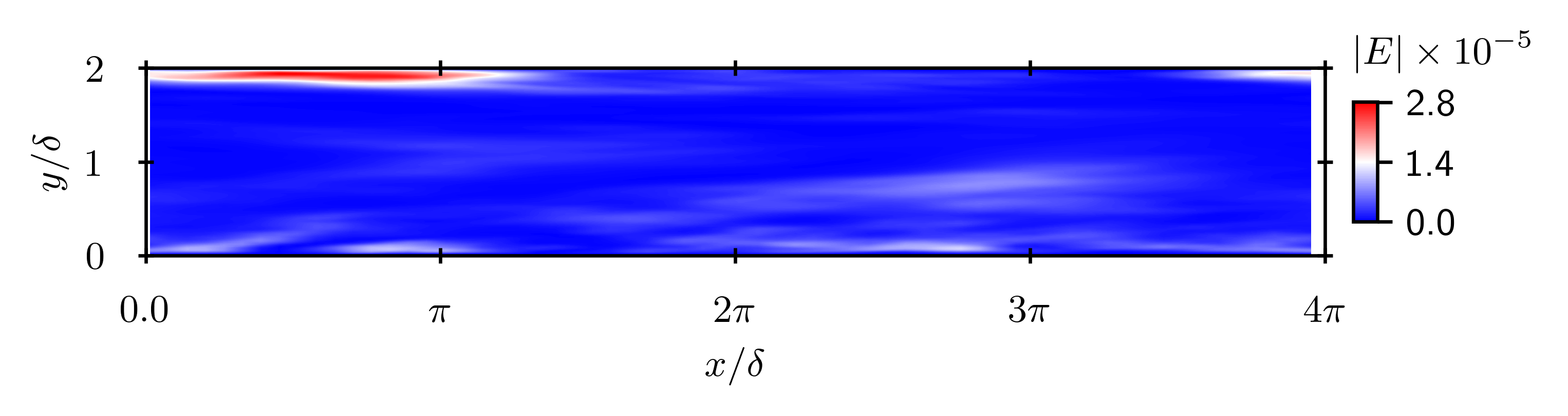}} \\ \vspace{-6mm}
    {\includegraphics[width=0.9\linewidth]{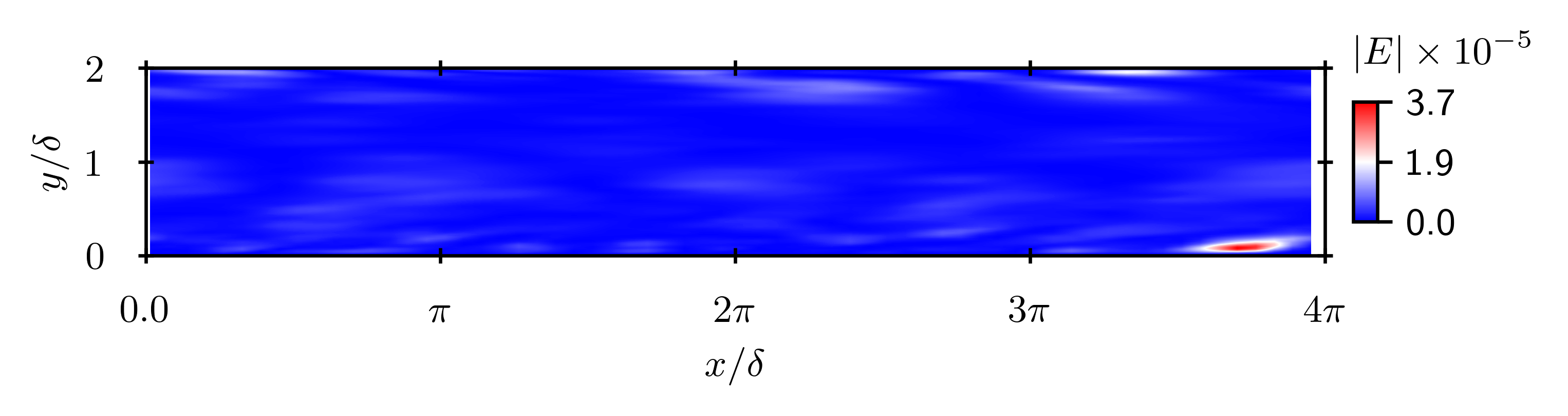}} \\ \vspace{-6mm}
    {\includegraphics[width=0.9\linewidth] {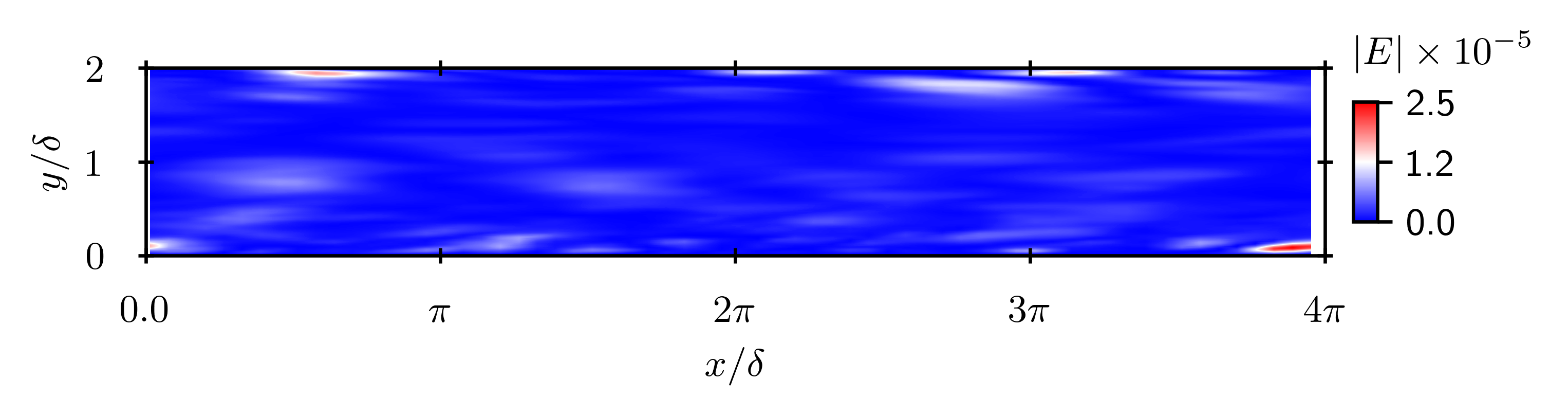}} \\ \vspace{-6mm}
    {\includegraphics[width=0.9\linewidth]{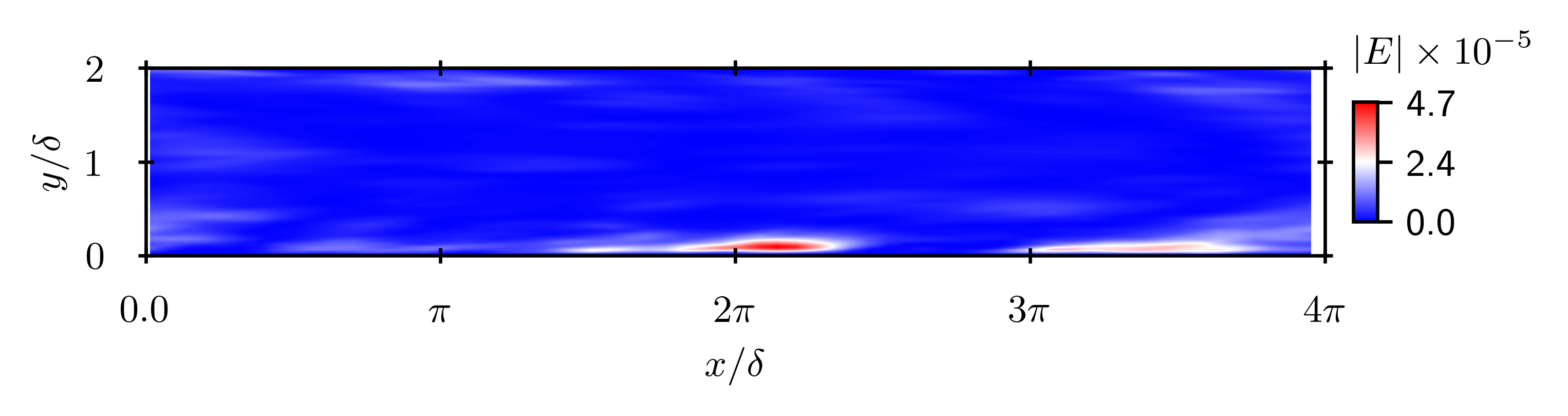}} \\ \vspace{-5mm}
	\caption{Normalized POD energy modes on the $x$-$y$ plane, shown from the sixth through tenth mode.} 
 \label{fig:POD_energy_mode_6_10}
\end{figure*}

To complement the phase-speed-filtered reconstructions and accurately identify spatial locations of peak energy amplification, POD has been performed on the full dataset. This approach enables precise quantification of the dominant energetic modes by first isolating regions of maximal energy content, followed by a detailed modal decomposition of the independent fluctuation variables ($\mathbf{q}$), thereby characterizing the coherent structures that govern the flow dynamics.
In this regard, Figures~\ref{fig:POD_energy_mode_1_5}–\ref{fig:POD_energy_mode_6_10} present the modal energy spectra for the $10$ leading POD eigenvalues, \review{defined in Eq.~\ref{eq:E_norm}}. \review{The color maps are normalized by the total energy such that values indicate fraction of total fluctuation energy contained in that mode across the spatial domain, linking the spectra to the velocity components and amplification mechanisms described in Sections~\ref{sec:resolvent_framework} and~\ref{sec:data_driven}}. The first mode captures the largest-scale, most energetic structures, predominantly localized near the cold/bottom wall, which is consistent with \review{classical} near-wall turbulence dynamics. This is evidenced by strong energy peaks in mode 1 concentrated at the cold wall, reflecting shear-driven motions typical of canonical turbulent channel flows. However, higher-order modes are preferred to isolate coherent features more relevant to turbulence dynamics and flow physics~\citep{Sirovich1987-A,Berkooz1993-A,Holmes2012-B}.
Notably, the modal shapes exhibit pronounced asymmetry about the channel centerline, arising directly from the thermodynamic and dynamic asymmetry induced by the proximity of the pseudo-boiling front near the hot/top wall. The vertical confinement and asymmetry of these energetic POD modes further emphasize the role of the pseudo-boiling region as a dynamic boundary-like layer that enhances thermofluid coupling.

\review{The fifth mode reveals a prominent energy peak near the hot/top wall, with its local energy relative to its own total mode energy exceeding that of other regions, indicating smaller-scale motions concentrated closer to the wall. In contrast, mode six shows energetic contributions at multiple wall-normal locations, with distinct high-energy patches spanning from the near-cold to near-hot wall. These discrete regions suggest the presence of large-scale, channel-spanning oscillations, or global modes, influenced by real-fluid thermodynamic effects, phenomena that are typically absent in isothermal wall-bounded flows.}
It is worth noting that global, channel-spanning modes have been reported in both isothermal channel~\citep{Kim1987-A} and Couette flows~\citep{Gibson2009-A,Tuckerman2011-A}; here, however, their amplitude and wall-to-wall coherence are markedly enhanced, highlighting the distinctive characteristics of the transcritical case.
Similarly, the seventh mode displays a pronounced energy peak near the hot/top wall, whereas the tenth mode concentrates its largest energy near the cold/bottom wall. These observations underscore the vertical heterogeneity of coherent motions throughout the channel, reflecting complex interactions between wall effects and transcritical thermodynamics.
Higher-order modes retain comparable spatial support but exhibit subtle phase shifts and modulations in their vertical extent, reinforcing the conclusion that a significant portion of the flow energy organizes into coherent structures adjacent to the pseudo-boiling region.

Collectively, the POD and phase-speed-filtered analyses underscore the emergence of energetically dominant structures that \review{classical} isothermal wall-bounded turbulence models fail to capture adequately. The coexistence of near-wall shear-driven modes (e.g., mode $1$), localized pseudo-boiling-driven structures (e.g., modes $5–7$), and channel-wide oscillations (mode $6$) reveals a flow regime where coherent organization arises from the complex interplay of wall-driven shear, and high-pressure transcritical thermodynamics. These findings imply that accurate modeling of transcritical fluid turbulence necessitates incorporation of localized production mechanisms and structural asymmetries, as conventional isothermal-based approaches are unlikely to represent the unique dynamics induced by the pseudo-boiling region.

\begin{figure*}
	\centering
	{\includegraphics[width=0.9\linewidth] {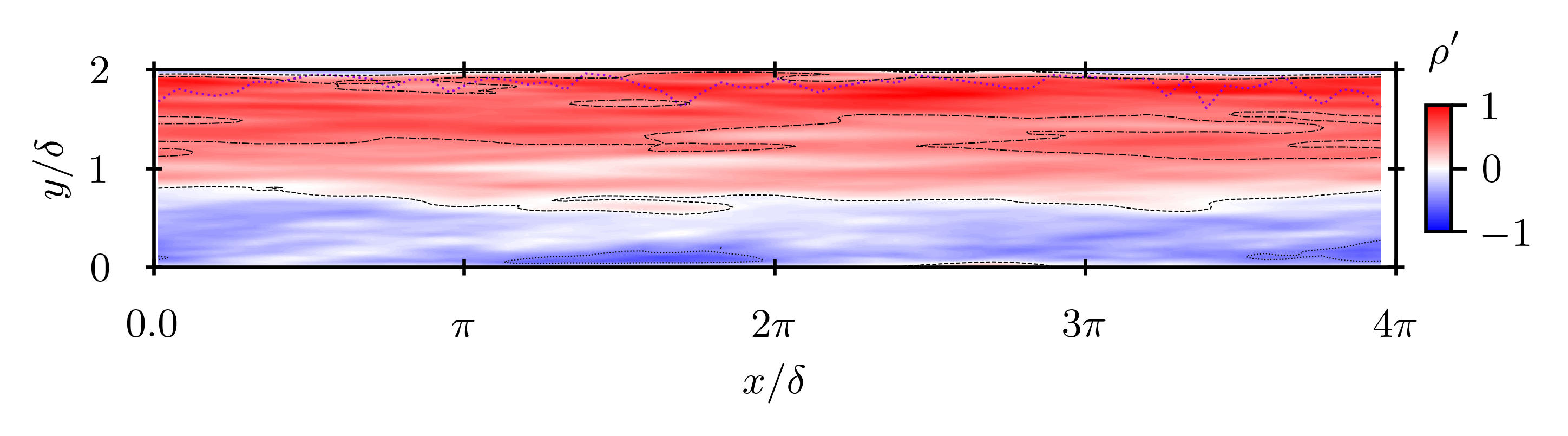}} \\ \vspace{-6.5mm}
    {\includegraphics[width=0.9\linewidth]{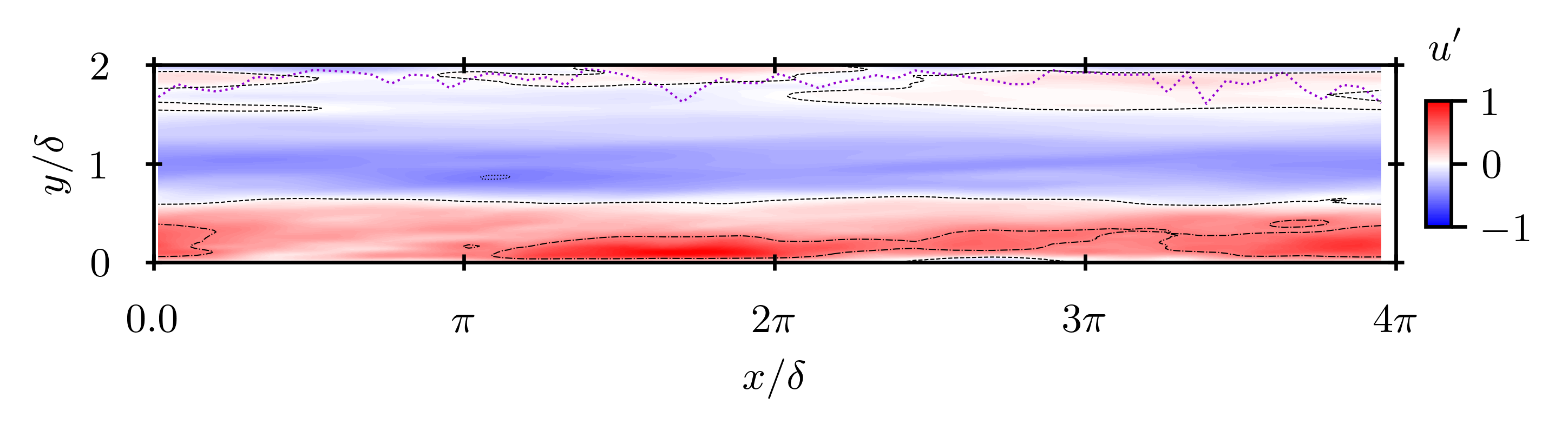}} \\ \vspace{-6mm}
    {\includegraphics[width=0.9\linewidth] {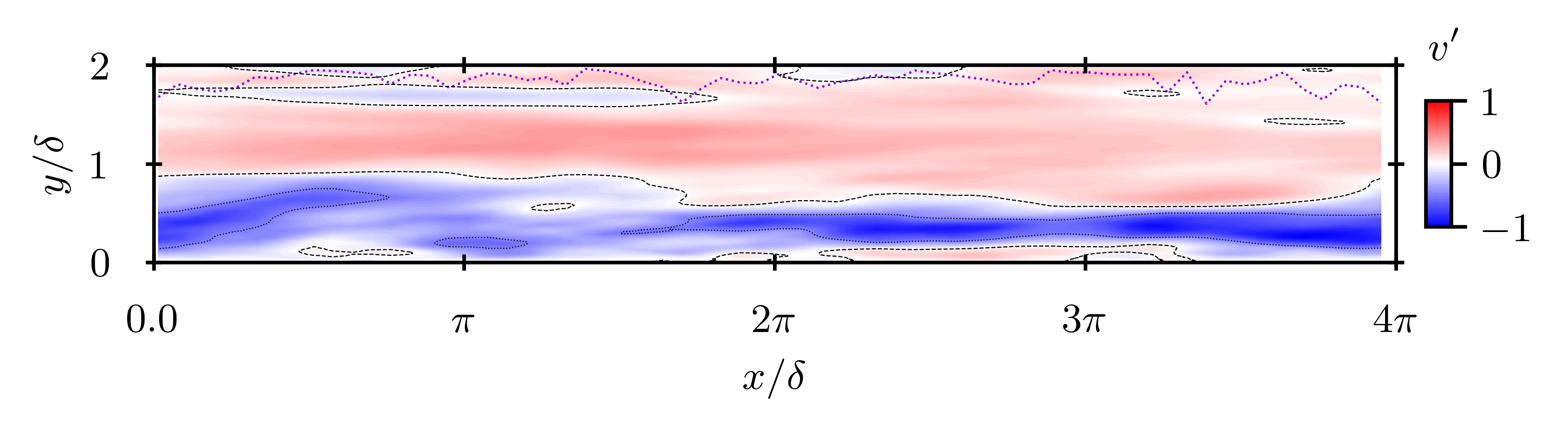}} \\ \vspace{-6mm}
    {\includegraphics[width=0.9\linewidth]{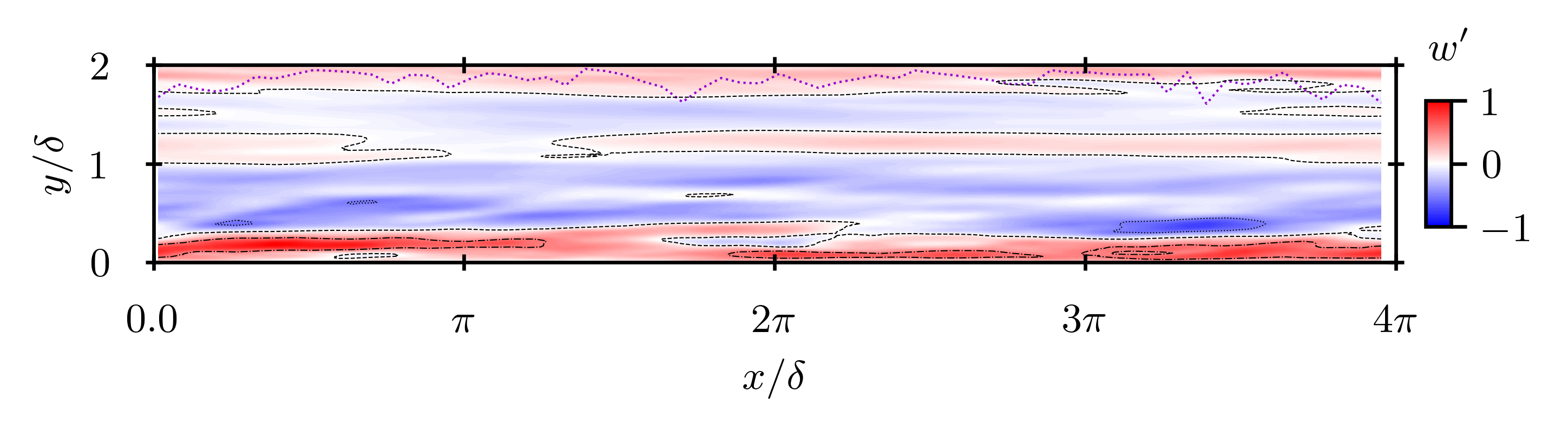}} \\ \vspace{-6mm}
    {\includegraphics[width=0.9\linewidth] {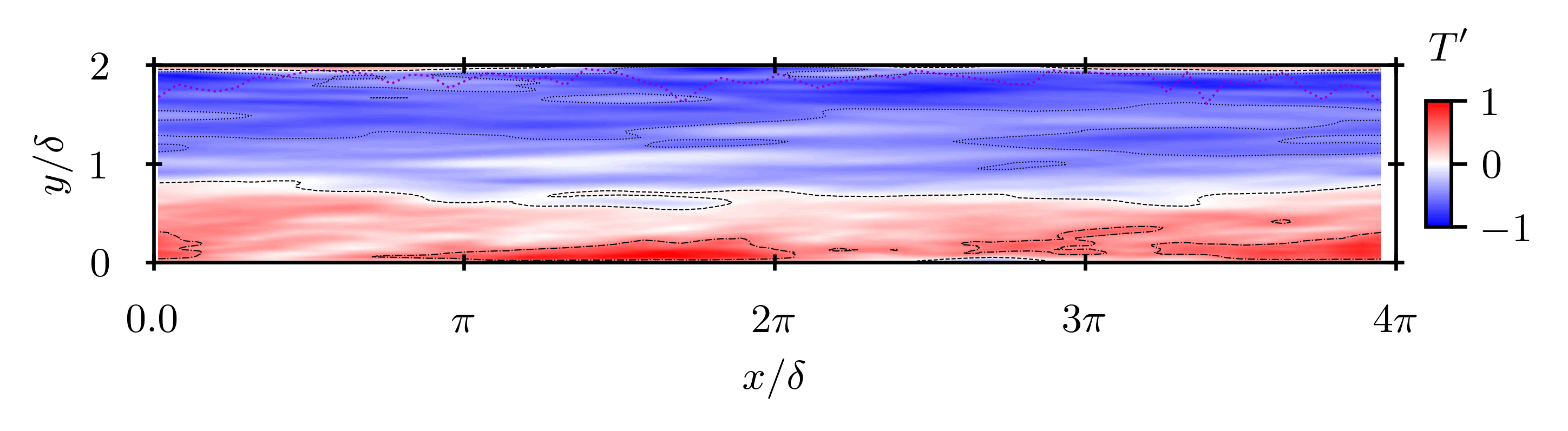}} \\ \vspace{-5mm}
	\caption{Normalized POD modes on the $x$-$y$ plane for the first mode, showing normalized density, streamwise, wall-normal, and spanwise velocities, as well as temperature. The instantaneous boiling line overlaid on each plot (purple dotted line) and selected isocontours at $-0.5$ (dotted), $0.0$ (dashed) and $0.5$ (dash-dotted) levels.} 
 \label{fig:POD_variable_mode_1}
\end{figure*}

\begin{figure*}
	\centering
    {\includegraphics[width=0.9\linewidth]{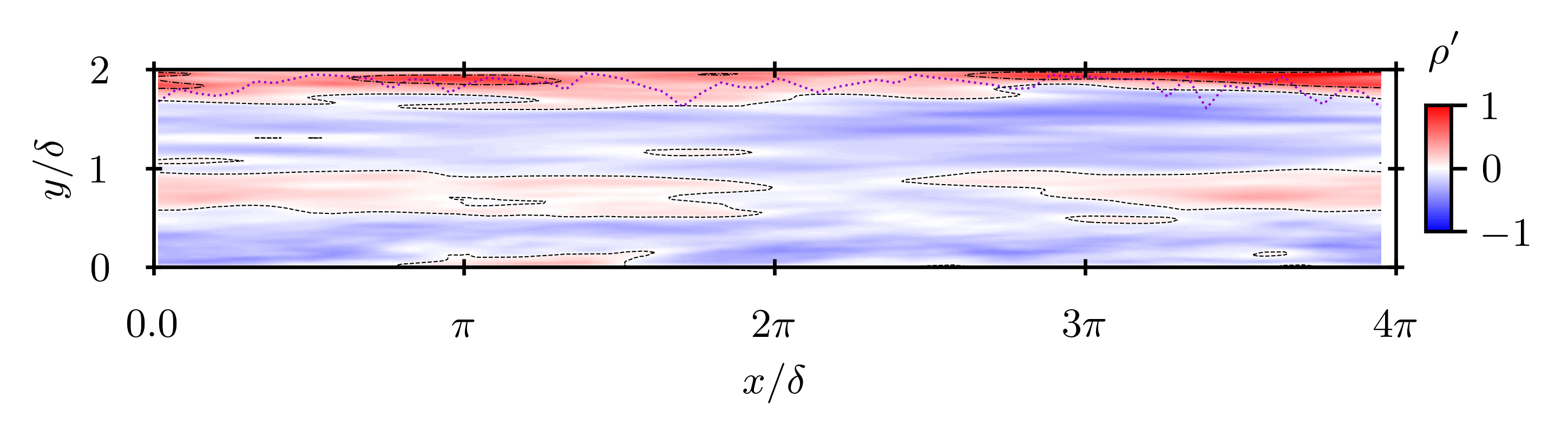}} \\ \vspace{-6.5mm}
    {\includegraphics[width=0.9\linewidth] {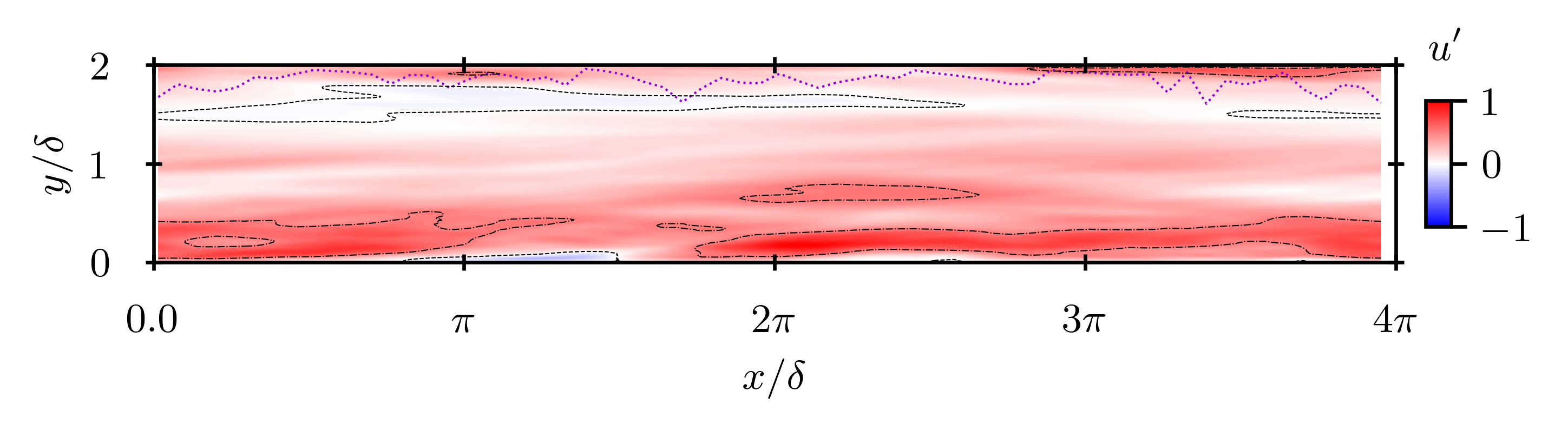}} \\ \vspace{-6mm}
    {\includegraphics[width=0.9\linewidth]{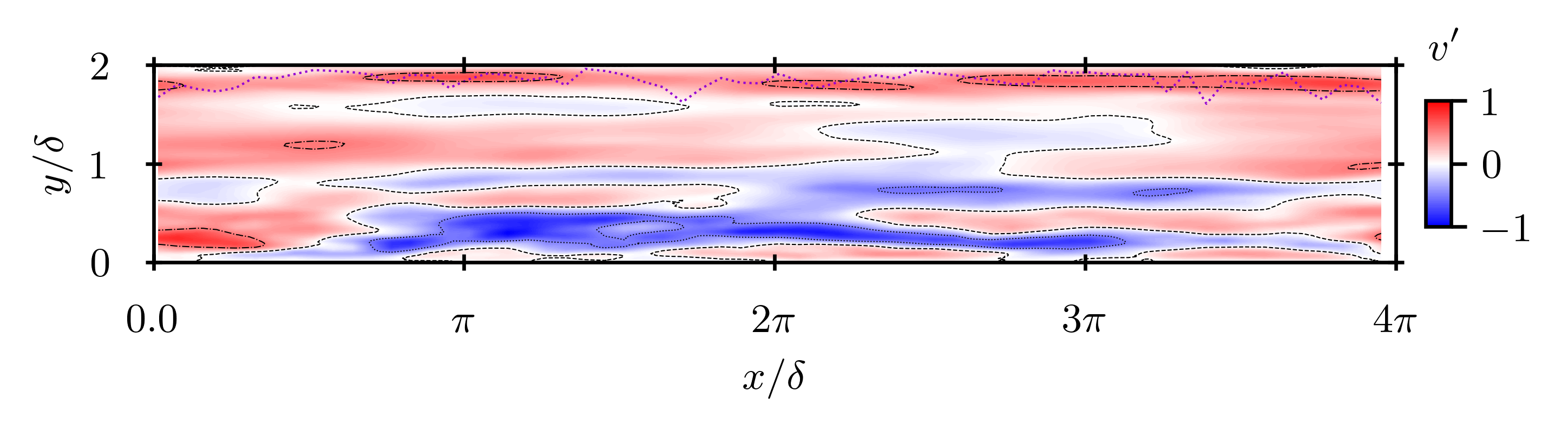}} \\ \vspace{-6mm}
    {\includegraphics[width=0.9\linewidth] {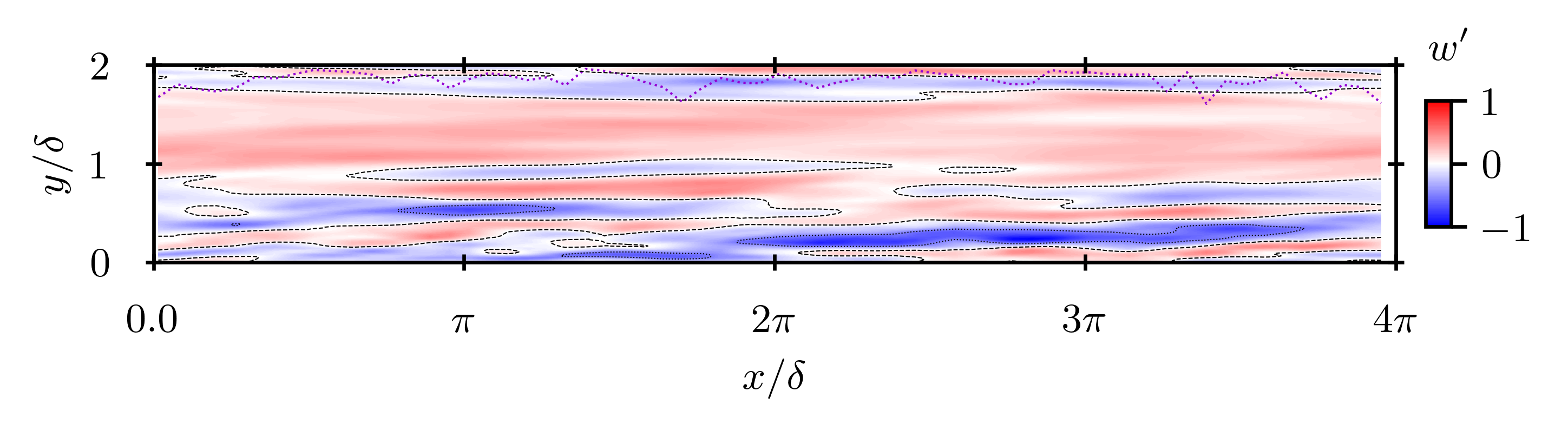}} \\ \vspace{-6mm}
    {\includegraphics[width=0.9\linewidth]{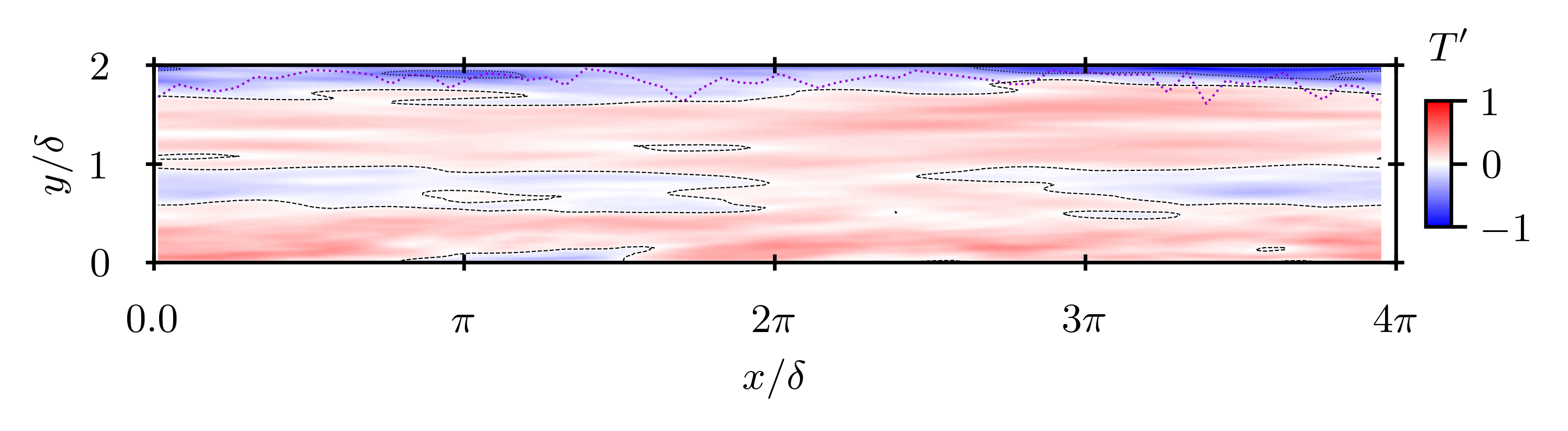}} \\ \vspace{-5mm}
	\caption{Normalized POD modes on the $x$-$y$ plane for the second mode, showing normalized density, streamwise, wall-normal, and spanwise velocities, as well as temperature. The instantaneous boiling line overlaid on each plot (purple dotted line) and selected isocontours at $-0.5$ (dotted), $0.0$ (dashed) and $0.5$ (dash-dotted) levels.} 
 \label{fig:POD_variable_mode_2}
\end{figure*}

Figures~\ref{fig:POD_variable_mode_1} and~\ref{fig:POD_variable_mode_2} present the spatial structures of the first and second POD modes, decomposed by fluctuation variables. Additional higher-order modes are provided in Appendix~\ref{sec:Appendix_B} for completeness.
The first mode, which carries the highest energy content and aligns closely with the mean flow, is primarily dominated by density and temperature fluctuations. These fields exhibit strong spatial asymmetry across the channel, reflecting the influence of the imposed thermal gradient and its impact on fluid properties. The concentration of fluctuations near the pseudo-boiling region underscores the coupling between energy transport and the local thermodynamic state.
Interestingly, despite its kinematic nature, the streamwise velocity fluctuation also displays a coherent structure in this region. Its localization between the hot/top wall and the pseudo-boiling line suggests an indirect modulation of momentum transport by thermal gradients, likely mediated by variations in local density and/or viscosity.
Furthermore, the wall-normal and spanwise velocity components exhibit alternating banded patterns near both walls; features reminiscent of \review{classical} ejection and sweep events. However, these structures are noticeably asymmetric, reflecting the strong thermal stratification inherent to the high-pressure transcritical flow regime.

The second mode further accentuates these features: viz. the spatial coherence of the thermodynamic fluctuations becomes more pronounced and increasingly confined near the pseudo-boiling region. This suggests the influence of a secondary mechanism, potentially linked to convective instabilities or intensified property gradients, that enhances coupling in this region. Additionally, the velocity components of the second mode exhibit greater phase complexity in the streamwise direction and finer-scale structures, particularly in the wall-normal and spanwise components, indicating the emergence of smaller, more localized energetic motions not captured by the leading mode.
These trends persist in the higher-order modes (Appendix~\ref{sec:Appendix_B}), where increased localization and sharper gradients in the fluctuation fields further underscore the critical role of near-wall thermodynamic phenomena in driving small-scale structures. Collectively, the POD results identify the pseudo-boiling region as a hotspot of coupled thermal–hydrodynamic activity across multiple energetic modes.
Moreover, the pronounced asymmetry of the modal structures about the channel centerline reflects the combined effects of thermodynamic stratification, pseudo-boiling-induced property variations, and wall-shear interactions. This asymmetry, which is absent in \review{classical} isothermal wall-bounded turbulence, reveals a fundamentally different organization of coherent structures in high-pressure transcritical channel flows.
\review{It should be noted that while resolvent analysis identifies modes with high amplification potential, the present study only provides qualitative validation using classical POD. A more rigorous, frequency-resolved assessment (e.g., via SPOD) could quantify the correspondence between resolvent-predicted amplification and the energetic content of DNS, and this represents an important avenue for future work.}

\section{Conclusions}   \label{sec:conclusions}

A resolvent analysis framework tailored for high-pressure transcritical wall-bounded flows has been developed and applied to a turbulent base flow obtained via an iterative solution of the linearized equations of fluid motion. This study, thus, aimed to bridge the gap between \review{classical} isothermal low-order decomposition methods and the complex dynamics of high-pressure transcritical flows through a comprehensive sensitivity analysis of resolvent operators, identifying the most responsive velocity modes to harmonic excitation. Singular value decomposition of the resolvent operator revealed that the largest amplification occurs at spanwise wavenumbers on the order of unity, consistent across laminar and turbulent regimes; however, the structure of forcings and responses differs markedly. Notably, the pseudo-boiling region plays a dominant role, particularly in the turbulent flow regime, where strong density and spanwise velocity forcings intensify response streaks that extend across the channel height. The identified scale motions highlight the constraining influence of the pseudo-boiling layer on large-scale structures, especially under low Reynolds number conditions. Complementary phase-speed-filtered analyses uncovered energetic turbulent structures propagating at anomalously low speeds, which are absent in incompressible wall-bounded turbulence. Furthermore, POD modes are predominantly localized between the hot/top wall and the pseudo-boiling region, indicating that this layer acts effectively as a pseudo-wall, modulating viscous damping and momentum transfer within the flow.

The sensitivity analysis of the resolvent operator confirmed that the transfer function exhibits a pronounced optimum at spanwise wavenumbers, while the streamwise wavenumber and temporal frequency act as low-pass filters favoring lower values for maximal amplification. The largest singular value ($\sigma_1$), evaluated at a phase speed corresponding to the bulk flow velocity, demonstrated parameter-space optima that behave analogously to transient growth with exponentially increasing energy amplification rates. Forcing structures are dominated by two pairs of counter-rotating vortices separated by the pseudo-boiling region: viz. weaker vortices reside above, while stronger, elongated vortices extend toward the cold/bottom wall. Based on the mean turbulent flow, the pseudo-boiling region significantly governs the sensitivity, where the optimal resolvent mode manifests as a high-energy ellipsoidal structure in the wall-normal direction.
Coherent structure analysis revealed the coexistence of large-scale motions near the wall and in the outer region, though their clear distinction was hindered by the relatively low friction Reynolds number. These scale motions remain confined by the pseudo-boiling layer, with large-scale motions exhibiting downstream tilt and vertical growth, while being narrower than very-large-scale motions. Data-driven spatio-temporal Fourier transform analyses corroborated these findings, highlighting energetic structures propagating at low phase speeds and confined between the hot/top wall and the pseudo-boiling zone, becoming progressively elongated in the streamwise direction.
POD further identified dominant energetic modes: while the most energetic structures are near the cold/bottom wall consistent with incompressible turbulence, lower-order modes localize high energy near the hot/top wall and pseudo-boiling region. Notably, the spatial coherence between streamwise velocity and density fluctuations underscored the interplay of hydrodynamic and thermodynamic mechanisms driving the flow dynamics.

Future investigations will focus on extending the \review{Fourier transform}-based methodology to fully three-dimensional domains to enable a comprehensive characterization of spanwise-elongated flow structures. Although three-dimensional POD has been applied in the present study, further detailed analysis of energetic and fluctuating modes (especially within the spanwise–wall-normal $y$–$z$ plane) remains an important avenue for exploration. To enhance the temporal resolution of coherent structures and validate POD results, advanced modal decomposition techniques, such as DMD and SPOD, could be employed.
\review{In particular, the present analysis serves as a foundational step toward applying resolvent analysis to high-pressure transcritical wall-bounded flows. Although the current evaluation is primarily qualitative, utilizing POD modes to characterize spatial organization, quantitative validation against SPOD modes is identified as a critical avenue for future research. Such an approach, capable of capturing both amplification and coherent forcing statistics, will facilitate a more rigorous assessment of the predictive capability of resolvent modes under the white-in-time forcing assumption and the influence of suboptimal modes.}
\review{In addition, the full distributions of resolvent singular values and POD eigenvalues could be examined to better quantify modal separation and energetic hierarchies, and to clarify how these trends differ from classical incompressible turbulence. Such analysis would strengthen the connection between resolvent predictions and observed flow structures, including the role of suboptimal modes.}
Concurrently, significant effort will be devoted to developing a data-driven resolvent analysis framework capable of addressing large-scale, nonlinear flow regimes. This initiative aims to extend and generalize existing linear frameworks to better predict and interpret complex flow dynamics in high-pressure transcritical conditions.
Ultimately, the goal is to optimize flow interactions along the pseudo-boiling region to improve mixing and heat transfer efficiency. To support this, new direct numerical simulation datasets will be generated under a wider range of operating conditions. These datasets could facilitate the construction of reduced-order models capable of real-time tracking of the instantaneous pseudo-boiling region through the superposition of resolvent mode iso-surfaces and scale-resolved flow structures, or via reconstruction using POD modes to obtain its representation within the full flow field.

\backsection[Supplementary data]{\label{SupMat}Supplementary material and movies are available at \url{https://github.com/marc-bernades/HPDDM/FFT/movies}.}

\backsection[Acknowledgements]{The authors acknowledge support from the \textit{Formaci\'o de Professorat Universitari} scholarship (FPU-UPC R.D 103/2019), the SGR program (2021-SGR-01045) of the \textit{Generalitat de Catalunya} (Catalonia), and the PID2023-150840OA-I00 grant of the \textit{Agencia Estatal de Investigación} (AEI, Spain).}

\backsection[Funding]{This work is funded by the European Union (ERC, SCRAMBLE, 101040379). Views and opinions expressed are however those of the authors only and do not necessarily reflect those of the European Union or the European Research Council. Neither the European Union nor the granting authority can be held responsible for them.}

\backsection[Declaration of interests]{The authors report no conflict of interest.}

\backsection[Data availability statement] 
{The analytical resolvent analysis presented in this work was conducted using an in-house \textsc{MATLAB} code named \textit{High-Pressure Resolvent Analysis} (\texttt{HPReAn}).  
The source code is openly available at: \url{https://github.com/marc-bernades/HPReAn}.
\texttt{HPReAn} is designed as a flexible tool for performing resolvent analyses under both ideal-gas and real-fluid thermodynamic conditions. It includes built-in wrappers for the \texttt{CoolProp} and \texttt{RefProp} libraries. The code is thoroughly commented for readability, and the repository contains a complete description, user instructions, and a setup guide.
The resolvent operator can be interrogated via a triplet-sweep (in temporal frequency and wavenumbers) or along a user-specified parameter space. While the code is currently configured for Poiseuille flow, it can be easily adapted to other wall-bounded flow configurations (e.g., Couette flow) by modifying the initial and boundary conditions.
The weighted operator formulation follows the definition presented in this paper, although the code is structured to allow easy customization if needed.
\texttt{HPReAn} depends on two additional packages that require installation beforehand: the \textit{High-Pressure Compressible Flow Solver} (\texttt{HPCFS}), available at \url{https://github.com/marc-bernades/HPCFS}, and the \textit{High-Pressure Linear Stability Analysis} (\texttt{HPLSA}), found at \url{https://github.com/marc-bernades/HPLSA}, which provide supporting functionality and thermodynamic models required by \texttt{HPReAn}.
In parallel, the data-driven methods (referred to as \texttt{HPDDM}) used in this study are implemented in \textsc{Python} and provided in a separate repository: \url{https://github.com/marc-bernades/HPDDM}. This package includes class definitions and functionalities for: (i) the data-driven resolvent operator; (ii) \review{Fourier transform}-based phase speed gating and generation of flow-field slices and movies; and (iii) POD techniques for mode extraction, eigenvalue/eigenvector analysis, and reconstruction of flow fields and energetic structures.}

\backsection[Author ORCIDs]{Marc Bernades, https://orcid.org/0000-0003-3761-2038; Jonathan M.~O. Massey, https://orcid.org/0000-0002-2893-955X; Beverley J. McKeon, https://orcid.org/0000-0003-4220-1583; Llu\'is Jofre, https://orcid.org/0000-0003-2437-259X.}

\backsection[Author contributions]{Marc Bernades: Conceptualization, Formal analysis, Investigation, Software, Visualization, Writing – original draft; Jonathan M.~O. Massey: Software; Beverley J. McKeon: Conceptualization, Investigation, Writing – review \& editing; Llu\'is Jofre: Conceptualization, Funding acquisition, Investigation, Writing – review \& editing.}


\appendix

\section{Pre-multiplied spectra of turbulent kinetic energy} \label{sec:Appendix_A}

\begin{figure*}
	\centering
	{\includegraphics[width=0.49\linewidth]{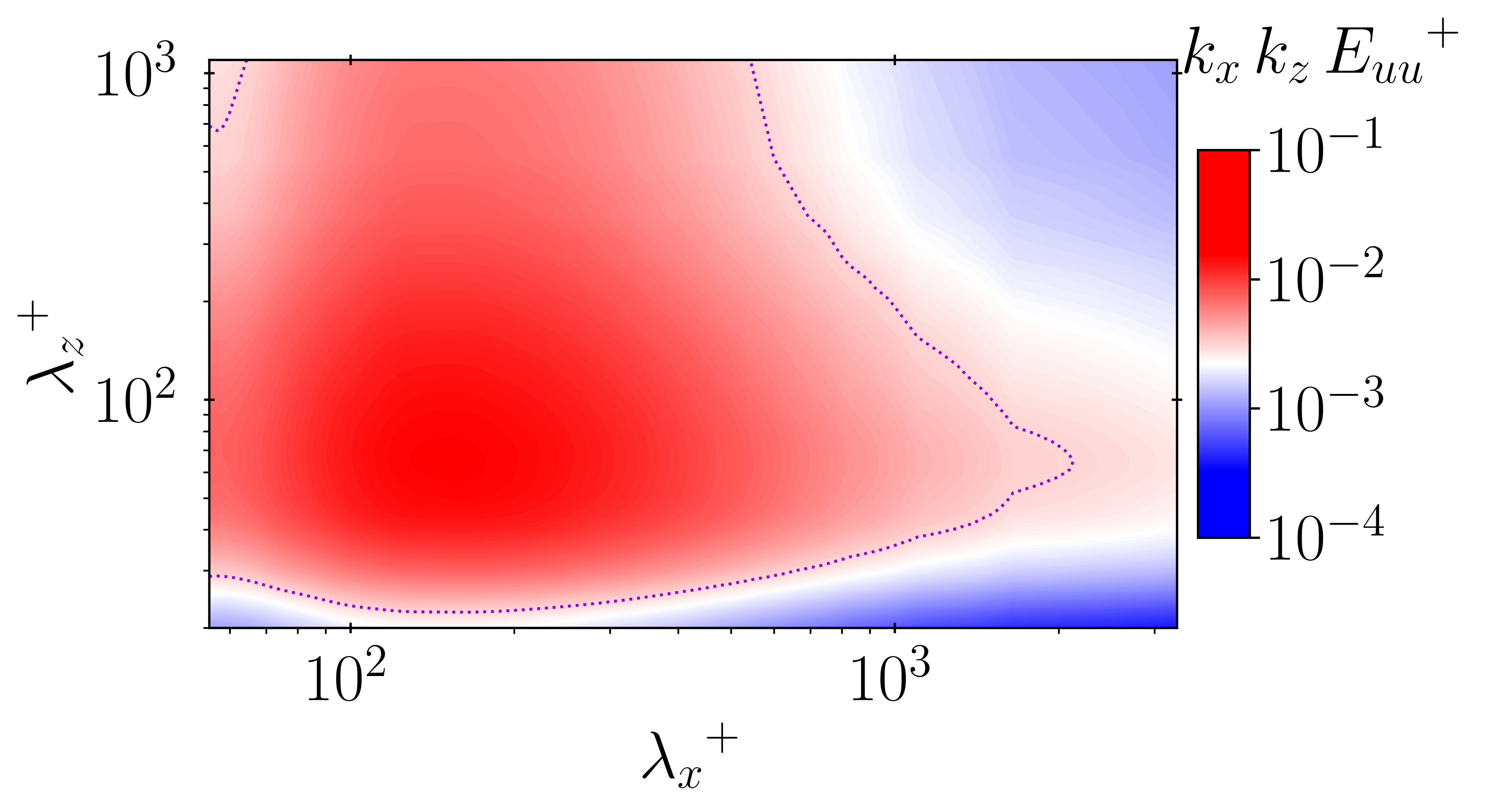}}
    {\includegraphics[width=0.49\linewidth]{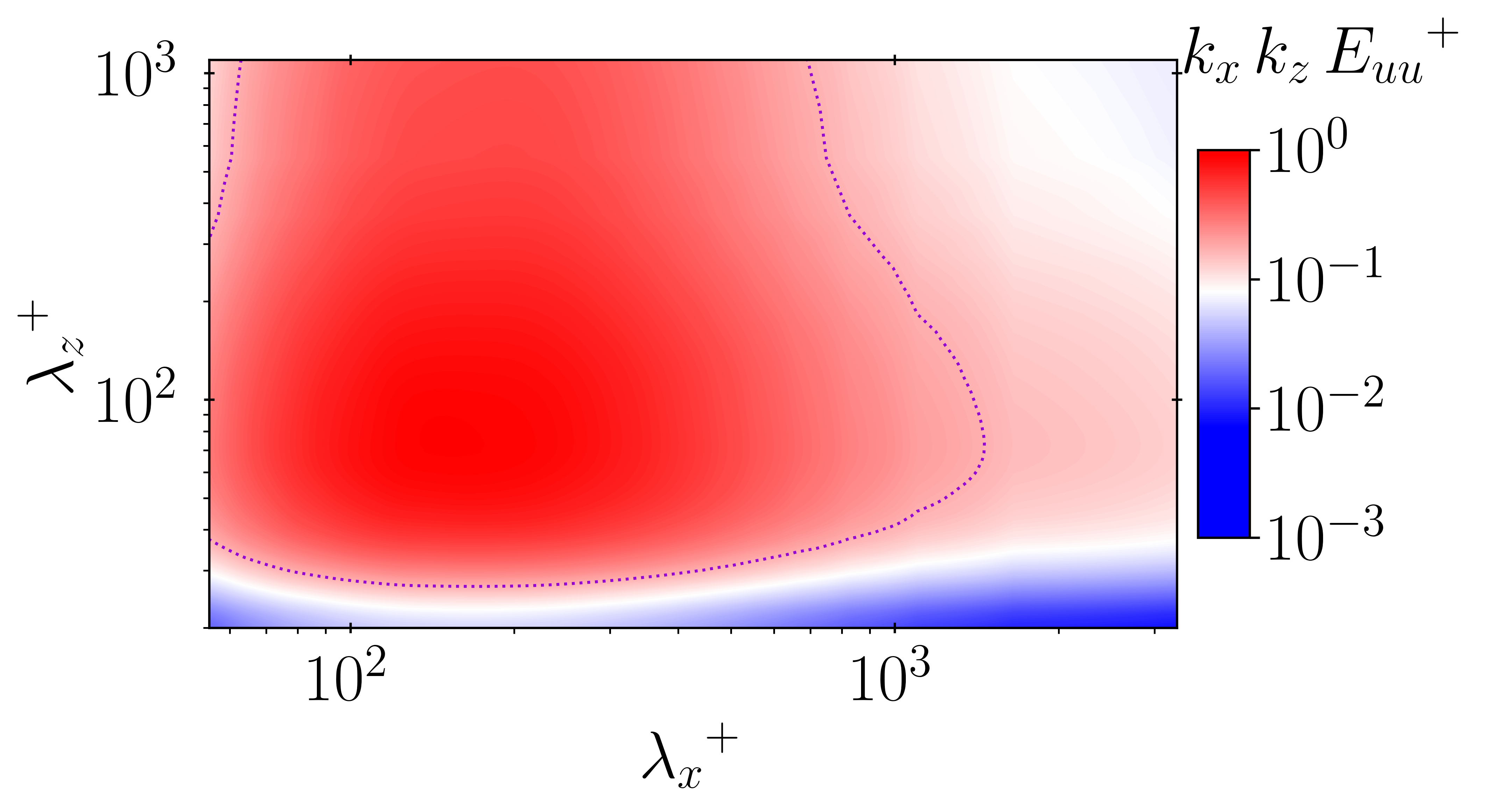}} \\
    {\includegraphics[width=0.49\linewidth]{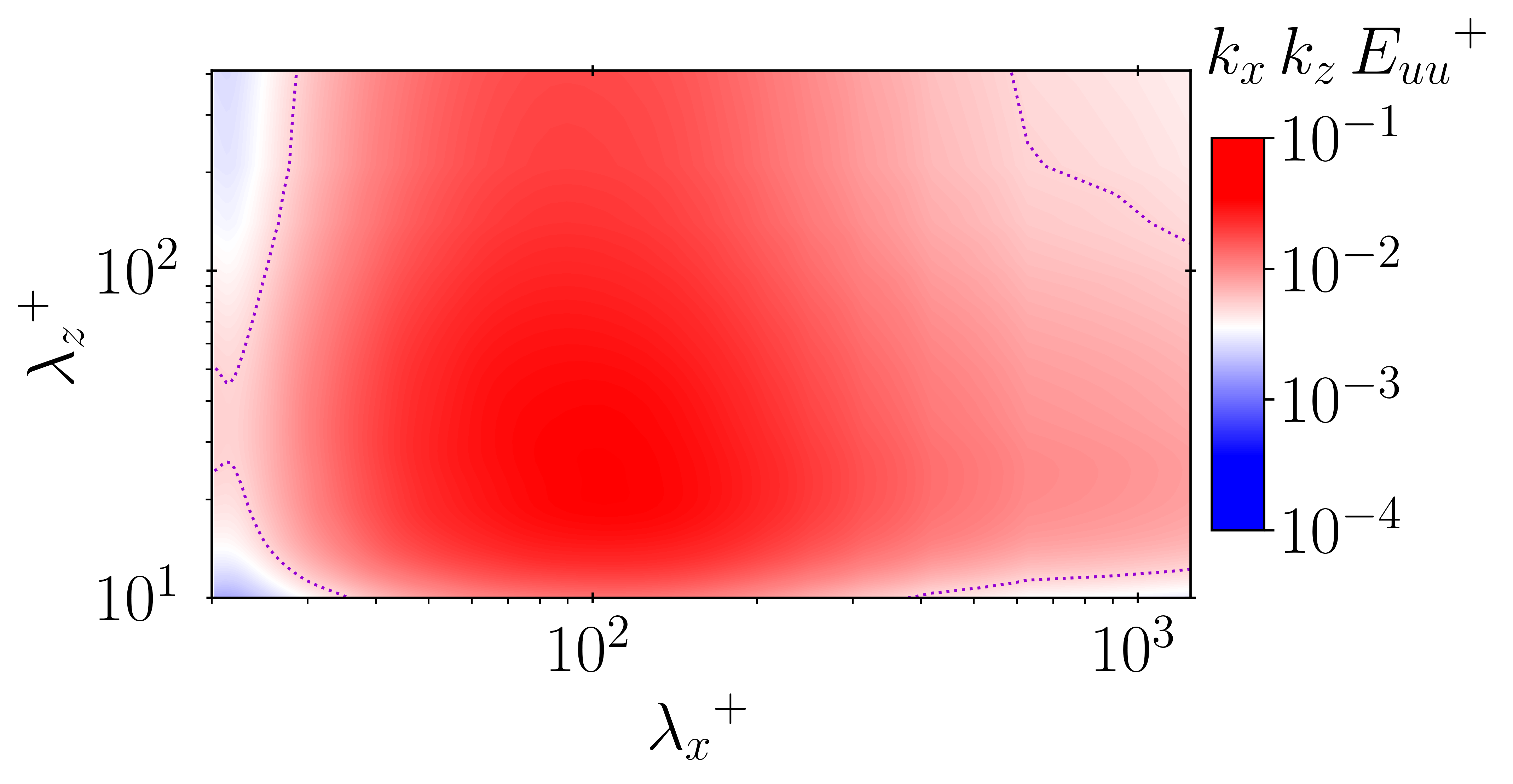}}
    {\includegraphics[width=0.49\linewidth]{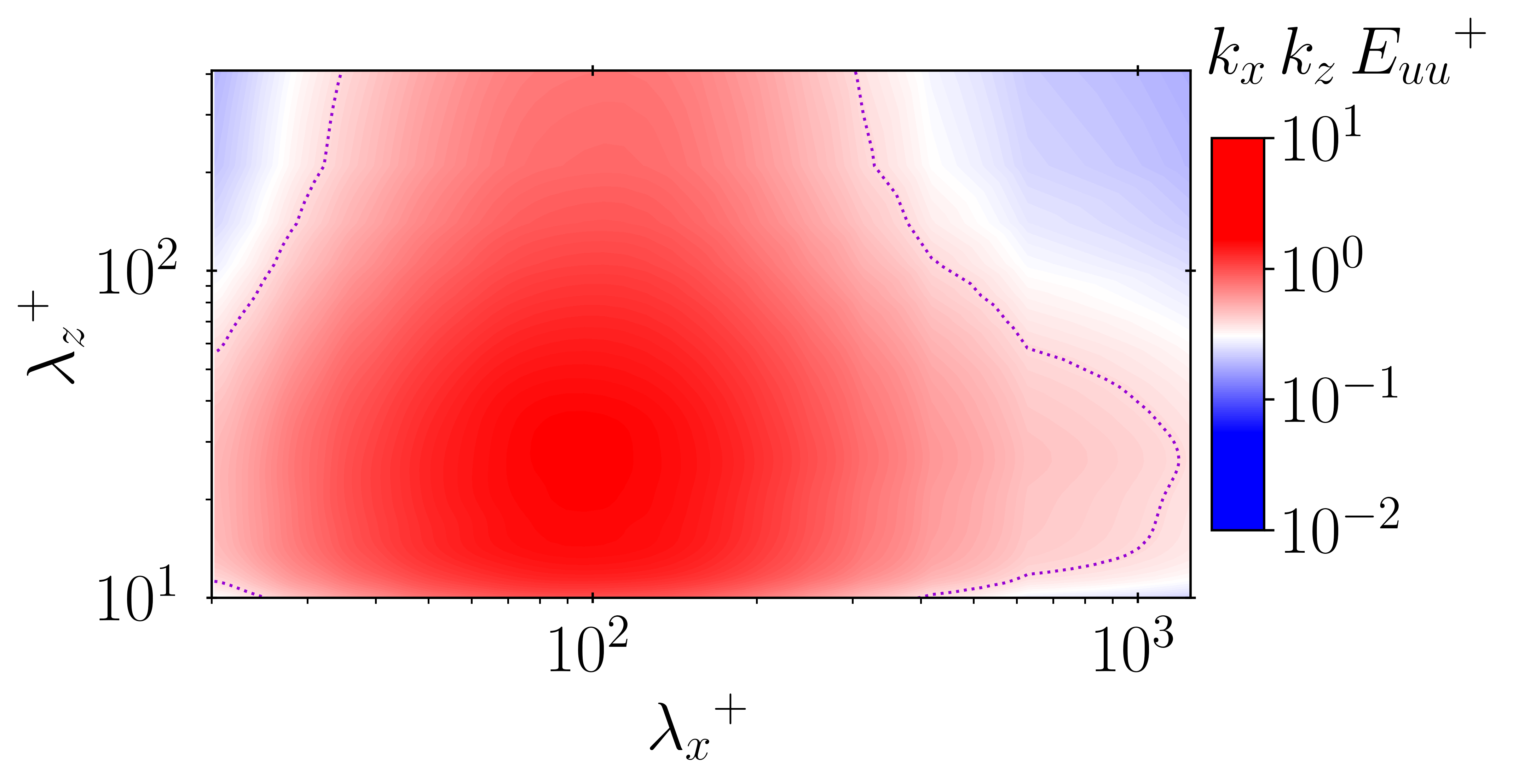}} \\ \vspace{2mm}
	\caption{Pre-multiplied energy spectra normalized in wall units at the hot/top and cold/bottom walls for $y^+ = 1$ (left) and $y^+ = 10$ (right), with contours indicating $90\%$ energy levels.} 
 \label{fig:premultiplied_energy_spectra}
\end{figure*}

Figure~\ref{fig:premultiplied_energy_spectra} depicts the ensemble-averaged pre-multiplied TKE spectra of the streamwise velocity component across the entire dataset for reference.

\begin{figure*}
	\centering
	{\includegraphics[width=0.495\linewidth]{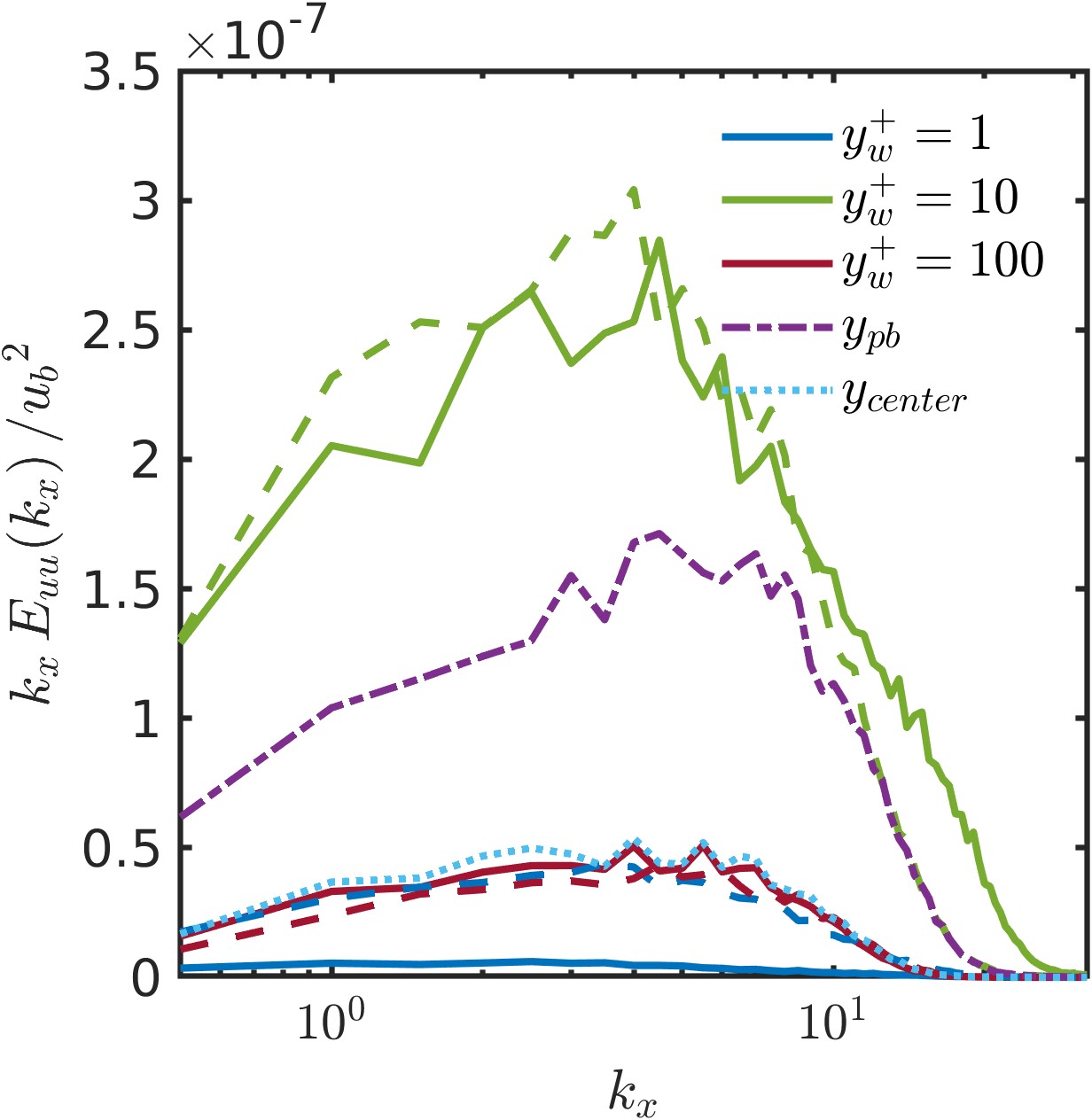}} \hfill
    {\includegraphics[width=0.49\linewidth]{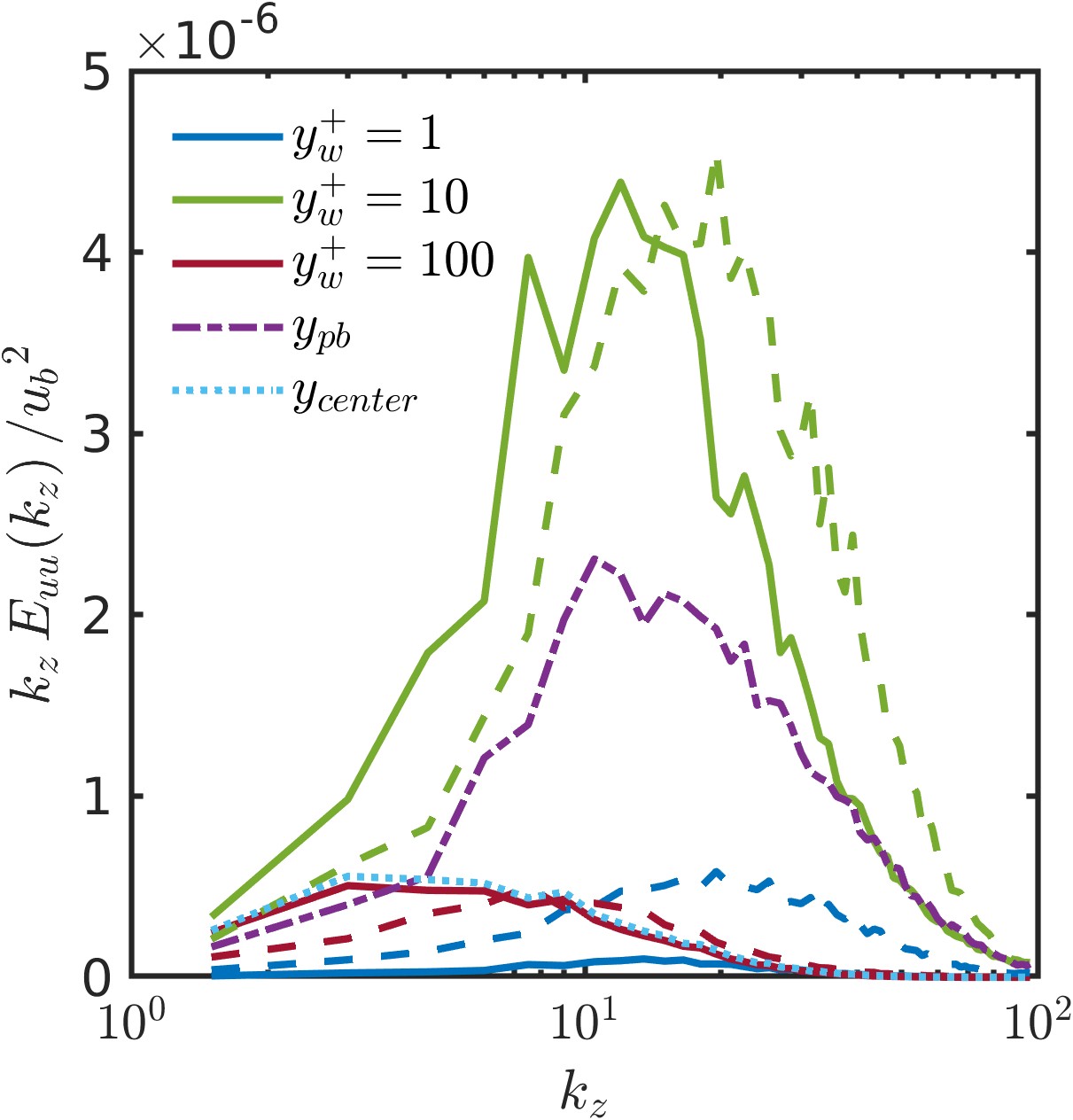}} \\ \vspace{-2mm}
	\caption{Premultiplied time averaged spatial normalized streamwise energy spectra ($E_{uu}$) at different $y$-planes on stream- (left) and spanwise direction (right). The solid- and dashed-lines denote $y_{bw}^+$ \review{(cold/bottom)} and $y_{tw}^+$ \review{(hot/top) walls}, respectively, with same color legend.} 
 \label{fig:Energy_spectra}
\end{figure*}

In addition, to serve as a basis for coherent structures, the most energetic spatial wavenumbers $k_x$ and $k_z$ of the turbulent channel flow are identified by means of the spectrum in the periodic dimensions of the channel. Multiple datasets have been time-averaged and the subsequent Fourier transformations for stream- and spanwise directions at different $y$-planes.
Thus, Figure~\ref{fig:Energy_spectra} illustrates the spectrum at various wall-normal locations for streamwise normalized kinetic energy component $E_{uu}$ (denoting spectral density). In brief, streamwise energy spectra are predominant, hence, further decomposition on $E_{vv}$ and $E_{ww}$ is found in~\citet{Bernades2024-B} along with the generalized turbulent kinetic energy spectra.
Moreover, Table~\ref{tab:energetic_wavenumbers} quantifies the wavenumbers and associated wavelengths ($\lambda = 2\pi / k$) corresponding to the most energetic regions of the spectra, utilized to compute resolvent modes of Section~\ref{sec:scale_motions}.
The phase speed of each mode is determined from the power spectral density (PSD) of temporal probe signals, combined with the corresponding mean velocity profile at each wall-normal location~\citep{Bernades2024-B}. Notably, the extracted physical parameters deviate from canonical scale hierarchies in wall-bounded turbulence~\citep{Saxton-Fox2017-A}.

To this extent, \review{as observed in Figure~\ref{fig:Energy_spectra}}, a distinct peak is noticeable in the near-wall region, specially at $y_{w}^+ = 10$ which approximately coincides with the turbulence production peak also, at wavelengths of ${\lambda_{x,cw}}^+ = 130$ and ${\lambda_{x,hw}}^+ = 430$. This indicates that most energetic streamwise velocity fluctuations are principally of this length due to the viscous-scaled near-wall structure of elongated high- and low-speed regions~\citep{Kline1967-A}.
At sufficient high-Reynolds-numbers, the length scales of the energetic peak at half-channel height are considerably larger, commonly referred as the outer peak~\citep{Hutchins2007b-A}. Nonetheless, this bimodal composition is not observed due to the broad nature of such low-Reynolds-regimes, which result in diminished energy levels in the outer site. Regardless, the near-wall wall structure obscures any prominent secondary peak.
The energy of the viscous sublayer is comparatively low relative to the buffer sublayer and log-law region, especially at the bottom wall.
Likewise, the wavelength shift between outer and inner peaks is noted in spanwise direction with length scale of $5\times$ at the channel center.
In this regard, based on these results, the triplet definition to reproduce the coherent structures presented in Section~\ref{sec:scale_motions} has been assigned as follows. First, although \textit{a priori} there is no distinction between LSM and VLSM, the resolvent operator has been interrogated, \review{in outer scales ($\lambda_x,\lambda_z,c$)}, for (i) LSM at $(1,0.5,0.96)$ corresponding to the phase speed at channel center, and (ii) VLSM $(10,1,1.09)$ noting the velocity at pseudo-boiling line height ($y/\delta \approx 1.9$). Next, the NWSM has been determined in viscous units ($\lambda_x^+,\lambda_z^+,c^+$) as $(200,50,0.5)$ and $(450,100,10)$ for cold/bottom and hot/top walls respectively.

\begin{table}
\centering
\tabcolsep=0.10cm
\renewcommand{\arraystretch}{1.2}
\begin{tabular}{c|ccc|ccc|cc} \hline
\centering
 & \review{NWSM} & \review{LSM} & \review{VLSM} & \review{NWSM} & \review{LSM} & \review{VLSM} & $-$ & $-$ \\
 & $y_{cw}^+ = 1$ & $y_{cw}^+ = 10$ & $y_{cw}^+ = 100$ & $y_{hw}^+ = 1$ & $y_{hw}^+ = 10$ & $y_{hw}^+ = 100$ & $y_{pb}$ & $y_{center}$ \\
\hline
$k_x$ & $3$ &  $5$ & $5$ & $4$ &  $4$ & $5$ & $5$ &  $5$ \\
$\lambda_x$ & $2.1$ &  $1.3$ & $1.3$ & $1.6$ &  $1.6$ & $1.3$ & $1.3$ &  $1.3$ \\
${\lambda_x}^+$ & $210$ &  $130$ & $130$ & $430$ &  $430$ & $350$ & $-$ &  $-$ \\ \hline
$k_z$ & $15$ &  $15$ & $3$ & $20$ &  $20$ & $7$ & $8$ &  $3$ \\
$\lambda_z$ & $0.4$ &  $0.4$ & $2.1$ & $0.3$ &  $0.3$ & $0.9$ & $0.8$ &  $2.1$ \\
${\lambda_z}^+$ & $40$ &  $40$ & $210$ & $80$ &  $80$ & $240$ & $-$ &  $-$ \\ \hline
$\omega$ & $4.8$ &  $9.5$ & $10.1-23.5$ & $8.6$ &  $10.0$ & $10.1-23.5$ & $8.8-20.5$ &  $10.0-23.5$ \\ \hline
\end{tabular}
\caption{Normalized stream- and spanwise wavenumbers and associated wavelengths of the most energetic the premultiplied $E_{uu}$ turbulent spectra.}\label{tab:energetic_wavenumbers}

\end{table}

\section{\review{Fourier transform}-filtered coherent structures and POD modes} \label{sec:Appendix_B}





\begin{figure*}
	\centering
	{\includegraphics[width=0.9\linewidth] {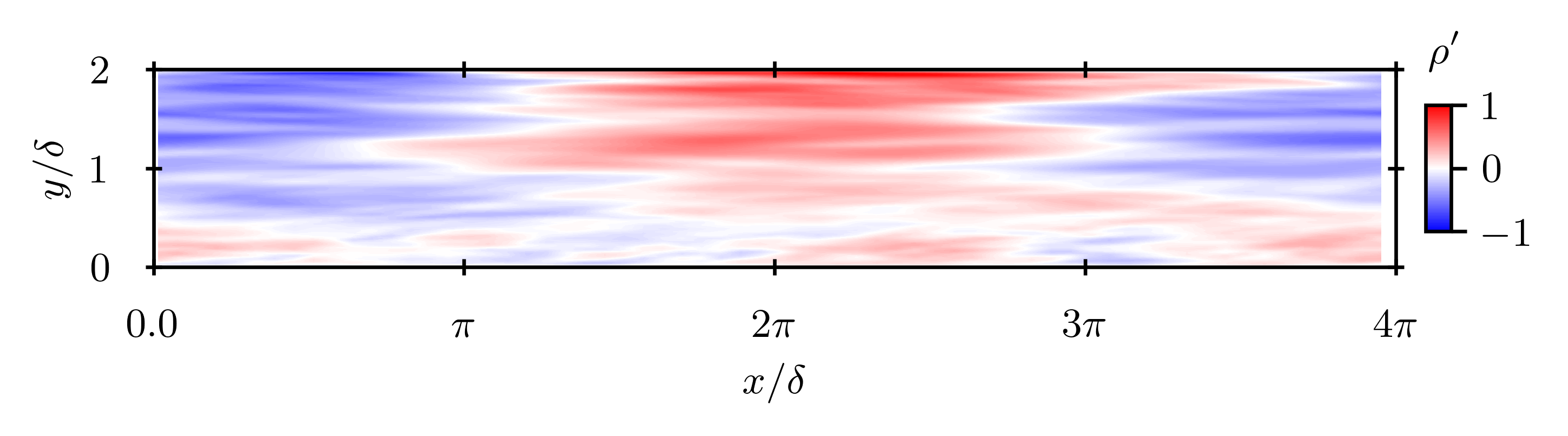}}  \\ \vspace{-6.5mm}
    {\includegraphics[width=0.9\linewidth]{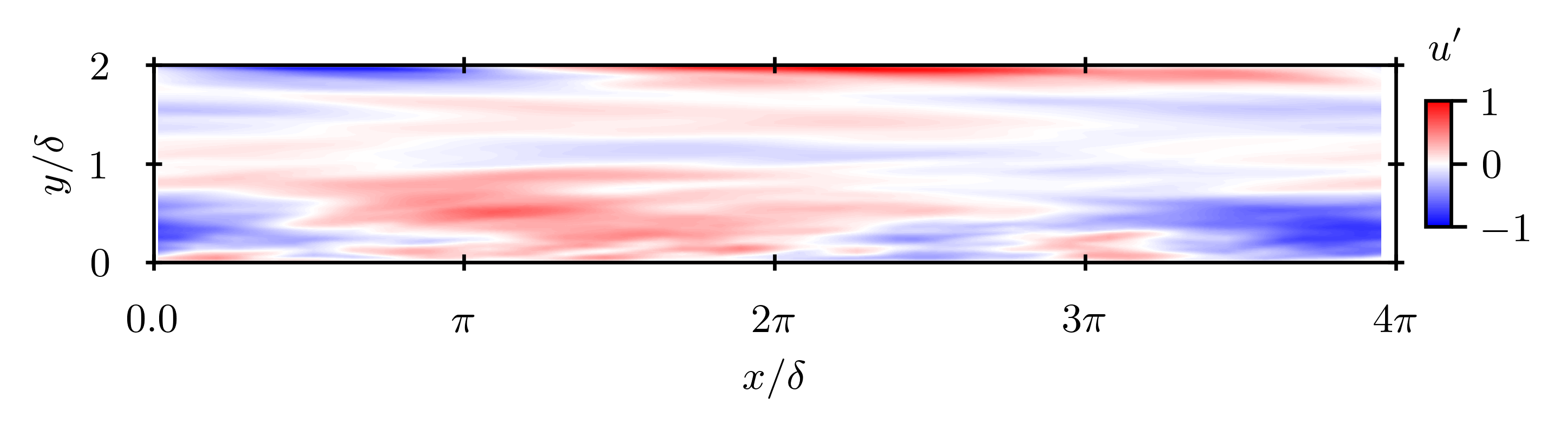}} \\ \vspace{-6mm}
    {\includegraphics[width=0.9\linewidth] {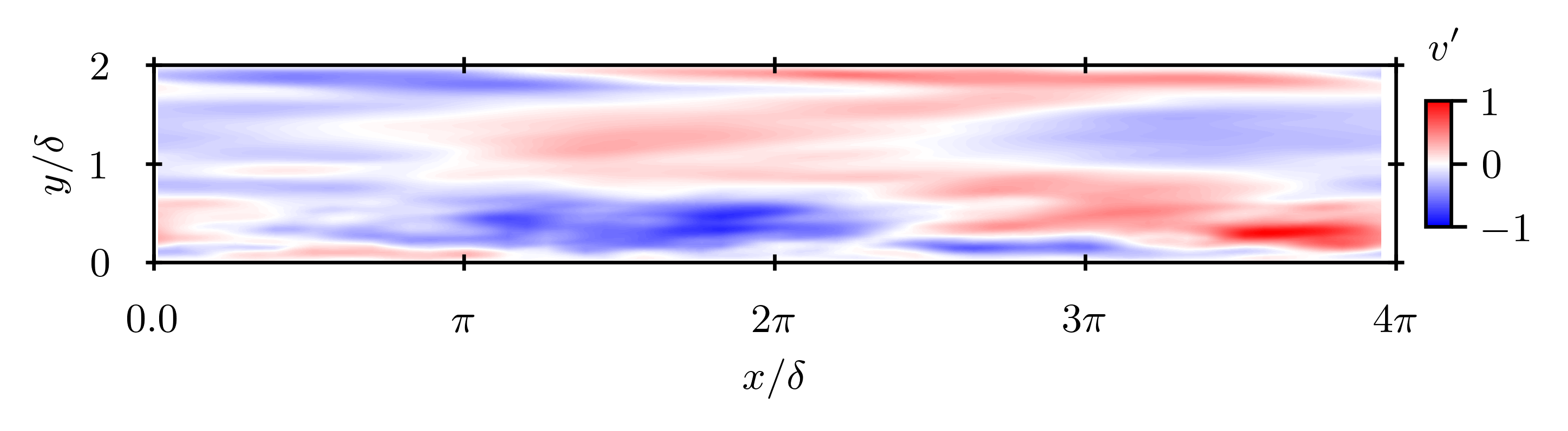}}  \\ \vspace{-6mm}
    {\includegraphics[width=0.9\linewidth]{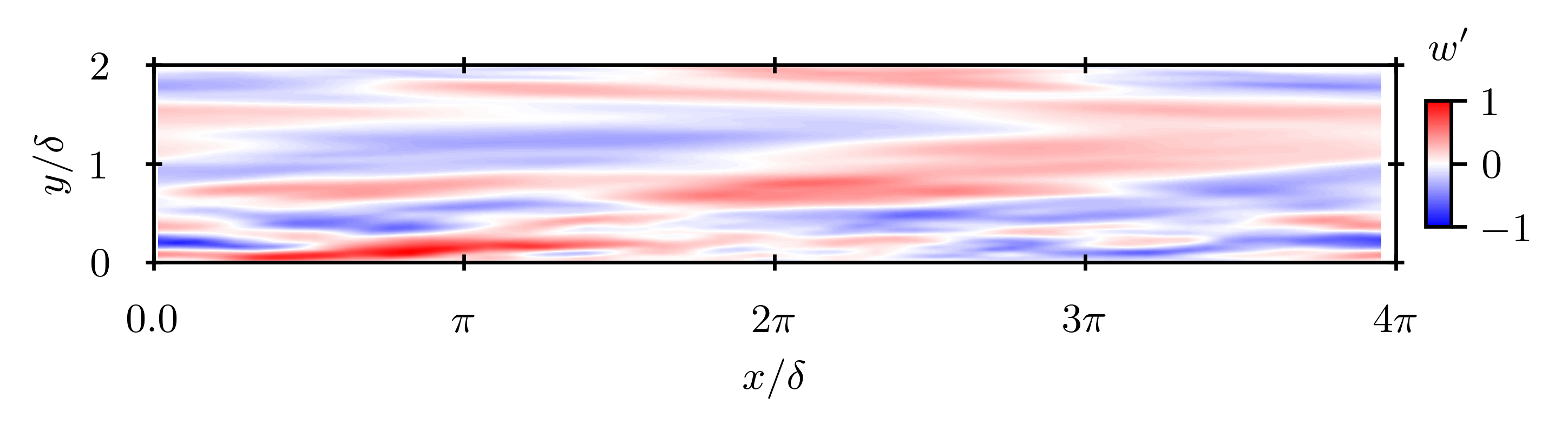}} \\ \vspace{-6mm}
    {\includegraphics[width=0.9\linewidth] {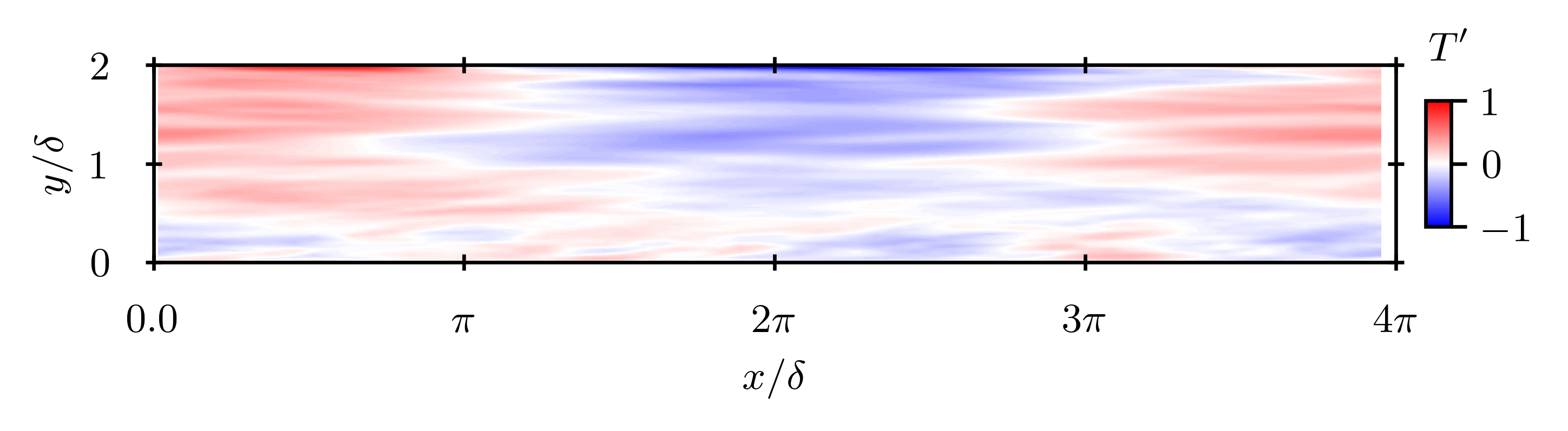}}  \\ \vspace{-5mm}
    \caption{Normalized POD modes on the $x$-$y$ plane for the third mode, showing normalized density, streamwise, wall-normal, and spanwise velocities, as well as temperature.} 
 \label{fig:POD_variable_mode_3}
\end{figure*}

\begin{figure*}
	\centering
	{\includegraphics[width=0.9\linewidth] {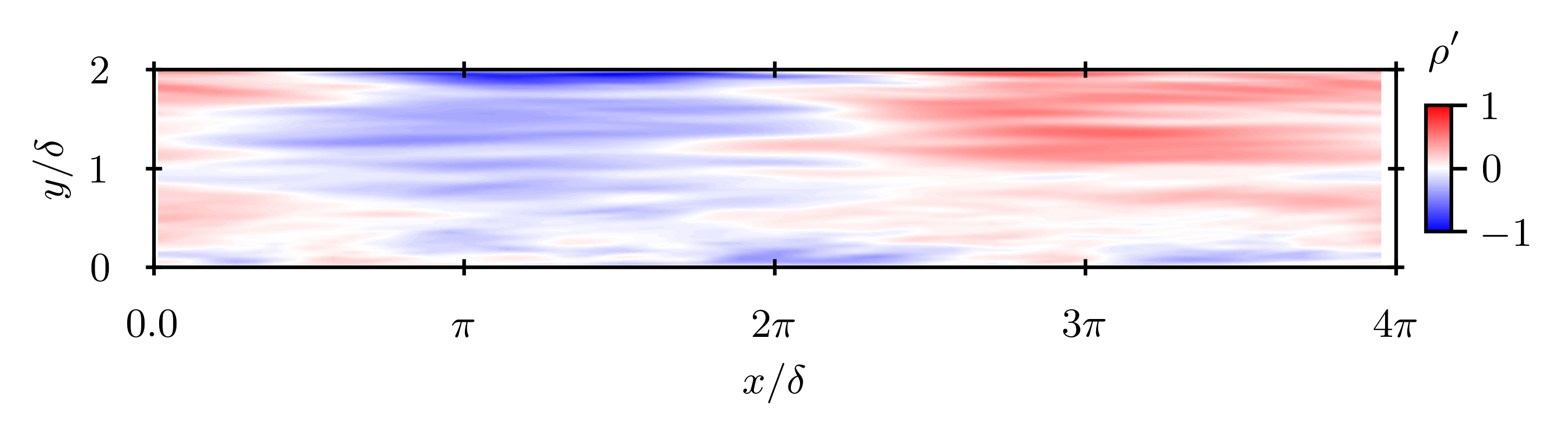}} \\ \vspace{-6.5mm}
    {\includegraphics[width=0.9\linewidth]{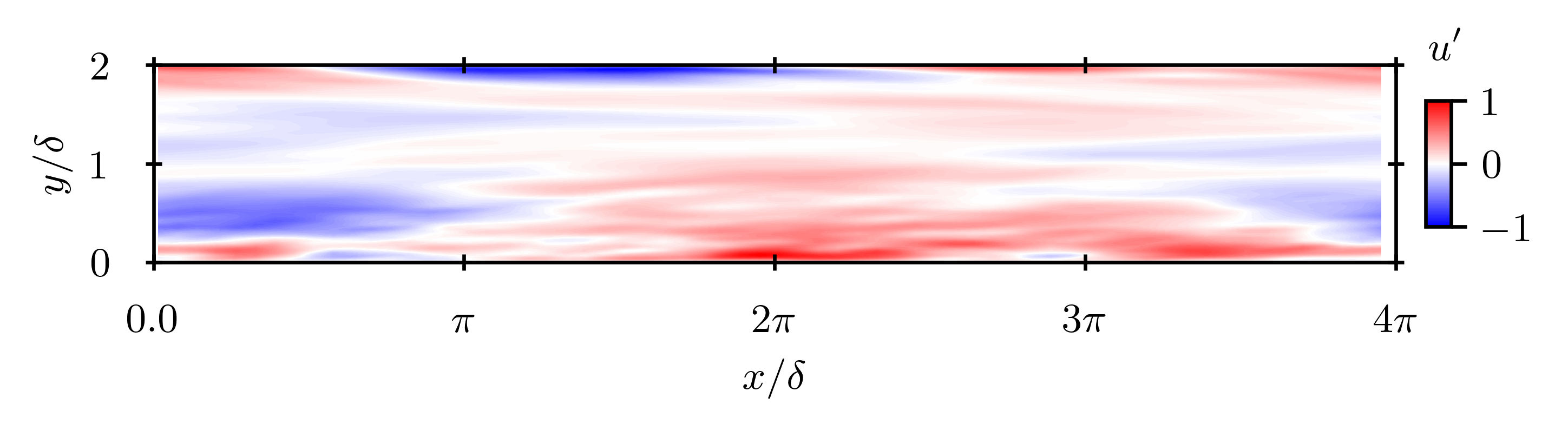}} \\ \vspace{-6mm}
    {\includegraphics[width=0.9\linewidth] {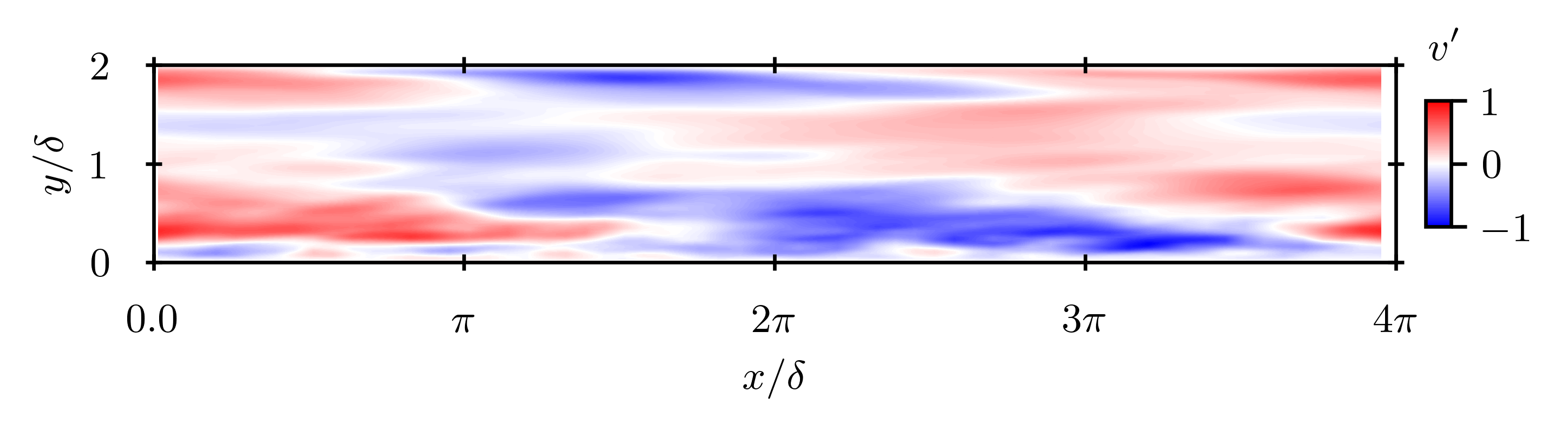}} \\ \vspace{-6mm}
    {\includegraphics[width=0.9\linewidth]{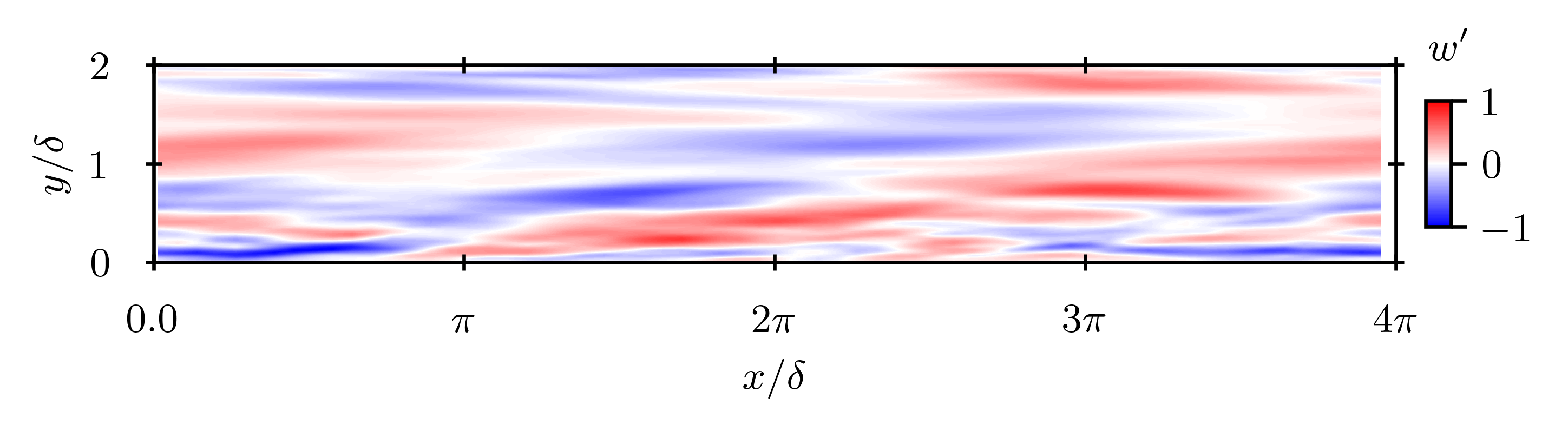}} \\ \vspace{-6mm}
    {\includegraphics[width=0.9\linewidth] {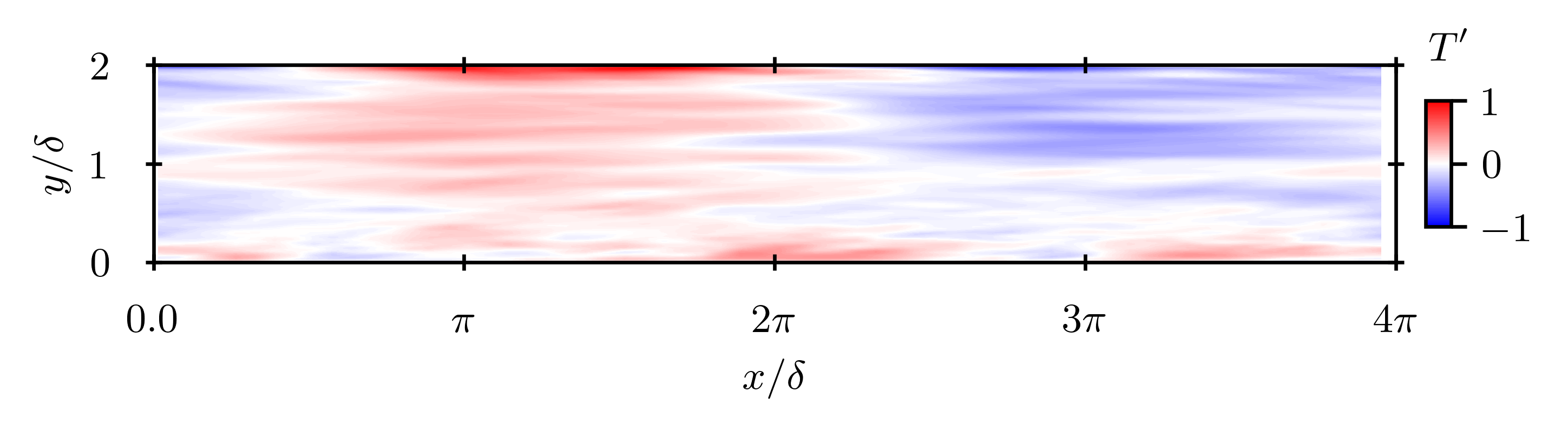}} \\ \vspace{-5mm}
    \caption{Normalized POD modes on the $x$-$y$ plane for the fourth mode, showing normalized density, streamwise, wall-normal, and spanwise velocities, as well as temperature.} 
 \label{fig:POD_variable_mode_4}
\end{figure*}

The following multi-panel figures present the third through sixth POD modes of the fluctuation variables, which were omitted from Section~\ref{sec:results} for brevity; only the first and second modes are shown in Figures~\ref{fig:POD_variable_mode_1}–\ref{fig:POD_variable_mode_2} in the main text.
Figures~\ref{fig:POD_variable_mode_3} and~\ref{fig:POD_variable_mode_4} display the third and fourth POD modes for each fluctuation variable. These higher-order modes exhibit increasingly fragmented and localized structures compared to the dominant modes discussed earlier. In particular, $\rho^\prime$ and $T^\prime$ reveal finer-scale patterns characterized by pronounced streamwise waviness and steeper gradients near the pseudo-boiling region. The velocity components (especially $u^\prime$ and $w^\prime$) demonstrate streak-like formations with noticeable phase shifts and increasing asymmetry across the channel height.
Collectively, these modes highlight the emergence of smaller-scale coherent motions and reinforce the dominant role of the pseudo-boiling region in shaping the energetic structure of the flow. The persistence of this influence across the modal hierarchy underscores the conclusion that near-wall and pseudo-boiling dynamics are key drivers of both large- and small-scale turbulence under wall-bounded high-pressure transcritical conditions.

\begin{figure*}
	\centering
    {\includegraphics[width=0.9\linewidth]{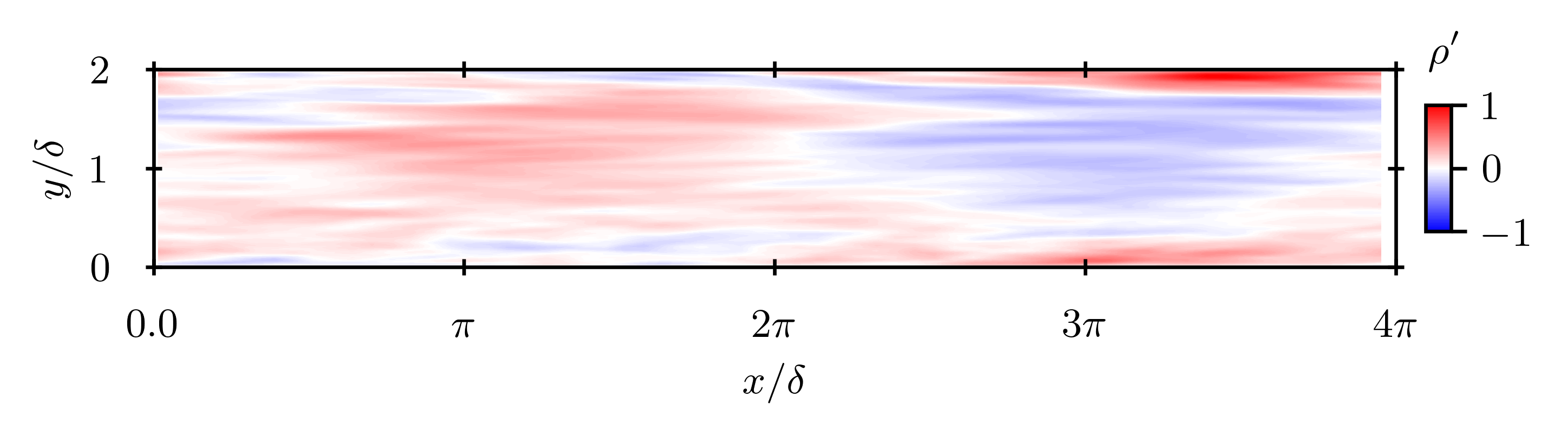}} \\ \vspace{-6.5mm}
    {\includegraphics[width=0.9\linewidth] {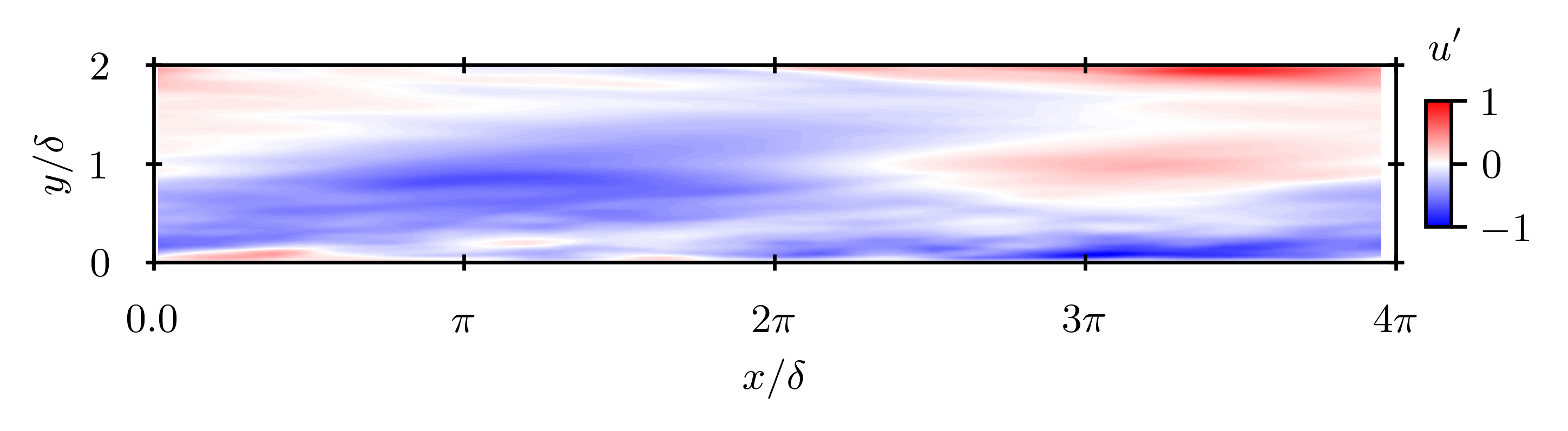}} \\ \vspace{-6mm}
    {\includegraphics[width=0.9\linewidth]{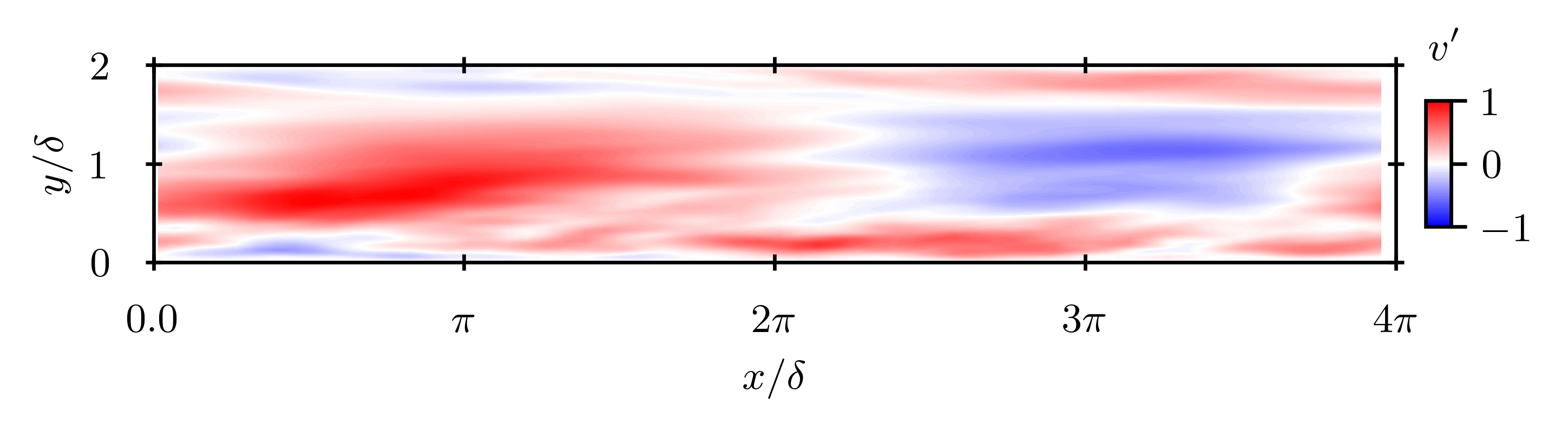}} \\ \vspace{-6mm}
    {\includegraphics[width=0.9\linewidth] {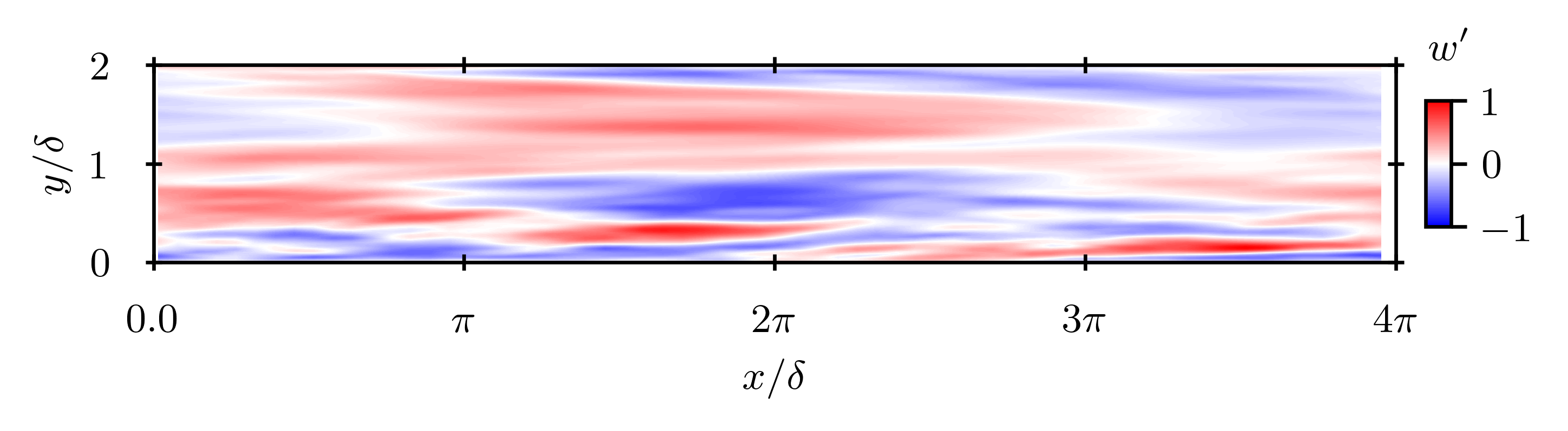}} \\ \vspace{-6mm}
    {\includegraphics[width=0.9\linewidth]{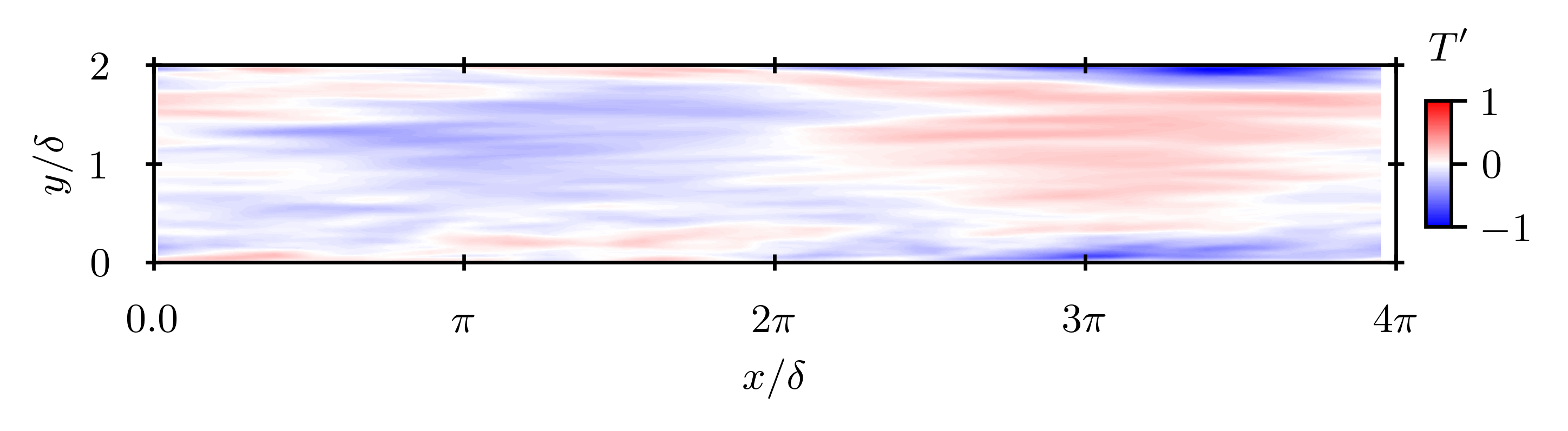}} \\ \vspace{-5mm}
	\caption{Normalized POD modes on the $x$-$y$ plane for the fifth mode, showing normalized density, streamwise, wall-normal, and spanwise velocities, as well as temperature.} 
 \label{fig:POD_variable_mode_5}
\end{figure*}

\begin{figure*}
	\centering
    {\includegraphics[width=0.9\linewidth]{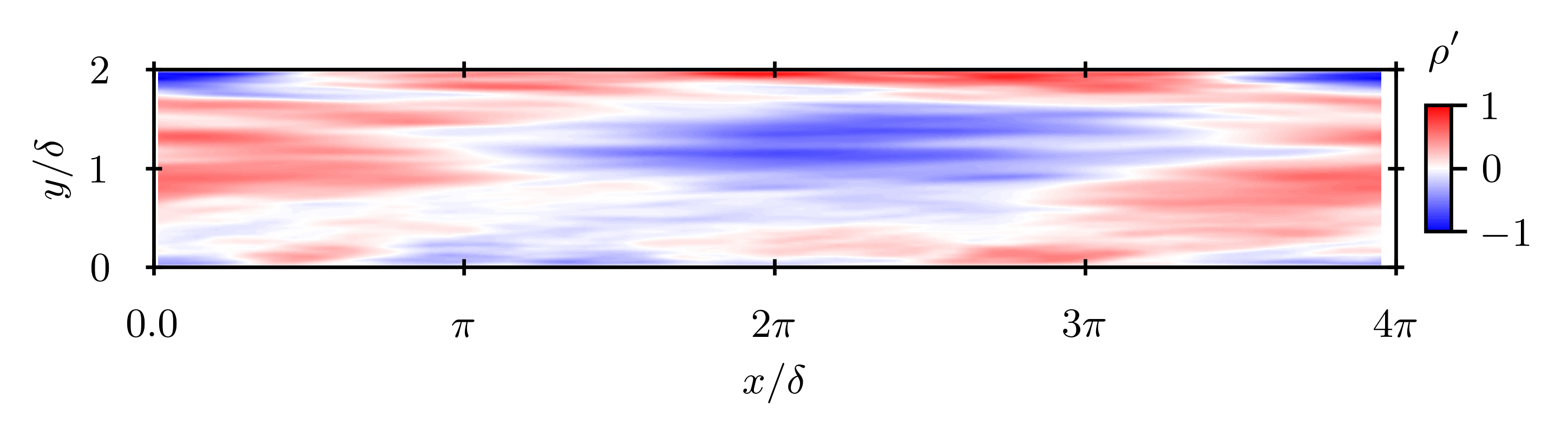}} \\ \vspace{-6.5mm}
    {\includegraphics[width=0.9\linewidth] {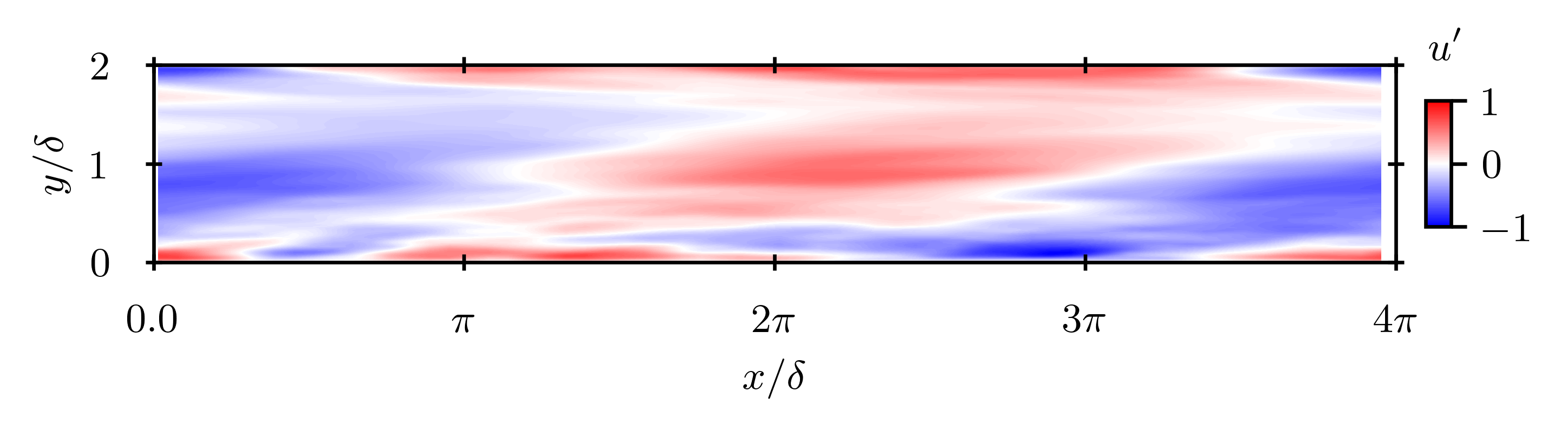}} \\ \vspace{-6mm}
    {\includegraphics[width=0.9\linewidth]{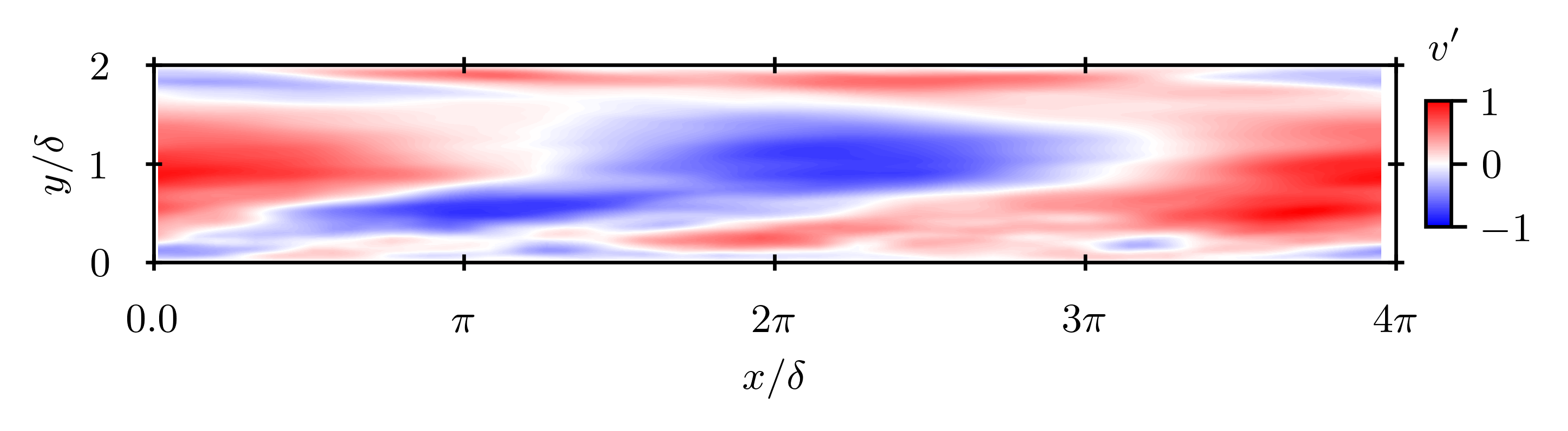}} \\ \vspace{-6mm}
    {\includegraphics[width=0.9\linewidth] {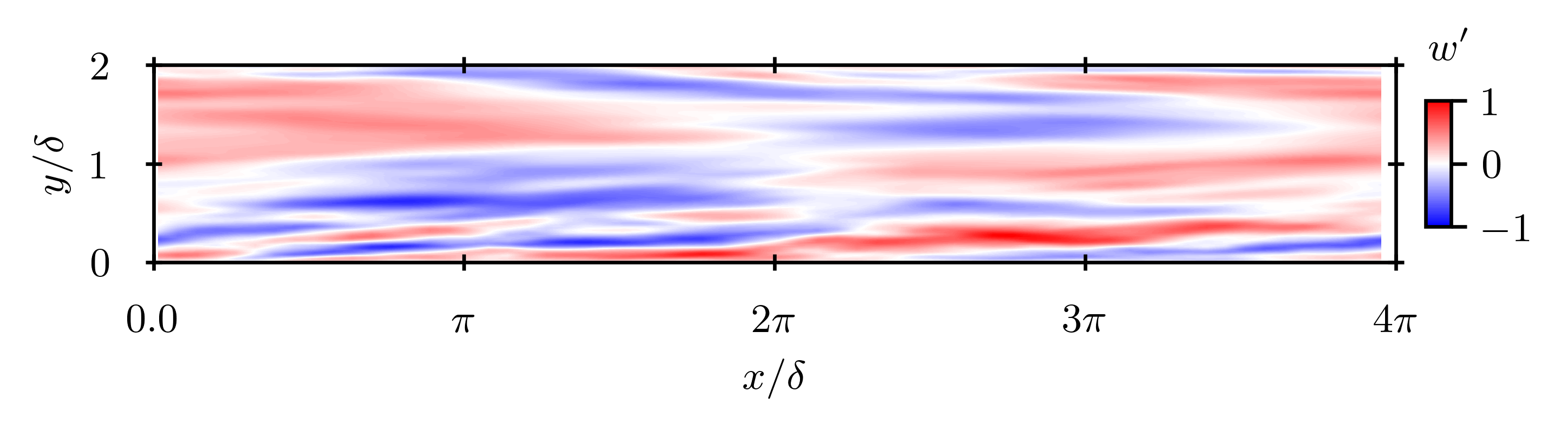}} \\ \vspace{-6mm}
    {\includegraphics[width=0.9\linewidth]{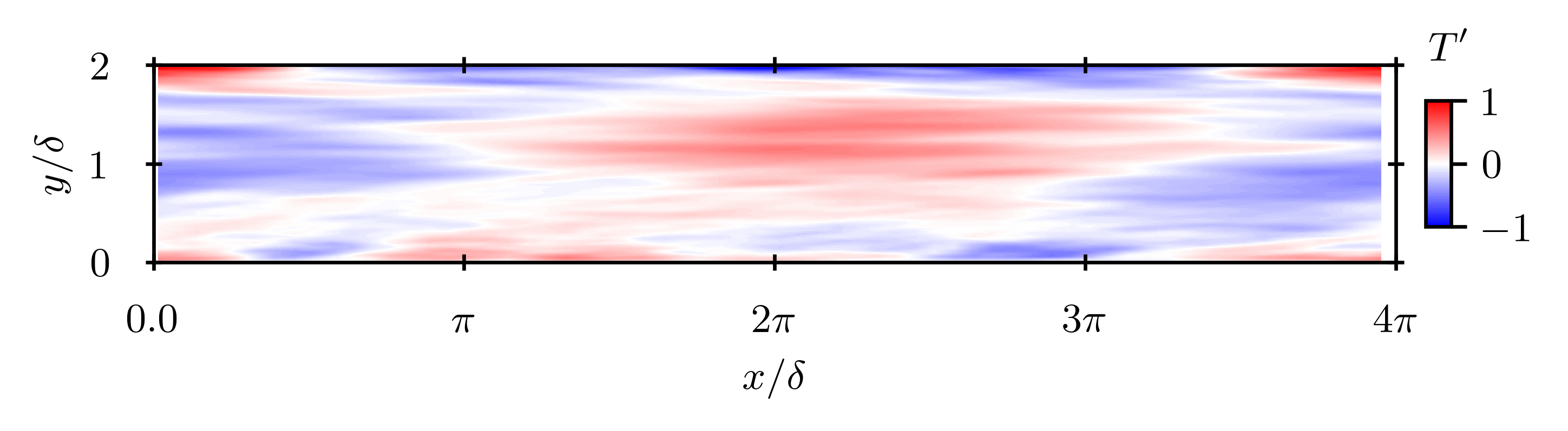}} \\ \vspace{-5mm}
	\caption{Normalized POD modes on the $x$-$y$ plane for the sixth mode, showing normalized density, streamwise, wall-normal, and spanwise velocities, as well as temperature.} 
 \label{fig:POD_variable_mode_6}
\end{figure*}

Moreover, Figures~\ref{fig:POD_variable_mode_5} and~\ref{fig:POD_variable_mode_6} present the fifth and sixth POD modes of the normalized fluctuation variables, which exhibit increasingly fragmented and spatially localized structures compared to modes $3$ and $4$. Notably, $\rho^\prime$ and $T^\prime$ continue to display pronounced asymmetry, with enhanced fine-scale patterns concentrated near the hot/top wall, reflecting the sustained influence of the pseudo-boiling region on density and thermal fluctuations. The velocity fluctuations, particularly $u^\prime$ and $v^\prime$, reveal complex streak- and wave-like structures characterized by shorter streamwise wavelengths and intensified phase shifts across the channel height. Moreover, $v^\prime$ and $w^\prime$ exhibit multiple zones of alternating positive and negative fluctuations, indicative of intricate wall-normal and spanwise motions likely linked to secondary instabilities or the breakdown of larger-scale structures seen in lower modes.
These observations confirm the progression toward richer small-scale dynamics deeper into the modal hierarchy, while underscoring the pseudo-boiling layer as a persistent organizing feature. Overall, modes $5$ and $6$ illustrate the ongoing cascade of coherent motions driven by near-wall and pseudo-boiling turbulence mechanisms, thereby extending and reinforcing the conclusions drawn from the leading energetic modes.

\section{Low-Rank POD modes and flow reconstruction} \label{sec:Appendix_C}

\begin{figure*}
	\centering
    \subfloat[]{\includegraphics[width=0.48\linewidth]{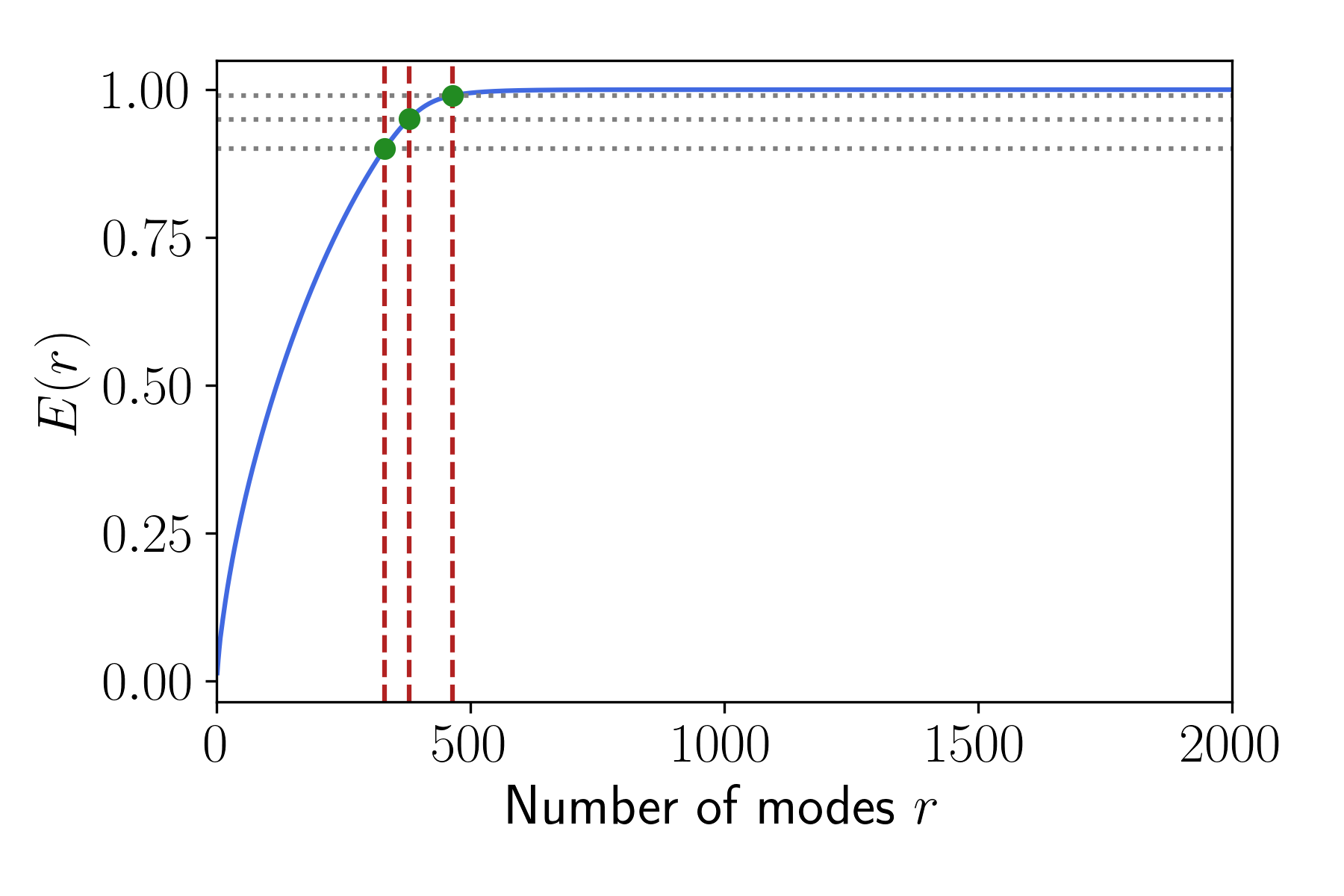}}
    \subfloat[]{\includegraphics[width=0.48\linewidth]{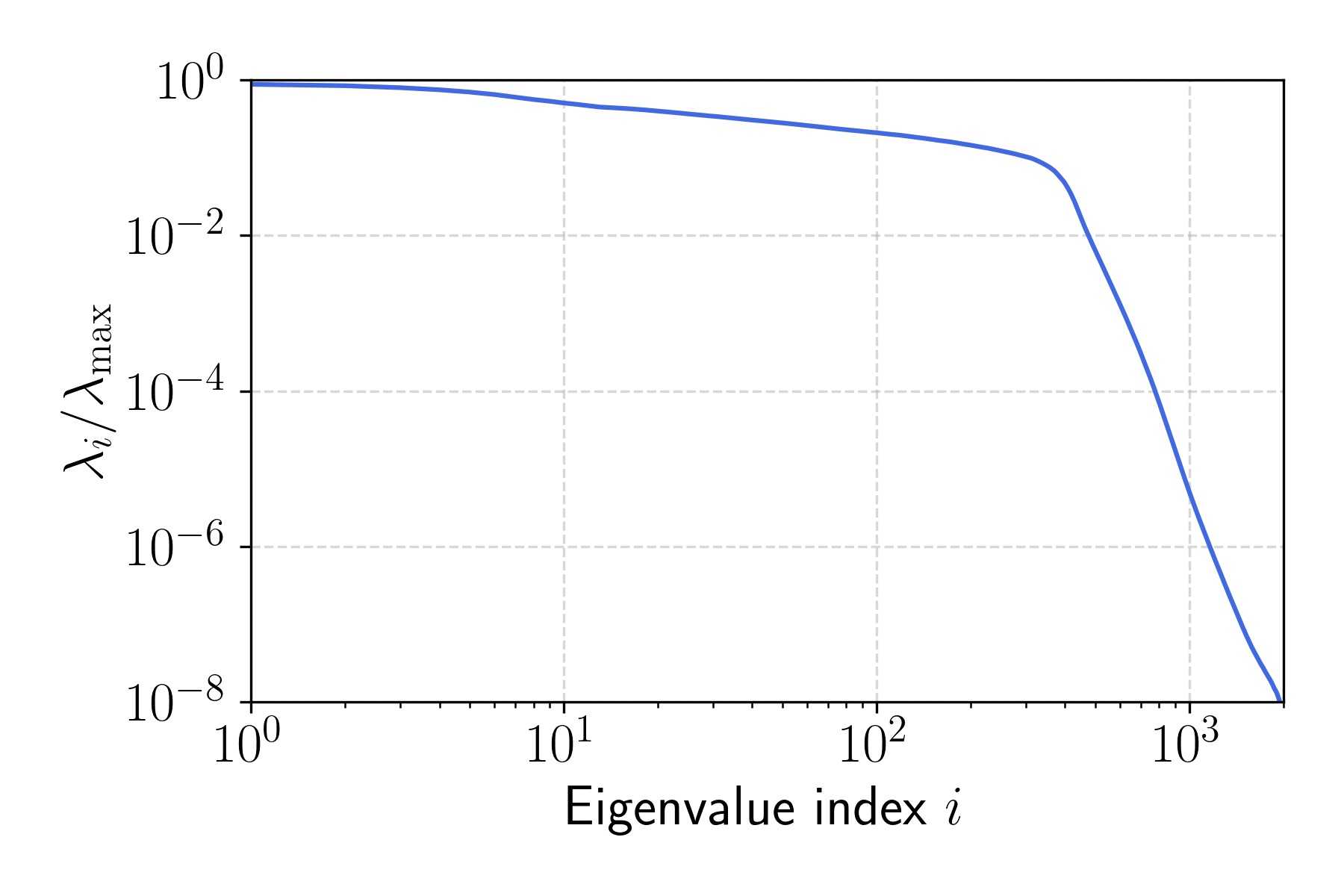}} \\  \vspace{-2mm}
	\caption{(a) Cumulative energy of the eigenvalues for all snapshots analyzed. The red vertical lines at $r = 331$, 379, 464 correspond to the horizontal gray lines at $E(r) = 0.90$, 0.95, 0.99, respectively, with their intersection points marked by green dots.(b) Eigenvalue spectrum.} 
 \label{fig:energy_convergence}
\end{figure*}

POD modes are obtained from the eigenvalues of the correlation, or covariance matrix, of the dataset. The energy associated with each mode is proportional to its eigenvalue, and the cumulative energy convergence is defined as
$\mathcal{E}(r) = (\sum_{i=1}^{r} \lambda_i)/(\sum_{i=1}^{N} \lambda_i)$, where $r$ is the truncation rank. Figure~\ref{fig:energy_convergence} illustrates this energy convergence, showing that $95\%$ of the total energy is captured by the first $r = 379$ modes.

\begin{figure*}
	\centering
	{\includegraphics[width=0.95\linewidth]{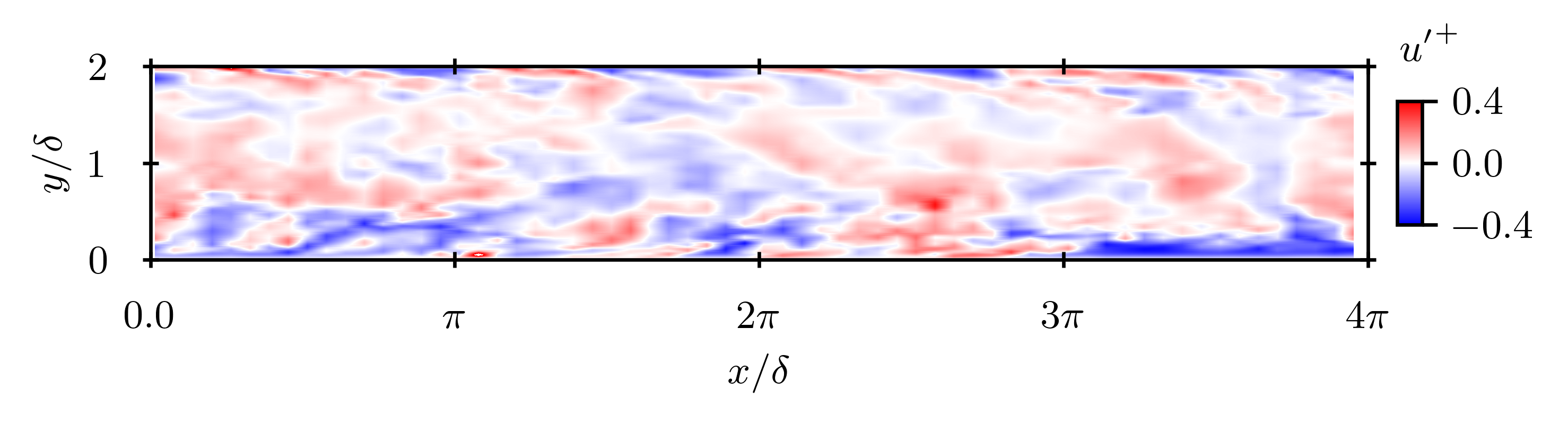}} \\ \vspace{-6mm}
    {\includegraphics[width=0.95\linewidth]{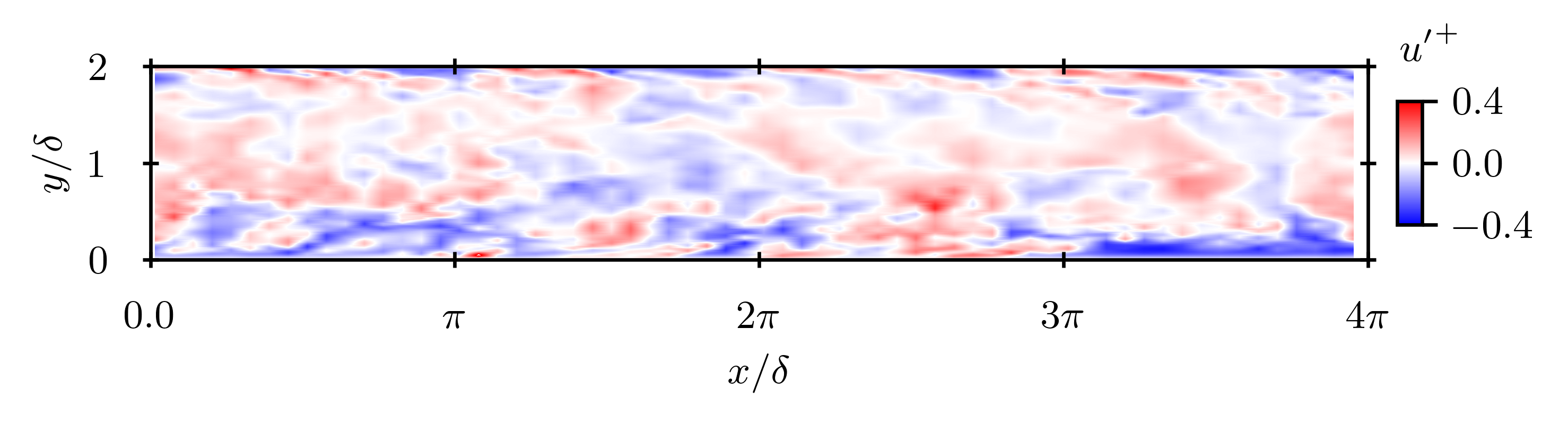}} \\ \vspace{-5mm}
	\caption{Original (top) and reconstructed (bottom) POD of the streamwise velocity on the $x$-$y$ plane for $r = 450$.} 
 \label{fig:original_reconstructed_POD}
\end{figure*}

Based on this truncation rank, the original POD field for a given snapshot is compared to the reconstructed field for the streamwise velocity component. As shown in Figure~\ref{fig:original_reconstructed_POD}, the reconstruction error remains within approximately $10\%$, which is expected given the level of energy captured at this low rank.
This discrepancy arises because energy content does not directly translate to reconstruction accuracy in turbulent flows: (i) small-scale structures, despite their low energy, contribute significantly to point-wise errors, and (ii) POD optimally minimizes the global $L^2$ projection error rather than local fluctuations or sharp gradients.
Specifically, while large-scale structures are well-represented, finer details are not fully captured. Consequently, reduced-order model (ROM) approaches are suitable for analyzing overall flow structures or performing reduced simulations but lack the precision needed to resolve instantaneous turbulent features accurately. Achieving a full-fidelity representation would require either a significantly higher number of modes, which may be computationally prohibitive, or the use of nonlinear techniques such as DMD for more efficient lower-rank reconstructions.

\bibliographystyle{jfm}

\end{document}